\documentclass[11pt,amssymb,amsfonts,amscd]{article}
\usepackage{amscd}
\usepackage{amssymb}
\usepackage{amsmath}
\usepackage{amsfonts}
\newtheorem{proposition}{Proposition}[section]
\newtheorem{lemma}[proposition]{Lemma}
\newtheorem{definition}[proposition]{Definition}
\newtheorem{theorem}[proposition]{Theorem}
\newtheorem{cor}[proposition]{Corollary}

\newtheorem{remark}[proposition]{Remark}

\newcounter{THNO}[section]
\newcounter{SNO}[section]
\newcounter{multieq}[equation]

\newcounter{tmp}

\def\h#1,#2{{\rm Hom}({#1}\:,\; {#2})}
\def\H#1,#2,#3,#4{{\rm Hom}^{#1}_{#2}({#3}\:,\; {#4})}
\def\E#1,#2,#3,#4{{\rm Ext}^{#1}_{#2}({#3}\:,\; {#4})}
\def\lto{\longrightarrow}

\def\da{\big\downarrow}

\def\bl{\boldsymbol}
\def\ot{\otimes}
\def\bt{\boxtimes}

\def\op{\oplus}

\def\wt#1{\widetilde{#1}}

\def\llongrightarrow{{\;\relbar\joinrel\relbar\joinrel\relbar\joinrel
\rightarrow\;}}

\def\A{{\mathcal{A}}}
\def\O{{\mathcal{O}}}
\def\E{{\mathcal{E}}}

\def\H{{\mathcal{H}}}

\def\L{{\mathcal{L}}}

\def\End{{\rm E}{\rm n}{\rm d}}
\def\Hom{{\rm H}{\rm o}{\rm m}}

\def\h{{\hbar}}

\def\Endo{\underline{{\mathcal E}nd}}
\def\X{{\mathcal{X}}}

\newcommand{\nn}{\nonumber}

\newcommand{\cU}{{\mathcal U}}
\def\be{\begin{equation}}
\def\ee{\end{equation}}

\newcommand{\CC}{{\mathbb C}}
\newcommand{\RR}{{\mathbb R}}
\newcommand{\ZZ}{{\mathbb Z}}

\newcommand{\NN}{{\mathbb N}}
\renewcommand{\SS}{{\mathbb S}}
\newcommand{\fH}{{\mathfrak H}}

\newcommand{\ra}{\rightarrow}

\newcommand{\id}{\mathbf{id}}

\newcommand{\Gi}{{\left(G^{-1}\right)}}
\newcommand{\vac}{{|vac\rangle}}
\newcommand{\ds}{\displaystyle}

\newcommand{\bc}{{\bar{c}}}
\newcommand{\bz}{{\bar{z}}}
\newcommand{\bw}{{\bar{w}}}
\newcommand{\bX}{{\bar{X}}}
\newcommand{\bT}{{\bar{T}}}
\newcommand{\ba}{{\bar{a}}}
\newcommand{\bj}{{\bar{j}}}
\newcommand{\bi}{{\bar{i}}}
\newcommand{\bn}{{\bar{n}}}
\newcommand{\bm}{{\bar{m}}}
\newcommand{\br}{{\bar{r}}}
\newcommand{\bq}{{\bar{q}}}
\newcommand{\bs}{{\bar{s}}}
\renewcommand{\bl}{{\bar{l}}}
\renewcommand{\bt}{{\bar{t}}}
\newcommand{\btheta}{{\bar{\theta}}}
\newcommand{\al}{{\alpha}}
\newcommand{\bal}{{\bar{\alpha}}}
\newcommand{\tpsi}{{\bar{\psi}}}
\newcommand{\hw}{{{W}}}
\newcommand{\hm}{{{M}}}
\newcommand{\bpartial}{{\bar{\partial}}}
\newcommand{\hp}{{{P}}}
\newcommand{\htp}{{{\bar{P}}}}

\newcommand{\grp}{{\mathbb C}\ [\Gamma\op\Gamma^*]}
\newcommand{\cA}{{\mathcal A}}
\newcommand{\cI}{{\mathcal I}}
\newcommand{\cJ}{{\mathcal J}}
\newcommand{\cG}{{\mathcal G}}
\newcommand{\cM}{{\mathcal M}}
\newcommand{\cN}{{\mathcal N}}
\newcommand{\cR}{{\mathcal R}}
\newcommand{\cB}{{\mathcal B}}
\newcommand{\cH}{{\mathcal H}}
\newcommand{\cT}{{\mathcal T}}
\newcommand{\cV}{{\mathcal V}}
\newcommand{\cF}{{\mathcal F}}
\newcommand{\cP}{{\mathcal P}}
\newcommand{\Be}{{\mathcal B}}

\newcommand{\tL}{{\bar{L}}}

\newcommand{\tc}{{\bar{c}}}
\newcommand{\tJ}{{\bar{J}}}
\newcommand{\tQ}{{\bar{Q}}}
\newcommand{\B}{B}
\newcommand{\bb}{{b}}
\newcommand{\rank}{{\rm rank}}

\newcommand{\bgamma}{{\bar\gamma }}
\newcommand{\bpsi}{{\bar\psi}}
\newcommand{\bx}{{\bar{x}}}
\newcommand{\bL}{{\bar L}}
\newcommand{\bQ}{{\bar Q}}
\newcommand{\bJ}{{\bar J}}
\newcommand{\cW}{{\mathcal W}}
\newcommand{\bN}{{\bar N}}
\newcommand{\bF}{{\bar F}}
\newcommand{\bY}{{\bar Y}}

\title{Vertex Algebras, Mirror Symmetry, And D-Branes:\\
The Case Of Complex Tori}

\author{Anton Kapustin\thanks{%
School of Natural Sciences, Institute for Advanced Study,
Olden Lane, Princeton, NJ 08540 ,
E-mail: kapustin@ias.edu }
\and Dmitri Orlov%
\thanks{%
Algebra Section, Steklov Mathematical Institute,
Russian Academy of Sciences,
 8 Gubkin str., GSP-1, Moscow 117966, Russia,
E-mail: orlov@mi.ras.ru}}
\date{}
\begin{document}

\begin{titlepage}

\maketitle

\begin{sloppypar}
\vspace{-7.5cm}
\centerline{{\small \hfill{IASSNS--HEP--00/81}}
}
\end{sloppypar}

\vspace{7.5cm}
\begin{abstract}

\noindent

A vertex algebra is an algebraic counterpart of a two-dimensional
conformal field theory.
We give a new definition of a vertex algebra which
includes chiral algebras as a special case, but allows for fields which
are neither meromorphic nor anti-meromorphic.
To any complex torus equipped with a flat K\"ahler metric and a closed 2-form we
associate an $N=2$ superconformal vertex algebra ($N=2$ SCVA) in the sense of our
definition.
We find a criterion for two
different tori to produce isomorphic $N=2$ SCVA's. We show that for algebraic
tori isomorphism of $N=2$ SCVA's implies the
equivalence
of the derived categories of coherent sheaves corresponding to the tori or
their noncommutative generalizations (Azumaya algebras over tori).
We also find a criterion for
two different tori to produce $N=2$ SCVA's related by a mirror
morphism. If the 2-form is of type $(1,1),$ this condition is identical to the one
proposed by Golyshev, Lunts, and Orlov, who used an entirely different approach
inspired by the Homological Mirror Symmetry Conjecture of Kontsevich. Our results
suggest that Kontsevich's conjecture must be modified: coherent sheaves must be
replaced with modules over Azumaya algebras, and the Fukaya category must be ``twisted''
by a closed 2-form. We also describe the implications of our results for BPS D-branes
on Calabi-Yau manifolds.

\end{abstract}

\end{titlepage}

\tableofcontents

\section{Introduction}

\subsection{Physicist's mirror symmetry}

{\it A physicist's Calabi Yau} is a triple $(X,G,\Be),$ where $X$
is a compact complex manifold with a trivial canonical bundle, $G$ is a Ricci-flat
K\"ahler metric on $X,$ and $\Be$ is a class in $H^2(X,\RR/\ZZ)$ which is in the kernel
of the Bockstein homomorphism $H^2(X,\RR/\ZZ)\ra H^3(X,\ZZ).$
The class $\Be$ can be lifted to a class $\bb\in H^2(X,\RR),$ and a closed
2-form $\B$ representing it is known as the $B$-field.

Physicists believe that there is a procedure which associates to any
such triple an $N=2$ superconformal vertex algebra ($N=2$ SCVA). The precise
definition of an $N=2$ SCVA is rather complicated and will be given in
Section~\ref{sec:general}. Roughly speaking, it is a Euclidean quantum field theory on a
two-dimensional manifold $\RR\times\SS^1$ whose Hilbert space is acted upon
by a unitary representation of the infinite-dimensional Lie super-algebra with
even generators $L_n,\bL_n,J_n,\bJ_n, n\in\ZZ,$ odd generators
$Q^\pm_r,\bQ^\pm_r, r\in\ZZ+\frac{1}{2},$
and the following nonvanishing Lie brackets:
\begin{equation}
\label{ntwovir}
\begin{array}{llll}
{}[L_m,L_n]&\multicolumn{3}{l}{\ds =(m-n)L_{m+n}+\frac{d}{4} (m^3-m)
\delta_{m,-n}, }\\
{}[\bL_m, \bL_n]&\multicolumn{3}{l}{\ds =(m-n)\bL_{m+n}+ \frac{d}{4} (m^3-m)
\delta_{m,-n},} \\
{}[L_m,J_n]&=-nJ_{n+m}, &
\hspace{3.9cm}[\bL_m,\bJ_n]&=-n\bJ_{n+m}\\
{}[J_m,J_n]&=dm\delta_{m,-n}, &
\hspace{3.9cm}[\bJ_m,\bJ_n]&=dm\delta_{m,-n}\\
{}[L_m, Q^\pm_r]&=\ds\left(\frac{m}{2}-r\right)Q^\pm_{r+m}, &
\hspace{3.9cm}[\bL_m, \bQ^\pm_r]&=\ds\left(\frac{m}{2}-r\right)\bQ^\pm_{r+m}\\
{}[J_m,Q^\pm_r]&=\pm Q^\pm_{r+m}, &
\hspace{3.9cm}[\bJ_m,\bQ^\pm_r]&=\pm \bQ^\pm_{r+m},
\\
\left\{Q^+_r,Q^-_s\right\}&
\multicolumn{3}{l}{\ds =
\frac{1}{4}L_{r+s}+\frac{1}{8}(r-s)J_{r+s}+
\frac{d}{8}\left(r^2-\frac{1}{4}\right)\delta_{r,-s},}\\
\left\{\bQ^+_r,\bQ^-_s\right\}
&\multicolumn{3}{l}{\ds =
\frac{1}{4}\bL_{r+s}+\frac{1}{8}(r-s)\bJ_{r+s}+
\frac{d}{8}\left(r^2-\frac{1}{4}\right)\delta_{r,-s}.}
\end{array}
\end{equation}
Here $d=\dim_\CC X,$ and $\{\cdot,\cdot\}$ denotes the anti-commutator.
This algebra is a direct sum of two copies of the celebrated
$N=2$ super-Virasoro algebra with central charge $c=3d.$ If one omits $J_n,\bJ_n$
and all the odd generators from the definition of the $N=2$ SCVA, one gets a structure
which we call a conformal vertex algebra (CVA), and which is also known as
a conformal field theory on $\RR\times\SS^1\cong\CC^*.$ Thus an $N=2$ SCVA is a conformal
field theory on $\CC^*$ with some additional structure.

Heuristically, the construction of an $N=2$ SCVA
from a triplet $(X,G,\Be)$ proceeds as follows. To any K\"ahler manifold $(X,G)$
equipped with a closed 2-form $\B$ one can
associate a two-dimensional classical field theory on $\RR\times\SS^1$, the so-called 
$N=2$ supersymmetric
$\sigma$-model. For reader's convenience, the definition of the $\sigma$-model
is given in Appendix~\ref{sigmamodel}.
The space of solutions of the corresponding classical equations of motion is
an infinite-dimensional symplectic supermanifold with a symplectic action of
two copies of the $N=2$ super-Virasoro algebra with zero central charge
(see~\cite{Polch,IASbook}, and Appendix~\ref{sigmamodel}).
It can be argued that consistent quantization of this classical field theory is
possible only for $c_1(T_X)\ge 0$, e.g. when $X$ is a Fano manifold or
a Calabi-Yau manifold. In the Fano case ($c_1(T_X)>0$) the quantized $\sigma$-model
is an $N=2$ field theory, but not a superconformal one, because only a finite-dimensional
subalgebra of the classical $N=2$ super-Virasoro algebra survives quantization.
The same happens if $c_1(T_X)=0$ but $G$ is not Ricci-flat.
If $c_1(T_X)=0$ and $G$ is Ricci flat, both $N=2$ super-Virasoro algebras survive
quantization (though the central charges become nonzero), and therefore
the quantized $\sigma$-model is an $N=2$ superconformal field
theory, i.e. an $N=2$ SCVA. One can also argue that this $N=2$ SCVA in fact depends only on the
image of $\B$ in $H^2(X,\RR/\ZZ),$ i.e. on $\Be.$

The actual quantization of the $\sigma$-model is
feasible only for very special $(X,G,\Be).$ In particular, if $X$ is a complex
torus, the corresponding $N=2$ SCVA can be constructed for any flat $G$ and any
$\Be\in H^2(X,\RR/\ZZ).$ The quantization of the $\sigma$-model for a flat complex
torus is sketched in Appendix~\ref{sigmamodel}.

Two physicist's Calabi-Yaus are said to be mirror if there exists an isomorphism
of the corresponding conformal vertex algebras
which acts on the algebra~(\ref{ntwovir}) as the so-called {\it mirror involution}:
\begin{align}\label{mirrorinv}
&L_n\ra L_n,\quad Q^\pm_r\ra Q^\mp_r,\quad J_n\ra -J_n,\\
&\bL_n\ra \bL_n, \quad \bQ^\pm_r\ra \bQ^\pm_r,\quad
\bJ_n\ra \bJ_n. \nn
\end{align}
Such a morphism of $N=2$ SCVA's will be called a {\it mirror
morphism.}

Mirror symmetry defined in this way acts pointwise on the moduli space of
physicist's Calabi-Yaus. If one drops $G$ and $B$
from the definition of a physicist's Calabi-Yau, then mirror symmetry becomes
a correspondence between two {\it families}
of K\"ahler manifolds with a trivial canonical bundle whose Hodge numbers are
related by $h^{p,q}={h'}^{d-p,q}.$
The latter notion of mirror symmetry is much weaker than the physicist's
mirror symmetry. Nevertheless, much of the mathematical work on mirror symmetry
up to now has focused on this weaker notion, since it proved hard to make sense
of the $\sigma$-model.

As mentioned above, the quantum $\sigma$-model is manageable when $X$ is a
complex torus, so one could hope to understand mirror symmetry in detail
in this particular case. This is what this paper aims to do.
Although from the physical point of view mirror symmetry for complex tori appears to
be rather trivial, we will see that its study sheds considerable light
on the Homological Mirror Symmetry Conjecture (HMSC), a subject to which we now turn.

\subsection{Homological mirror symmetry}

String theory makes highly nontrivial predictions about
the enumerative geometry of a Calabi-Yau $X$ in terms of its mirror $X'.$
The success of these predictions seems quite mysterious
from a purely mathematical standpoint.
In an insightful paper~\cite{K}, M.~Kontsevich formulated a conjecture which relates
the properties of a Calabi-Yau with those of its mirror and suggested that it
captures the essence of mirror symmetry. Subsequently this conjecture was reinterpreted in
physical terms~\cite{Vafa}. In this subsection we remind the main
features of Kontsevich's conjecture.

Let $X$ be a complex algebraic variety (or a complex manifold). Denote
by $\O_{X}$ the sheaf of regular functions
(or the sheaf of holomorphic functions). Recall that a coherent
sheaf is a sheaf of $\O_X$\!--modules that locally can be
represented as a cokernel of a morphism of holomorphic vector bundles.
Coherent sheaves form an abelian category which will be denoted by $Coh(X).$
To any abelian category we can associate a certain triangulated category
called the bounded derived category.
We denote by $D^b(X)$ the bounded derived category of coherent sheaves on $X.$
Roughly speaking, the category $D^b(X)$ is a factor-category
of the category of bounded complexes of coherent sheaves by
the subcategory of acyclic complexes (i.e. complexes with
trivial cohomology sheaves).

On the other hand, it has been proposed~\cite{Fukaya,K} that to
any compact symplectic manifold $Y$ one can associate a certain
category whose objects are (roughly speaking)
vector bundles on Lagrangian submanifolds equipped
with unitary flat connections.
The morphisms in this category have been defined when Lagrangian
submanifolds  intersect transversally. This conjectural category
is called the Fukaya category and denoted $\cF(Y).$
The category $\cF(Y)$ is not an
abelian category; rather, it is supposed to be an $A_\infty$\!-category
equipped with a shift functor. For an introduction to $A_\infty$\!-categories
see~\cite{Ainf}.
An $A_{\infty}$\!--category is not a category in the usual
sense, because the composition of morphisms is not associative.
The set of morphisms between two objects in an $A_{\infty}$\!--category is a differential
graded vector space.
To any $A_{\infty}$\!--category one can associate a true category
which has the same objects but the space of morphisms between
two objects is the $0$\!--th
cohomology group of the morphisms in $A_{\infty}$\!--category.
Applying this construction to $\cF(Y),$ we obtain a true category
$\cF_0(Y)$ which is also called the Fukaya category. Kontsevich~\cite{K}
also constructs a certain triangulated category $D\cF_0(Y)$ out of $\cF(Y).$
We will call it the derived Fukaya category.
Conjecturally, the category $\cF_0(Y)$ is a full subcategory of
$D\cF_0(Y).$

A physicist's Calabi-Yau $(X,G,\Be)$ is both a complex manifold and a symplectic
manifold (the symplectic form being the K\"ahler form $\omega=GI$). Thus we can
associate to it a pair of triangulated categories $D^b(X)$ and
$D\cF_0(X).$ The Homological Mirror Symmetry Conjecture (HMSC) asserts
that if two algebraic Calabi-Yaus $(X,G,\Be)$ and $(X',G',\Be')$ are
mirror to each other, then $D^b(X)$ is equivalent to $D\cF_0(X'),$
and vice versa.

The Homological Mirror Symmetry Conjecture can be reinterpreted in physical terms.
To every $N=2$ superconformal field theory one can associate the set of BPS D-branes,
or more precisely two sets: the set of A-type D-branes and the set of B-type D-branes.
This is reviewed in more detail in Section~\ref{modHMSC}.
These sets are equipped with a rather intricate
algebraic structure: that of an $A_\infty$\!--category. This structure encodes the
properties of correlators in a topological open string theory~(see \cite{topstring} and
references therein).
A mirror morphism between
two $N=2$ superconformal field theories identifies the A-type D-branes of the first
theory with the B-type D-branes of the second theory, and vice versa.
Now suppose that an $N=2$ superconformal field theory originates from a physicist's
Calabi-Yau $(X,G,\Be).$ In this case there is evidence that A-type D-branes are closely related
to objects of the Fukaya category, while B-type D-branes are
related to coherent sheaves on $X.$ To prove the Homological Mirror Symmetry Conjecture
it would be sufficient to show that the derived Fukaya category of $(X,G,\Be)$ (resp.
the derived category of $X$) can be recovered from the $A_\infty$\!-category of A-type D-branes
(resp. B-type D-branes). Conversely, proving the HMSC would likely result in an improved
understanding of BPS D-branes.

So far the Homological Mirror Symmetry Conjecture (with some important modifications,
see Section~\ref{modHMSC} for details) has been proved only for
$\dim_\CC X=\dim X'_\CC=1,$ i.e. for the elliptic curve~\cite{PoliZas}. Two features make
this case particularly manageable.
First, the $N=2$ SCVA for the elliptic curve is known, so one knows the precise
conditions under which $(X,G,\Be)$ is mirror to $(X',G',\Be').$ Second, all
objects and morphisms in the Fukaya category can be explicitly described.

In this paper we perform a check of the HMSC for the case when both $X$ and
$X'$ are algebraic tori of arbitrary dimension.
We will see that for algebraic tori of dimension higher than one the HMSC as formulated
by Kontsevich can not be true in general. The main reason is that both the derived
category of coherent sheaves and the derived Fukaya category do not depend on the
B-field, while in the physical mirror symmetry it plays an essential role.
However, a certain modification of the HMSC which takes into
account the B-field passes our check and has a good chance to be correct.
This modification is suggested
both by our results on the $N=2$ SCVA for complex tori, and by consideration of
BPS D-branes. The modified HMSC conjecture is formulated in
Section~\ref{modHMSC}. It reduces to the original HMSC when
the B-field vanishes for both manifolds related by the mirror morphism.

The implications of our results for BPS D-branes on Calabi-Yau manifolds are briefly
described in Section~\ref{results} and in more detail in Section~\ref{modHMSC}.

\subsection{Vertex algebras and chiral algebras}

Vertex algebras play a key role in physicist's mirror symmetry. A vertex algebra
is an algebraic counterpart of a two-dimensional conformal field theory.
In the mathematical literature the terms vertex algebra and chiral algebra
are used interchangeably. Roughly speaking, a chiral algebra is a vector superspace
$V$ together with a map $Y: V\ra \End V[[z,z^{-1}]]$ satisfying a number of
properties~\cite{Kac}. One says that $Y$ maps states to quantum fields.
The definition of a chiral algebra first appeared in the
work of Borcherds~\cite{Borch}, but its origins go back to the classic paper
of Belavin, Polyakov, and Zamolodchikov~\cite{BPZ} where an algebraic approach
to two-dimensional conformal field theory was proposed.

{}From a physical viewpoint, chiral algebras are conformal
field theories such that all fields are meromorphic (do not depend on $\bz$).
Only very special conformal field theories have this property. Moreover,
a generic conformal field theory does not factorize as a tensor product of
two chiral algebras, one depending on $z$ and another on $\bz,$ despite
some claims to the contrary in the physics literature.  For example,
the quantization of the $\sigma$-model associated to a flat torus yields a
conformal field theory which factorizes in this
manner only for very special values of $G$ and $B.$

Thus in order to give a precise meaning to physicist's mirror symmetry,
we need to find a sufficiently general definition of a vertex algebra
allowing for fields which depend both on $z$ and $\bz.$ To avoid
confusion, we will refer to these more general objects as vertex algebras,
while vertex algebras in the sense of~\cite{Kac} will be called
chiral algebras.

Once both $z$ and $\bz$ are allowed, they need not enter only in
integer powers, so $Y$ will take values in a space of ``fractional power series
in $z$ and $\bz$ with coefficients in $\End(V)$'', rather than in
$\End(V)[[z,\bz,z^{-1},\bz^{-1}]].$ The necessity  of fractional powers can
be seen by inspecting the conformal field theories associated to flat tori.
Because of this, the definition of a vertex algebra is not a trivial extension
of the definition of a chiral algebra.

We hope that our definition of a vertex algebra will be of some interest to physicists
as well as mathematicians. Its advantage over the more standard definitions of conformal
field theory
is that it is purely algebraic and  based on the notion of Operator Product
Expansion (OPE). In contrast, other rigorous definitions take Wightman axioms as a
starting point. These axioms have an analytic flavor and do not make reference to
OPE. In fact, the existence of OPE does not follow from Wightman axioms (except
in some very special cases), and has to be postulated separately. Another advantage of our
definition is that it does not require an inner product on the state space. Thus it is
capable of describing "non-unitary" conformal field theories which find applications
in statistical mechanics.

\section{Summary of results}\label{results}

\subsection{Physicist's mirror symmetry for complex tori}

Let $T$ be a $2d$-dimensional real torus $U/\Gamma,$ where $U\cong\RR^{2d}$ is
a real vector space, and $\Gamma\cong\ZZ^{2d}$ is a lattice in $U.$ Let $I$ be a
(constant) complex structure on $T,$
$G$ be a flat K\"ahler metric on $T,$ and $\bb\in H^2(T,\RR).$
We will represent $\bb$
by a constant 2-form $B$ which is uniquely determined by $\bb.$ In this simple case
there is
a well-known explicit construction of the corresponding $N=2$ SCVA which we
denote $Vert(\Gamma,I,G,\B).$ We review this
construction in Section~\ref{torusSCVA}.
The relation of this construction to the quantized $\sigma$-model is
explained in Appendix~\ref{sigmamodel}.

Our first result describes when two different quadruples $(\Gamma,I,G,\B)$ and
$(\Gamma',I',G',\B')$ yield isomorphic $N=2$ SCVA's. To state it,
we first introduce some notation. Let $\Gamma^*=\Hom(\Gamma,\ZZ)$ be the dual lattice
in $U^*,$ and $T^*$ be the dual torus $U^*/\Gamma^*.$
There is natural pairing $l:\Gamma\op\Gamma^*\ra \ZZ.$ There is also
a natural $\ZZ$-valued symmetric bilinear form $q$ on $\Gamma\op\Gamma^*$ defined by
$$
q((w_1,m_1),(w_2,m_2))=l(w_1,m_2)+l(w_2,m_1), \qquad
w_{1,2}\in\Gamma,\ m_{1,2}\in\Gamma^*.
$$
Given $G,I,\B,$ we can define
two complex structures on $T\times T^*$:
\begin{align}
\cI(I,\B)&=\begin{pmatrix} I & 0 \\ \B I+I^t\B & -I^t \end{pmatrix}, \\
\cJ(G,I,\B)&=\begin{pmatrix} -IG^{-1}\B & IG^{-1} \\ GI-\B IG^{-1}\B & \B IG^{-1}
\end{pmatrix}.
\end{align}
The notation here is as follows. We regard $\cI$ and $\cJ$ as endomorphisms of
$U\op U^*,$ and write the corresponding matrices in the basis in which the first $2d$
vector span $U,$ while the remaining vectors span $U^*.$ In addition, $G$ and $B$ are
regarded as elements of $\Hom(U,U^*),$ and $I^t$ denotes the endomorphism of $U^*$
conjugate to $I.$

It is easy to see that $\cJ$ depends on $G,I$ only in the combination $\omega=GI,$
i.e. it depends only on the symplectic structure on $T$ and the B-field.
There is also a third natural complex structure $\tilde{\cI}$ on $T\times T^*,$
which is simply the
complex structure that $T\times T^*$ gets because it is a Cartesian product of two
complex manifolds:
$$
\tilde{\cI}=\begin{pmatrix} I & 0 \\ 0 & -I^t \end{pmatrix}.
$$
This complex structure will play only a minor role in what follows.
Note that $\cI$ coincides with $\tilde{\cI}$ if and only if $\B^{(0,2)}=0.$

\begin{theorem}\label{ntwoiso}
$Vert(\Gamma,I,G,\B)$ is isomorphic to $Vert(\Gamma',I',G',\B')$ if and only if
there exists an isomorphism of lattices $\Gamma\op\Gamma^*$ and $\Gamma'\op\Gamma^{'*}$
which takes $q$ to $q',$ $\cI$ to $\cI',$ and $\cJ$ to $\cJ'.$
\end{theorem}

Our second result describes when $(T,I,G,\B)$ is mirror to $(T',I',G',\B').$
\begin{theorem}\label{ntwomirror}
$Vert(\Gamma,I,G,\B)$ is mirror to $Vert(\Gamma',I',G',\B')$ if and only if
there is an isomorphism of lattices $\Gamma\op\Gamma^*$ and $\Gamma'\op\Gamma^{'*}$
which takes $q$ to $q',$ $\cI$ to $\cJ',$ and $\cJ$ to $\cI'.$
\end{theorem}

\subsection{Applications to homological mirror symmetry}

Let us now explain the implications of these results for the HMSC.
First, note that if both $\B$ and $\B'$ are of type $(1,1),$ the criterion for mirror
symmetry is identical to the one proposed in~\cite{GLO}. In that work,
this criterion was taken as a {\it definition} of mirror symmetry for algebraic
tori. We now see that this definition agrees with the physical notion of
mirror symmetry and can be generalized to non-algebraic tori and
arbitrary $B$-fields.

Second, Theorem~\ref{ntwoiso} allows us to make a check
of the HMSC. Suppose the
tori $(T_1,I_1,G_1,\B_1)$ and $(T_2,I_2,G_2,\B_2)$
are both mirror to $(T',I',G',\B').$
Then  $Vert(\Gamma_1,I_1,G_1,\B_1)$ is isomorphic to
$Vert(\Gamma_1,I_1,G_1,\B_1),$
and by Theorem~\ref{ntwoiso} there is an isomorphism of lattices
$\Gamma_1\op\Gamma_1^*$ and $\Gamma_2\op\Gamma_2^*$
which intertwines $q_1$ and
$q_2,$ $\cI_1$ and $\cI_2,$ and $\cJ_1$ and $\cJ_2.$

On the other hand, if we assume that both $(T_1,I_1)$
and $(T_2,I_2)$ are algebraic, then
HMSC implies that $D^b((T_1,I_1))$ is equivalent to $D^b((T_2,I_2)).$
The criterion for this equivalence is known~\cite{Po,Or}: it requires the
existence of an isomorphism
of $\Gamma_1\op\Gamma_1^*$ and $\Gamma_2\op\Gamma_2^*$ which intertwines $q_1$
and $q_2,$ and $\tilde{\cI_1}$ and $\tilde{\cI_2}.$ Clearly, since $\cI\neq
\tilde{\cI}$ in general, this condition is in conflict
with the one stated in the end of the previous paragraph.
Instead, we only have the following result:
\begin{cor}\label{oneone}
If $Vert(\Gamma_1,I_1,G_1,\B_1)$ is isomorphic to $Vert(\Gamma_2,I_2,G_2,\B_2),$ both
$(T_1,I_1)$ and $(T_2,I_2)$ are algebraic,
and both $\B_1$ and $\B_2$ are of type $(1,1),$ then $D^b((T_1,I_1))$ is equivalent
to $D^b((T_2,I_2)).$
\end{cor}
In Section~\ref{morphisms} we also prove the following result.
\begin{theorem}
Let $(T_1,I_1,G_1,B_1)$ be a complex torus equipped with a flat K\"ahler metric and a 
B-field of type
$(1,1).$ Let $(T_2,I_2)$ be another complex torus. 
Let $\tilde{\cI}_1$ and $\tilde{\cI}_2$ be the product complex structures
on $T_1\times T_1^*$ and $T_2\times T_2^*$. Suppose there exists an isomorphism of 
lattices $g:\Gamma_1\op\Gamma_1^*\ra \Gamma_2\op\Gamma_2^*$ mapping $q_1$ to $q_2$ and 
$\tilde{\cI}_1$ to $\tilde{\cI}_2.$ Then on $T_2$ there exists a K\"ahler metric $G_2$
and a B-field $B_2$ of type $(1,1)$ such that 
$Vert(\Gamma_1,I_1,G_1,\B_1)$ is isomorphic to $Vert(\Gamma_2,I_2,G_2,\B_2)$ as an $N=2$ SCVA.
\end{theorem}
Combining this with Theorem~\ref{ntwoiso} and the criterion for the equivalence of $D^b((T_1,I_1))$ 
and $D^b((T_2,I_2)),$ we obtain a result converse to Corollary~\ref{oneone}.
\begin{cor}
Let $(T_1,I_1,G_1,B_1)$ be an algebraic torus equipped with a flat K\"ahler metric and a 
B-field of type $(1,1).$ Let $(T_2,I_2)$ be another algebraic torus. 
Suppose $D^b((T_1,I_1))$ is equivalent to 
$D^b((T_2,I_2)).$ Then on $T_2$ there exists a K\"ahler metric $G_2$ and a B-field $B_2$ of 
type $(1,1)$ such that $Vert(\Gamma_1,I_1,G_1,\B_1)$ is isomorphic to 
$Vert(\Gamma_2,I_2,G_2,\B_2)$ as an $N=2$ SCVA.
\end{cor}

If $\dim_\CC T=1,$ then the B-field is automatically of type $(1,1).$ Therefore the
HMSC passes the check in this special case. Of course, this
is what we expect, since the HMSC is known to be true for the elliptic
curve~\cite{PoliZas}.
On the other hand, for $\dim_\CC T>1$ we seem to have a problem.

Not all is lost however, and a simple modification of the HMSC passes our check.
The modification involves replacing $(T,I)$ with a noncommutative
algebraic variety, or more precisely, replacing the structure sheaf of $(T,I)$
with an Azumaya algebra over $(T,I).$

%

Let us remind the definition and basic facts about Azumaya algebras.
Let $\A$ be an $\O_X$\!--algebra which is coherent
as a sheaf $\O_X$\!--modules. Denote by $Coh(\A)$ the abelian
category of sheaves of (right) $\A$\!--modules which are coherent
as sheaves of $\O_X$\!--modules, and by $D^b(\A)$ the
bounded derived category of $Coh(\A).$

We will be interested in a simple case of this situation when
$\A$ is an Azumaya algebra.  Recall that $\A$ is called an Azumaya
algebra if it is locally free as a sheaf of $\O_X$\!--modules,
and for any point $x\in X$ the restriction
$\A(x):=\A\otimes_{\O_X} \CC(x)$ is isomorphic to a matrix algebra
$M_r(\CC).$

A trivial Azumaya algebra is an algebra of the form $\Endo(E)$ where
$E$ is a vector bundle. Two Azumaya algebras $\A$ and $\A'$ are called
similar (or Morita equivalent)
if there exist vector bundles $E$ and $E'$ such that
$$
\A\otimes_{\O_X}\Endo(E)\cong \A'\otimes_{\O_X}\Endo(E')
$$
It is easy to see that in this case the categories $Coh(\A)$ and
$Coh(\A')$ are equivalent, and therefore the derived categories
$D^b(\A)$ and $D^b(\A')$ are equivalent as well.

Azumaya algebras modulo Morita equivalence generate a group
with respect to tensor product. This group is called the Brauer group
of the variety and is denoted by $Br(X).$

There is a natural map
$$
Br(X)\lto H^2(X, \O^*_X).
$$
This map is an embedding
and its image is contained in the torsion subgroup of $H^2(X, \O_X^*).$ The latter
group is denoted by $Br'(X)$ and called the cohomological
Brauer group of $X.$ The well-known Grothendieck conjecture
asserts that  the natural map $Br(X)\lto Br'(X)$
is an isomorphism for smooth varieties.
This conjecture was proved for abelian varieties~\cite{GroAbelian}; we
will assume that it is true in general.

Let $X$ be an algebraic variety over $\CC,$ and let $\Be\in H^2(X,\RR/\ZZ).$
Let $\beta: H^2(\RR/\ZZ)\ra H^2(X, \O^*_X)$ be the homomorphism
induced by the canonical map $\RR/\ZZ\lto \O^*_X.$ We have the following
commutative diagram of sheaves:
$$
\begin{array}{ccccccccc}
0&\lto&\ZZ&\lto&\RR&\llongrightarrow&\RR/\ZZ&\lto&0\\
&&\big\|&&\da&&\da\rlap{\scriptsize{$\beta$}}&&\\
0&\lto&\ZZ&\lto&\O_X&\stackrel{\exp(2\pi i\cdot)}
{\llongrightarrow}&\O^*_X&\lto&0
\end{array}
$$

Suppose $\beta(\Be)$ is a torsion element of $H^2(X, \O^*_X),$ and consider
an Azumaya algebra $\A_{\Be}$ which corresponds to this element.
The derived category $D^b(X,\A_{\Be})$ does not depend on the choice of
$\A_{\Be}$ because all these algebras are Morita equivalent. Thus we can denote
it simply $D^b(X,\Be).$

\begin{remark}\label{gerbe}
It appears that
a similar triangulated category can be defined even when $\beta(\Be)$ is not
torsion.
Any element $a\in H^2(X, \O_X^*)$ gives
us an $\O^*_X$ gerbe $\X_a$ over $X.$
Consider the derived category  $D^b_{Qcoh}(\X_{\beta(\Be)})$
of quasicoherent sheaves on this gerbe. Now our triangulated category
can be defined as a full subcategory of $D^b_{Qcoh}(\X_{\beta(\Be)})$
consisting of weight-1 objects  with some condition of finiteness,
which replaces coherence.
\end{remark}

A sufficient condition for the equivalence of $D^b(X_1,\Be_1)$ and
$D^b(X_2,\Be_2)$ for the case of algebraic tori is provided by the following
theorem~\cite{Po}.
\begin{theorem}\label{Poli}
Let $(T_1,I_1)$ and $(T_2,I_2)$ be two algebraic tori.
Let $\Be_1\in H^2(T_1,\RR/\ZZ)$
and $\Be_2\in H^2(T_2,\RR/\ZZ),$ and suppose $\beta$ maps both $\Be_1$
and $\Be_2$ to torsion elements. If there exists an isomorphism of lattices
$\Gamma_1\op\Gamma_1^*$ and $\Gamma_2\op\Gamma_2^*$ which maps $q_1$
to $q_2,$ and $\cI_1$ to $\cI_2,$ then $D^b((T_1,I_1),\Be_1)$
is equivalent to $D^b((T_2,I_2),\Be_2).$
\end{theorem}

\begin{remark}
It appears plausible that this is also a necessary condition for
$D^b((T_1,I_1),\Be_1)$ to be equivalent to $D^b((T_2,I_2),\Be_2).$
\end{remark}

\begin{remark}
It appears plausible that the theorem remains true even when $\beta(\Be_1)$
and $\beta(\Be_2)$ have infinite order, see Remark~\ref{gerbe}.
\end{remark}

Combining Theorem~\ref{Poli} with our Theorem~\ref{ntwoiso}, we
obtain the following result.
\begin{cor}\label{cortwo}
Suppose $Vert(\Gamma_1,I_1,G_1,\B_1)$ is isomorphic to
$Vert(\Gamma_2,I_2,G_2,\B_2),$
both $(T_1,I_1)$ and $(T_2,I_2)$ are algebraic, and both $\Be_1$ and $\Be_2$ are
mapped by $\beta$ to torsion elements.
Then $D^b((T_1,I_1),\Be_1)$ is equivalent to $D^b((T_2,I_2),\Be_2).$
\end{cor}

This corollary suggests that we modify the HMSC by replacing $D^b(X)$
with $D^b(X,\Be).$ Once we decided to include the B-field, it seems unnatural
to assume that the Fukaya category is independent of it.
D-brane considerations suggest a particular way to ``twist'' the Fukaya
category with a B-field (see Section~\ref{modHMSC}). Let us denote
this ``twisted'' category by $\cF(Y,\Be).$ Here $Y$ is a compact symplectic manifold,
and $\Be\in H^2(Y,\RR/\ZZ)$ is in the kernel of the Bockstein homomorphism
$H^2(Y,\RR/\ZZ)\ra H^3(Y,\ZZ).$ The modified HMSC asserts that if
$(X,G,\Be)$ is mirror to $(X',G',\Be'),$ then $D^b(X,\Be)$ is equivalent to
$D\cF_0(X',\Be').$ Corollary~\ref{cortwo} shows that
this conjecture passes the check which the original HMSC fails.

If both $\Be$ and $\Be'$ vanish, the modified HMSC reduces to the original HMSC.
Thus one could ask if it is possible to set the B-field to zero once and for all
and work with the original HMSC. This is highly unnatural for the following
reason. Suppose we have a mirror pair of physicist's Calabi Yaus which both
happen to have zero B-fields. Now let us start varying the complex structure
of the first Calabi-Yau. It can be seen in the case of complex tori and can be
argued in general that the corresponding deformation of the second Calabi-Yau
generally involves both the K\"ahler form and the B-field. Thus if we have a
family of Calabi-Yaus with zero B-field and varying complex structure, the
mirror family of Calabi-Yaus will have nonzero B-field for almost all values of
the parameter.

For example, in the case of the elliptic curve, the usual Teichm\"uller
parameter $\tau$
takes values in the upper half-plane. The mirror elliptic curve has vanishing $\Be$
if and only if $\tau$ can be made purely imaginary by a modular transformation.

In the case of the elliptic curve, the effect of the B-field on the HMSC
is relatively minor. It has no effect on the derived
category of coherent sheaves because $h^{0,2}=0.$ The objects of the Fukaya
category are also unmodified in this case (see Section~\ref{modHMSC}),
and the only change in the definition of morphisms is to complexify the
symplectic form. For higher-dimensional varieties, the modification of the
Fukaya category is more serious.

\subsection{Physical applications}

Transformations of the target space metric and the $B$-field which leave the
conformal field theory unchanged are known as T-duality transformations.
For a real torus $T^{n}=\RR^{n}/\Gamma,$ $\Gamma\cong \ZZ^{n},$ such transformations
form a group isomorphic to $O(n,n,\ZZ)$~\cite{Polch,LT}.
The main novelty of this work is that we consider complex tori, and study transformations
of $G,B,$ and the complex structure which leave the $N=2$ superconformal field theory unchanged
or induce a mirror morphism.

Our results have implications for the study of BPS D-branes on Calabi-Yau manifolds,
a subject which received much attention recently (see~\cite{Douglas} and references therein).
They suggest that BPS D-branes of type B are best thought
of as objects of the derived category of coherent sheaves when the B-field is zero.
When the B-field is nonzero but the corresponding class in $H^2(X,\O^*_X)$ is
a torsion class, the derived category of coherent sheaves should be replaced
with the derived category of a certain noncommutative algebraic variety (an Azumaya
algebra over $X$). When the class of the B-field in $H^2(X,\O^*_X)$ has infinite
order, it appears that B-type D-branes should be regarded as objects of the derived
category of "coherent" sheaves on a gerbe over $X.$

Note a similarity with the results of~\cite{Kap,BM} where it was shown that
in the presence of a B-field D-brane charges on a smooth manifold $X$ are classified by
the K-theory of an Azumaya algebra over $X,$ or more generally by the K-theory of a
Dixmier-Douady algebra over $X.$ The main differences are that
Refs.~\cite{Kap,BM} work in a $C^\infty$-category, the D-branes are not required to be BPS,
and the focus is on D-brane charges rather  on D-branes themselves.

In Section~\ref{modHMSC} we describe the effect of a closed B-field on BPS D-branes of type A
(the ones associated to flat unitary bundles on special Lagrangian submanifolds in a Calabi-Yau).
This subject was previously studied by Hori et al.~\cite{HoriIqVafa} for the case of a
single D-brane, i.e. when the rank of the bundle is one. Hori et al.
find that the restriction of the B-field to the Lagrangian submanifold must vanish.
We find that this restriction is too strong: it is sufficient to require the restriction
of the B-field to have integer periods. For the higher rank case
we argue that in general the unitary bundle on the Lagrangian submanfold is projectively flat
rather than flat. Correspondingly, the restrictions on the B-field are even weaker.

\section{Superconformal vertex algebras}
\label{sec:general}

\subsection{Quantum fields}

Let $V$ be a vector superspace over $\CC.$ The parity of an element $a\in V$ is
denoted $p(a)$ and takes values in integers modulo 2.
\begin{definition}
The space of quantum fields in one formal variable with values
in $\End(V)$ is a vector superspace whose elements have the form
$$
\sum_{h\in J}\ \sum_{n, \bn\in \ZZ}
C_{(h + n, h+\bn)} z^{-h-n} \bz^{-h-\bn},
$$
where $J$ is some subset of $[0,1)$ (different for different elements),
$C_{(h+n,h+\bn)}\in \End(V),$ and the following conditions are satisfied:

\begin{itemize}
\item[{\rm a)}]
the set $J$ is countable;
\item[{\rm b)}]
for any element $v\in V$ there is a finite subset $J_v\subset J$
such that
$$
C_{(h+n,h+\bn)}(v)=0
$$
for all
$h\in J\backslash J_v$ and all $n,\bn\in\ZZ$;
\item[{\rm c)}]
for any element $v\in V$ there is an integer $N$
such that
$ C_{h+n,h+\bn}(v)=0$
for all $h\in J$ if $n> N$ or $\bn> N.$
\end{itemize}
The space of quantum fields in one formal variable with values in $\End(V)$ is
denoted $QF_1(V).$
\end{definition}
Given an element $A(z,\bz)$ of $QF_1(V),$ we will denote the coefficient of
$z^{-h-n}\bz^{-h-\bn}$ in $A(z,\bz)$ by $A_{(h+n,h+\bn)}.$

The intersection of
$QF_1(V)$ with $\End (V)[[z,z^{-1}]]$ (resp. $\End (V)[[\bz,\bz^{-1}]]$)
will be called the space of
meromorphic (resp. anti-meromorphic) fields. We will denote by $A(z)$ (resp.
$A(\bz)$) meromorphic (resp. anti-meromorphic) fields.
The coefficient of $z^{-n}$ in $A(z)$ (resp. the coefficient of $\bz^{-n}$
in $A(\bz)$) will be denoted $A_{(n)}.$

\begin{definition}
The space of quantum fields in two formal variables with values in $\End(V)$
is a vector superspace whose elements have the form
$$
\sum_{(h,g)\in J}\ \sum_{n, \bn, m, \bm\in \ZZ}
C_{(h + n, h+\bn,g+m, g+\bm)} z^{-h-n} \bz^{-h-\bn} w^{-g-m} \bw^{-g-\bm},
$$
where $J\in [0,1)^2,$ $C_{(h + n, h+\bn,g+m, g+\bm)}\in \End(V),$ and the
following conditions are satisfied:

\begin{itemize}
\item[${\rm a')}$]
the set $J$ is countable;
\item[${\rm b')}$]
for any element $v\in V$ there is a finite subset $J_v\subset J$
such that
$$
C_{(h+n,h+\bn,g+m,g+\bm)}(v)=0
$$
for all
$(h,g)\in J\backslash J_v$ and all $n,\bn,m,\bm\in\ZZ$;
\item[${\rm c')}$]
for any element $v\in V$ and any $(h,g)\in J,$
there is an integer $N$ such that
\begin{align}\nn
C_{(h+n,h+\bn,g+m,g+\bm)}(v)=0,\\ \nn
C_{(h+m,h+\bm,g+n,g+\bn)}(v)=0,
\end{align}
for $n>N$ and any $\bn,m,\bm\in\ZZ,$
and
\begin{align}\nn
C_{(h+n,h+\bn,g+m,g+\bm)}(v)=0,\\ \nn
C_{(h+m,h+\bm,g+n,g+\bn)}(v)=0,
\end{align}
for $\bn>N$ and any $n,m,\bm\in\ZZ.$
\end{itemize}
The space of quantum fields in two formal variables with values
in $\End(V)$ is denoted $QF_2(V).$
\end{definition}

Item ${\rm (c')}$ in the definition of $QF_2(V)$ ensures that
given an element $C(z,\bz,w,\bw)$of $QF_2(V),$ one can substitute
$z=w,\bz=\bw$ and get a well-defined element of $QF_1(V).$
This element will be denoted $C(w,\bw,w,\bw).$ Note that in general a product
of two fields $A(z,\bz)\in QF_1(V)$ and $B(w,\bw)\in QF_1(V)$ does not belong
to $QF_2(V),$ precisely because ${\rm(c')}$ is not satisfied. In this situation
one says that the product of $A(z,\bz)$ and $B(w,\bw)$ has a singularity
for $z=w,\bz=\bw.$

If an element $A(z,\bz,w,\bw)\in QF_2(V)$ does not contain nonzero powers of
$\bz$ (resp. $z$) we will say that this field is meromorphic (resp.
anti-meromorphic) in the first variable, and write it as $A(z,w,\bw)$
(resp. $A(\bz,w,\bw)$). Fields in two variables (anti-)meromorphic in the
second variable are defined in a similar way.

\subsection{The definition of a vertex algebra}

We set
\begin{align}            \nn
i_{z,w}\frac{1}{(z-w)^h}&=\sum_{j=0}^\infty
\binom{-h}{j}
(-1)^j w^j z^{-j-h},
&
i_{\bz,\bw}\frac{1}{(\bz-\bw)^h}&=\sum_{j=0}^\infty
\binom{-h}{j}
(-1)^j \bw^j \bz^{-j-h},
\\ \nn
i_{w,z}\frac{1}{(z-w)^h}&=\sum_{j=0}^\infty
\binom{-h}{j}
e^{-i\pi h}(-1)^j z^j w^{-j-h},& \nn
i_{\bw,\bz}\frac{1}{(\bz-\bw)^h}&=\sum_{j=0}^\infty
\binom{-h}{j}
e^{i\pi h}(-1)^j \bz^j \bw^{-j-h},
\end{align}
where
$$
\binom{-h}{j}= \frac{(-h)(-h-1)\cdots(-h-(j-1))}{j!}.
$$
These are formal power series expansions of the functions $(z-w)^{-h}$ and $(\bz-\bw)^{-h}$
in the regions $|z|>|w|,|z|<|w|$ and $|\bz|>|\bw|,|\bz|<|\bw|.$
\begin{definition}\label{VAdef}
A vertex algebra structure on a vector superspace $V$
consists of the following
data:
\begin{itemize}
\item[{\rm(i)}]
 an even vector $\vac\in V.$
\item[{\rm(ii)}]
 a pair $T,\bT$ of commuting even endomorphisms of $V$ annihilating
$\vac.$
\item[\rm{(iii)}]
a parity-preserving linear map
$$Y:V\to QF_1(V),\quad Y:a\mapsto Y(a)=a(z,\bz).$$
\end{itemize}
These data must satisfy the following requirements.
\begin{itemize}
\item[{\rm 1.}]
$Y(\vac)=id\in \End (V).$
\item[{\rm 2.}]
$[T,a(z,\bz)]=\partial a(z,\bz),\ [\bT,a(z,\bz)]=\bpartial a(z,\bz).$
\item[{\rm 3.}]
$a(z,\bz)\vac = e^{zT+\bz\bT}a.$
\item[{\rm 4.}]
For any $a,b\in V$ there are integers $N,M,$ real numbers $h_j\in[0,1), j=1,\ldots,M,$
and quantum fields $C_j(z,\bz,w,\bw)\in QF_2(V), j=1,\ldots,M,$ such that
\begin{align}\label{OPEaxiomone}
a(z,\bz)b(w,\bw)=\sum_{j=1}^M i_{z,w}\frac{1}{(z-w)^{h_j+N}}\ i_{\bz,\bw}
\frac{1}{(\bz-\bw)^{h_j+N}}\ C_j(z,\bz,w,\bw),\\ \label{OPEaxiomtwo}
(-1)^{p(a)p(b)}b(w,\bw)a(z,\bz)=\sum_{j=1}^M i_{w,z}\frac{1}{(z-w)^{h_j+N}}\
i_{\bw,\bz}\frac{1}{(\bz-\bw)^{h_j+N}}\ C_j(z,\bz,w,\bw).
\end{align}
\end{itemize}

The map $Y$ is called the state-operator correspondence. The coefficient of
$z^{-\alpha} \bz^{-\beta}$ in
$Y(a)$ is called the $(\alpha,\beta)$ component of $Y(a)$ and denoted
by $a_{(\alpha,\beta)}.$
\end{definition}

The last requirement in the definition of a vertex algebra
is called the Operator Product Expansion (OPE) axiom. It contains two important
ideas. The equality~(\ref{OPEaxiomone}) says that the product of two
fields in the image of $Y$ has only power-like singularities for $z=w,\bz=\bw.$
The difference of~(\ref{OPEaxiomone}) and (\ref{OPEaxiomtwo}) means, roughly speaking,
that the fields in the
image of $Y$ are mutually local, in the sense that their supercommutator
vanishes when $z\neq w$ and $\bz\neq\bw.$
This is particularly clear when all $h_i$ are
equal to zero. Then the supercommutator of $a(z,\bz)$ and $b(w,\bw)$
is proportional to
\begin{multline}
\frac{1}{((N-1)!)^2} \delta^{(N-1)}(z-w)\delta^{(N-1)}(\bz-\bw)+
\frac{1}{(N-1)!} \delta^{(N-1)}(z-w)\ i_{\bz,\bw}\frac{1}{(\bz-\bw)^N}\\
+\frac{1}{(N-1)!} \delta^{(N-1)}(\bz-\bw)\ i_{z,w}\frac{1}{(z-w)^N},
\end{multline}
where $\delta^{(k)}(z-w)$ is the $k^{\rm th}$ derivative of the
formal delta-function defined as a formal power series
$$
\delta(z-w)=z^{-1}\sum_{n\in\ZZ} \left(\frac{z}{w}\right)^n.
$$
Given any two elements of $QF_1(V),$ we will say that they are mutually
local if for their products the OPE
 formulas~(\ref{OPEaxiomone},\ref{OPEaxiomtwo})
hold for some $N,M\in \ZZ,$ $h_j\in [0,1),$ $j=1,\ldots, M,$ and
$C_j\in QF_2(V),$ $j=1,\ldots, M.$

Vertex algebras as defined above are a generalization of chiral
algebras as defined in~\cite{Kac} in the following sense. First, any
chiral algebra is automatically a vertex algebra, with $\bT=0$
and the image of $Y$ consisting of meromorphic fields only.
Second, if we consider
the subspace in $V$ consisting of vectors which are mapped to meromorphic fields,
the restriction of $T$ and $Y$ to this subspace specifies on it the structure
of a chiral algebra. Similarly, the restriction of $\bT$ and $Y$ to the
anti-meromorphic
sector yields another chiral algebra. Moreover, all meromorphic fields
supercommute with all
anti-meromorphic fields. Thus any vertex algebra contains a pair
of commuting chiral subalgebras. All these facts are proved in Appendix~\ref{vertexandchiral}.

The OPE formulas simplify when one of the fields is meromorphic or anti-meromorphic.
For example, the OPE of a meromorphic field $a(z), a\in V,$ with a general field
$b(w,\bw),b\in V,$ has the following form (see Appendix~\ref{vertexandchiral} for proof):
\begin{align}\label{specOPE}
a(z)b(w,\bw) & =\sum_{j=1}^N i_{z,w}\frac{1}{(z-w)^j}\ D_j(w,\bw)+:a(z)b(w,\bw):,\\ \nn
(-1)^{p(a)p(b)}b(w,\bw)a(z) & =\sum_{j=1}^N i_{w,z}\frac{1}{(z-w)^j}\ D_j(w,\bw)+:a(z)b(w,\bw):.
\end{align}
Here $N$ is some integer, $D_j(w,\bw)\in QF_1(V),$ and $:a(z)b(w,\bw):$ is an element of
$QF_2(V)$ defined as follows:
$$
:a(z)b(w,\bw):=a(z)_+ b(w,\bw)+(-1)^{p(a)p(b)}b(w,\bw) a(z)_-,
$$
where we set
\begin{equation}\label{posnegfreq}
a(z)_{+}=\sum_{n\leq 0} a_{(n)} z^{-n},\qquad  a(z)_{-}=\sum_{n> 0} a_{(n)} z^{-n}.
\end{equation}
The field $:a(z)b(w,\bw):$ is called the normal ordered product of $a(z)$ and
$b(w,\bw).$ Since it belongs to $QF_2(V),$ one can set $z=w$ and get
a well-defined field in one variable $:a(w)b(w,\bw):.$
The difference between
the right-hand side of~(\ref{specOPE}) and $:a(z)b(w,\bw):$ is called the
singular part of the OPE.

Similarly, one can define the normal ordered product of
an anti-meromorphic field with a general field. The normal ordered product of
two general fields is not defined.

Let us consider now the OPE of two meromorphic fields $a(z)$ and $b(z).$ We
already mentioned that meromorphic fields form a chiral algebra, thus the OPE
~(\ref{specOPE}) simplifies even further:
\begin{align}\nn
a(z)b(w) & =\sum_{j=1}^N i_{z,w}\frac{1}{(z-w)^j}\ D_j(w)+:a(z)b(w):,\\ \nn
(-1)^{p(a)p(b)}b(w)a(z) & =\sum_{j=1}^N i_{w,z}\frac{1}{(z-w)^j}
\ D_j(w)+:a(z)b(w):.
\end{align}
Here $D_j(w),j=1,\ldots,N,$ are meromorphic elements of $QF_1(V).$ Exchanging
$a(z)$ and $b(w)$ we get
\begin{align}\nn
b(w)a(z) & =\sum_{j=1}^N i_{w,z}\frac{1}{(w-z)^j}\ C_j(z)+:b(w)a(z):,\\ \nn
(-1)^{p(a)p(b)}a(z)b(w) & =\sum_{j=1}^N i_{z,w}\frac{1}{(w-z)^j}
\ C_j(z)+:b(w)a(z):,
\end{align}
where $C_j(z),j=1,\ldots,N,$ are meromorphic elements of $QF_1(V).$

In general, the normal ordered product is not supercommutative, i.e.
$$
:a(z)b(w):\,\neq\, (-1)^{p(a)p(b)}:b(w)a(z):.
$$
Neither is it associative, in the sense that in general
$$
:a(z):b(z)c(z)::\ \neq\ ::a(z)b(z):c(z):.
$$
We will define
the normal ordered product of more than two (anti-)meromorphic fields inductively
from right to left:
$$
:a_1(z)a_2(z)\ldots a_n(z):=:a_1(z):a_2(z)\ldots a_n(z)::.
$$
An important special case where the normal ordered product of meromorphic fields
is supercommutative is when the fields $D_j(w)$ do not depend
on $w,$ i.e. are constant endomorphisms of $V.$
This follows directly from the above OPE formulas.
One can also show that
if pairwise OPE's of meromorphic fields $a(z),b(z),$ and $c(z)$ have this
property, then their normal ordered product is associative~\cite{DiF}.
For example, the normal ordered product of free fermion and free boson
fields is supercommutative and associative~\cite{Kac,DiF}.

Another important special case is the OPE of a meromorphic field and an
anti-meromorphic field. In this case one can also define two normal ordered
products, $:a(z)b(\bw):$ and $:b(\bw)a(z):.$ But it follows easily from the
equations~(\ref{specOPE}) and analogous equations for the OPE of an anti-meromorphic
field and a general field, that in this case the singular part of the OPE
vanishes, the normal ordered product coincides with the ordinary product,
and that consequently all meromorphic fields supercommute with all
anti-meromorphic fields. Thus
$$
:a(z)b(\bw):=(-1)^{p(a)p(b)}:b(\bw)a(z):.
$$
This is discussed in more detail in Appendix~\ref{vertexandchiral}.

The singular part of the OPE of two meromorphic fields $a(z)$ and $b(z)$ completely
determines and is determined by the supercommutators of $a_{(n)}$ and $b_{(m)}$
for all $n,m\in\ZZ.$ Explicit formulas which enable one to pass from the OPE
to the supercommutators and back can found in~\cite{Kac}.

When writing the OPE of two meromorphic fields
we will use a shortened notation in which only the singular part of the OPE
is shown. To indicate this, the equality sign is replaced by $\sim.$ In addition,
we will only write the first of the OPE's in~(\ref{specOPE}), and correspondingly
will omit the symbol $i_{z,w},$ as is common in the physics literature. Similar
notation is used for the OPE of two anti-meromorphic fields. Thus instead of
$$
a(z)b(w)= \sum_{j=1}^N i_{z,w} \frac{1}{(z-w)^j}\ D_j(w)+:a(z)b(w):
$$
we will write
$$
a(z)b(w)\sim \sum_{j=1}^N \frac{D_j(w)}{(z-w)^j}.
$$

We conclude this subsection by defining morphisms of vertex algebras.
A morphism from a vertex algebra $(V,\vac,T,\bT,Y)$ to a vertex algebra
$(V',\vac',T',\bT',Y')$ is a morphism of superspaces $f:V\ra V'$ such that
$$
f(\vac)=\vac', \qquad fT=T'f,\qquad f\bT=\bT'f,
$$
and
$$
Y'(f(a))f(b)=f(Y(a)b) \qquad \forall a,b\in V.
$$

\subsection{Conformal vertex algebras}

\begin{definition} Let $\cV=(V,\vac,T,\bT,Y)$ be a vertex algebra.
Conformal structure on $\cV$ is a pair
of even vectors $L,\tL\in V$ such that
\begin{equation}
\begin{array}{rll}
{\rm(i)}
&\ds L(z,\bz)=L(z)=\sum_{n\in \ZZ} L_n z^{-n-2},&
\ds\tL(z,\bz)=\tL(\bz)=\sum_{n\in \ZZ} \tL_n \bz^{-n-2}.\\
{\rm (ii)}
& L_{-1}=T, \tL_{-1}=\bT.\\
{\rm (iii)}&
\multicolumn{2}{l}{
\label{Vir}
\ds L(z)L(w)\sim \frac{c/2}{(z-w)^4}+\frac{2L(w)}{(z-w)^2}+
\frac{\partial L(w)}{z-w},}
\\
&
\multicolumn{2}{l}{\ds
\tL(\bz)\tL(\bw)\sim \frac{\tc/2}{(\bz-\bw)^4}+\frac{2\tL(\bw)}{(\bz-\bw)^2}+
\frac{\bpartial \tL(w)}{\bz-\bw}.}\\
{\rm (iv)}
&\text{for any $a\in V$}&\\
&\ds [L_0,a(z,\bz)]=z\partial a(z,\bz)+(L_0 a)(z,\bz),&
\ds [\tL_0,a(z,\bz)]=\bz\bpartial a(z,\bz)+(\tL_0 a)(z,\bz).
\end{array}
\end{equation}
Here $c,\bc\in\CC.$ A vertex algebra with a conformal structure is called a
conformal vertex algebra
(CVA).
\end{definition}

The numbers $c$ and $\tc$ are called the holomorphic and anti-holomorphic
central charges of the CVA. The reason for this name
is the following. The OPE's~(\ref{Vir}) are equivalent
to the following commutation relations for all $n,m\in\ZZ$~\cite{Kac}:
\begin{align}
{}[L_m,L_n]&= (m-n)L_{m+n}+c\frac{m^3-m}{12}\delta_{m,-n},\nn\\
{}[\tL_m,\tL_n]&= (m-n) \tL_{m+n}+\tc \frac{m^3-m}{12}\delta_{m,-n},\nn\\
{}[L_n,\tL_m]&= 0.\nn
\end{align}
Hence the components of $L(z)$ and $\tL(z)$ form two
commuting Virasoro algebras. The Virasoro algebra is the unique
central extension of the Witt algebra (the algebra of the infinitesimal
diffeomorphisms of a circle). In the present case the central charges of
the two Virasoro algebras are $c$ and $\bc.$

Note that axiom~3 in the definition of a vertex algebra implies that
both $L_n$ and $\bL_n$ annihilate $\vac$ for all $n\geq -1.$

A morphism $f$ from a CVA $(V,\vac,Y,L,\bL)$ to a CVA $(V',\vac',Y',L',\bL')$
is a morphism of the underlying vertex algebras which satisfies
$$
f(L)=L',\qquad f(\bL)=\bL'.
$$

A conformal vertex algebra is almost the same as a conformal field theory.
Namely, a physically acceptable conformal field theory is a conformal vertex
algebra whose state space $V$ is equipped with a positive-definite Hermitian
inner product,
and the following additional constraints are satisfied:

\begin{itemize}
\item[(v)]
The space $V$ splits as a direct sum
of the form
$$
\op_{j\in J} {\mathcal W}_j\ot \bar{{\mathcal W}}_j,
$$
where $J$ is a countable set, and ${\mathcal W}_j$ and $\bar{\mathcal W}_j$ are unitary
highest-weight modules over the meromorphic and anti-meromorphic Virasoro algebras,
respectively.
\item[(vi)]
The vacuum vector is the only vector in $V$ annihilated by both $L_0$ and $\bL_0.$
\end{itemize}

The conformal vertex algebras we will be working with satisfy these
constraints and therefore are honest conformal field theories. However, we
prefer not to stress the ``real'' aspects of conformal field theories in this paper.

Furthermore, in order for a conformal field theory to admit a string-theoretic
interpretation, it must be defined on a Riemann surface of arbitrary genus.
(The above axioms define a conformal field theory in genus zero.)
This does not
require new data, but imposes additional, so-called sewing, constraints.
We will work in genus zero only, and therefore will neglect the sewing
constraints.

\subsection{N=1 superconformal vertex algebras}

\begin{definition}
Let ${\mathcal V}=(V,\vac,Y,L,\bL)$ be a conformal vertex algebra with
central charges $c,\bc.$ $N=1$
superconformal structure on ${\mathcal V}$ is a pair of odd vectors
$Q,\bQ\in V$ such that
$$
\begin{array}{rll}
{\rm (i)}& \ds
Q(z,\bz)=Q(z)=\mathop{\sum}\limits_{r\in \ZZ+\frac{1}{2}}
\frac{Q_r}{z^{r+3/2}},&\qquad \ds
\bQ(z,\bz)=\bQ(\bz)=\mathop{\sum}\limits_{r\in \ZZ+\frac{1}{2}}
\frac{\bQ_r}{\bz^{r+3/2}}.\\
{\rm (ii)}&
\multicolumn{2}{l}{\text{
The following OPE's hold true:}}\\
&\multicolumn{2}{c}{%
\begin{array}{rcl}\label{openone}
L(z)Q(w)&
\sim \ds\frac{3}{2}\frac{Q(w)}{(z-w)^2}+\frac{\partial Q(w)}{(z-w)},\\ \nn
Q(z)Q(w)&\sim\ds\frac{c/6}{(z-w)^3}+\frac{1}{2}\frac{L(w)}{(z-w)},
\end{array}}\\
&\multicolumn{2}{l}{\text{%
and similar OPE's for the anti-meromorphic fields with $z,w,c,\partial$
replaced with }}
\\
&\multicolumn{2}{l}{\bz,\bw,\bc,\bpartial.}
\end{array}
$$
The fields $Q(z)$ and $\bQ(\bz)$
are called left-moving and right-moving supercurrents, respectively.
A CVA with an $N=1$ superconformal structure is called an $N=1$ superconformal
vertex algebra ($N=1$ SCVA).
\end{definition}

$N=1$ superconformal structure is also known as $(1,1)$ superconformal
structure. Omitting $\bQ,$ one obtains the
definition of $(1,0)$ superconformal structure.
Morphisms of $N=1$ SCVA's are defined in an obvious way.

The OPE's of $Q(z),\bQ(\bz)$ with themselves and $L(z),\bL(\bz)$ are equivalent to the
following commutation relations:
\begin{align*}
[L_m,Q_r]&=\left(\frac{m}{2}-r\right) Q_{r+m}, &
[\bL_m,\bQ_r]&=\left(\frac{m}{2}-r\right) \bQ_{r+m}, \\ \nn
\{Q_r,Q_s\}&=\frac{1}{2}L_{r+s}+\frac{c}{12}\left(r^2-\frac{1}{4}\right)
\delta_{r,-s},&
\{\bQ_r,\bQ_s\}&=\frac{1}{2}\bL_{r+s}+\frac{\bc}{12}\left(r^2-\frac{1}{4}
\right)
\delta_{r,-s}.
\end{align*}
As usual, the barred generators supercommute with the unbarred ones. Thus $L_n,\bL_n,
Q_r,\bQ_r$ form an infinite-dimensional Lie super-algebra which is
a direct sum of two copies of the $N=1$ super-Virasoro algebra with central
charges $c$ and $\bc.$

\subsection{N=2 superconformal vertex algebras}

\begin{definition}\label{ntwo}

Let ${\mathcal V}=(V,\vac,Y,L,\bL)$ be a conformal vertex algebra with
central charges $c,\bc.$ $N=2$
superconformal structure on ${\mathcal V}$ is a pair of even vectors
$J,\tJ\in V$ and four odd vectors $Q^+,Q^-,\tQ^+,\tQ^-\in V$ such that

$$
\begin{array}{rll}
{\rm (i)}&
\ds
J(z,\bz)=J(z)=\ds\mathop{\sum}\limits_{n\in \ZZ} \frac{J_n}{z^{n+1}},
&
\ds
\qquad
\tJ(z,\bz)=\tJ(\bz)= \ds
\mathop{\sum}\limits_{n\in \ZZ} \frac{\tJ_n}{\bz^{n+1}},
\\
&
\ds
 Q^+(z,\bz)=Q^+(z)=\ds\mathop{\sum}\limits_{r\in\ZZ+\frac{1}{2}}
\frac{Q^+_r}{z^{r+3/2}},
&\qquad
\ds
\tQ^+(z,\bz)=\tQ^+(\bz)=\ds\mathop{\sum}_{r\in\ZZ+\frac{1}{2}}
\frac{\tQ^+_r}{\bz^{r+3/2}},
\\
&

\ds
Q^-(z,\bz)=Q^-(z)=\ds\mathop{\sum}\limits_{r\in\ZZ+\frac{1}{2}}
\frac{Q^-_r}{z^{r+3/2}},

&

\ds
\qquad
\tQ^-(z,\bz)=\tQ^-(\bz)=\ds\mathop{\sum}\limits_{r\in\ZZ+\frac{1}{2}}
\frac{\tQ^-_r}{\bz^{r+3/2}};
\\
{\rm (ii)}&
\multicolumn{2}{l}{\text{ the following OPE's hold true:}}\\
&
\multicolumn{2}{c}
{
\begin{array}{llll}
&\ds L(z)Q^\pm (w)&
\sim &\ds\frac{3}{2}\frac{Q^\pm(w)}{(z-w)^2}+\frac{\partial Q^\pm(w)}{(z-w)},
\nn\\
&L(z)J(w)&
\sim& \ds\frac{J(w)}{(z-w)^2}+\frac{\partial J(w)}{(z-w)},\nn\\
&J(z)J(w)&\sim&\ds \frac{c/3}{(z-w)^2},\nn\\
&J(z) Q^\pm(w)&\sim&\ds \pm \frac{Q^\pm(w)}{(z-w)},\nn\\
&Q^+(z)Q^-(w)&
\sim&
\ds \frac{c/12}{(z-w)^3}+\frac{1}{4}\frac{J(w)}{(z-w)^2}+\frac{1}{8}
\frac{\partial J(w)+2L(w)}{(z-w)},\nn\\
&Q^\pm(z)Q^\pm(w)&\sim & 0,\nn
\end{array}
}\\
&\multicolumn{2}{l}{
\text{and similar OPE's for the anti-meromorphic fields with $z,w,c,\partial$
replaced with}}\\
&\multicolumn{2}{l}{\text{$\bz,\bw,\tc,\bpartial.$}}
\end{array}
$$

The fields $J(z)$ and $\tJ(\bz)$ are called left-moving and right-moving R-currents,
the fields $Q^\pm(z)$ and $\tQ^\pm(\bz)$ are called left-moving and right-moving
supercurrents, respectively.
A CVA with $N=2$ superconformal structure is called an $N=2$ superconformal
vertex algebra ($N=2$ SCVA).
\end{definition}
The above OPE's together with the OPE's for $L(z),\bL(z)$ are equivalent to
the commutation relations~(\ref{ntwovir}) if we set $c=\bc=3d.$

$N=2$ superconformal structure is also known as a $(2,2)$ superconformal
structure. If one omits the anti-meromorphic currents $\tJ(\bz),
\tQ^\pm(\bz),$ one gets the definition of a $(2,0)$ superconformal
structure.

Given an $N=2$ SCVA, one can obtain an $N=1$ SCVA by setting $Q=Q^++Q^-,$
$\bQ=\bQ^++\bQ^-.$ Thus an $N=2$ SCVA can be regarded as an $N=1$ SCVA
with additional structure.

Morphisms of $N=2$ SCVA's are defined in an
obvious way. {\it A mirror morphism} between two $N=2$ SCVA's is an
isomorphism between the underlying $N=1$ SCVA's which induces the following
map on $Q^\pm,\bQ^\pm,J,\bJ$:
\begin{align*}
&f(Q^+)=Q^{-'},\ f(Q^-)=Q^{+'},\ f(J)=-J',\\
&f(\bQ^+)=\bQ^{+'},\
f(\bQ^-)=\bQ^{-'},\ f(\bJ)=\bJ'.
\end{align*}
This map acts as an outer automorphism on the algebra~(\ref{ntwovir}).
A composition of two mirror morphisms is an isomorphism of $N=2$ SCVA's.

\section{N=2 SCVA of a flat complex torus}
\label{torusSCVA}

The purpose of this section is to describe an $N=2$ SCVA
canonically associated to a complex torus endowed with a flat K\"ahler metric and
a constant 2-form. None of this material is new, and everything can be found,
in one form or another, in standard string theory textbooks~\cite{LT,Polch}.
We simply translate
these standard constructions into the language of vertex algebras.

\subsection{Vertex algebra structure}

Let $U$ be a real vector space of dimension $2d.$
Let $\Gamma\cong \ZZ^{2d}$ be a lattice in $U.$ Let $\Gamma^*\subset U^*$ be the dual
lattice $\Hom(\Gamma,\ZZ).$ Let $T=U/\Gamma,\ T^*=U^*/\Gamma^*.$
Let $G$ be a metric on $U,$ i.e. a positive symmetric bilinear form on $U.$
Let $\B$ be a real skew-symmetric bilinear form on $U.$
Let $l$ be the natural pairing $\Gamma\times \Gamma^*\to\ZZ.$ The natural
pairing $U\times U^*\to\RR$ will be also denoted $l.$ Let $\ZZ^*$ be the set
of nonzero integers. Let the vectors $e_1,\ldots,e_{2d}\in U$ be the generators
of $\Gamma.$ The components of an element $w\in \Gamma$ in this basis will be denoted
by $w^i,i=1,\ldots,2d.$ The components of an element $m\in\Gamma^*$ in the dual basis
will be denoted by $m_i, i=1,\ldots,2d.$ We also denote by $G_{ij},\ \B_{ij}$
the components of $G,\ \B$ in this basis. It will be apparent that the superconformal 
vertex algebra which we construct does not depend on the choice of basis in $\Gamma.$ 
In the physics literature $\Gamma$ is sometimes
referred to as the winding lattice, while $\Gamma^*$ is called the momentum
lattice.

Consider a triple $(T,G,\B).$ To any such triple we will associate a superconformal vertex
algebra $\cV$ which may be regarded as a quantization of the supersymmetric
$\sigma$-model described in Appendix~\ref{sigmamodel}.

The state space of the vertex algebra $\cV$ is
$$
V=\cH_b\ot_\CC\cH_f\ot_\CC \grp.
$$
Here $\cH_b$ and $\cH_f$ are bosonic and fermionic Fock spaces
defined below, while $\grp$ is the group algebra of $\Gamma\op\Gamma^*$ over $\CC.$

To define $\cH_b,$ consider an algebra over $\CC$ with generators
$\al^i_s,\bal^i_s,\ i=1,\ldots,2d, s\in \ZZ^*$
and relations
\begin{equation}\label{CCR}
[\al^i_s,\al^j_p]=s\Gi^{ij}\delta_{s,-p}, \quad
[\bal^i_s,\bal^j_p]=s\Gi^{ij}\delta_{s,-p}, \quad
[\al^i_s,\bal^j_p]=0.
\end{equation}
If $s$ is a positive integer, $\al^i_{-s}$ and $\bal^i_{-s}$ are called left
and right bosonic creators, respectively,
otherwise they are called left and right bosonic annihilators. Either creators
or annihilators are referred to as oscillators.

The space $\cH_b$ is defined as the space of polynomials
of even variables $a^i_{-s},\ba^i_{-s}, i=1,\ldots,2d, s=1,2,\ldots,$
The bosonic oscillator algebra~(\ref{CCR})
acts on the space $\cH_b$ via
$$
\begin{array}{lllllll}
\al^i_{-s}&\mapsto& a^i_{-s}\cdot, &\qquad& \bal^i_{-s}&\mapsto&
\ba^i_{-s}\cdot, \\
\al^i_s &\mapsto&\ds s\Gi^{ij}\frac{\partial}{\partial a^j_{-s}}, &
\qquad&
\bal^i_s&\mapsto&\ds s\Gi^{ij}\frac{\partial}{\partial \ba^j_{-s}},
\end{array}
$$
for all positive $s.$ This is
the Fock-Bargmann representation of the bosonic oscillator algebra.
The vector $1\in \cH_b$ is annihilated by all bosonic annihilators
and will be denoted $|vac_b\rangle.$

The space $\cH_b$ will be regarded as a $\ZZ_2$-graded vector
space with a trivial (purely even) grading. It is clear that $\cH_b$ can be decomposed
as $\fH_b\ot \bar{\fH}_b,$ where $\fH_b$ (resp. $\bar{\fH}_b$) is the bosonic Fock
space defined using only the left (right) bosonic oscillators.

To define $\cH_f,$ consider an algebra over $\CC$ with generators
$\psi^i_s,\tpsi^i_s,\ i=1,\ldots,2d, s\in\ZZ+\frac{1}{2}$ subject to relations
\begin{equation}\label{CAR}
\{\psi^i_s,\psi^j_p\}=\Gi^{ij}\delta_{s,-p}, \quad
\{\bpsi^i_s,\bpsi^j_p\}=\Gi^{ij}\delta_{s,-p}, \quad
\{\psi^i_s,\bpsi^j_p\}=0.
\end{equation}
If $s$ is positive, $\psi^i_{-s}$ and $\tpsi^i_{-s}$ are called left and right
fermionic creators respectively, otherwise they are called left and right fermionic
annihilators. Collectively they are referred to as fermionic oscillators.

The space $\cH_f$ is defined as the space of skew-polynomials
of odd variables $\theta^i_{-s},\btheta^i_{-s}, i=1,\ldots,2d, s=1/2,3/2,\ldots,$
The  fermionic oscillator algebra~(\ref{CAR})
acts on $\cH_f$ via
$$
\begin{array}{lllllll}
\psi^i_{-s}&\mapsto& \theta^i_{-s}\cdot, &
\qquad& \bpsi^i_{-s}&\mapsto& \btheta^i_{-s}\cdot,\\
\psi^i_s&\mapsto&\ds \left(G^{-1}\right)^{ij}\frac{\partial}
{\partial \theta^j_{-s}}, &
\qquad&
\bpsi^i_s&\mapsto& \ds\left(G^{-1}\right)^{ij}\frac{\partial}{\partial \btheta^j_{-s}},
\end{array}
$$
for all positive $s\in\ZZ+\frac{1}{2}.$
This is the Fock-Bargmann representation of the fermionic oscillator algebra.
The vector $1\in \cH_f$ is annihilated by all fermionic annihilators and will be denoted
$|vac_f\rangle.$
The fermionic Fock space has a natural $\ZZ_2$ grading such that
$|vac_f\rangle$ is even.
It can be decomposed as $\fH_f\ot \bar{\fH}_f,$
where $\fH_f$ (resp. $\bar{\fH}_f$) is constructed using only the left (right) fermionic
oscillators.

For $w\in \Gamma,\ m\in \Gamma^*$ we will denote the vector $w\op m\in \grp$ by
$(w,m).$ We will also use a shorthand $|vac, w,m\rangle,$ for
$$
|vac_b\rangle\ot|vac_f\rangle\ot (w,m).
$$

To define $\cV,$ we have to specify the vacuum vector, $T,\bT,$
and the state-operator
correspondence $Y.$
But first we need to define some auxiliary objects. We define the operators
$\hw:V\to V\ot\Gamma$ and $\hm:V\to V\ot\Gamma^*$ as follows:
$$
\hw^i: b\ot f\ot (w,m)\mapsto w^i(b\ot f\ot(w,m)),
\quad \hm_i: b\ot f\ot(w,m)\mapsto m_i(b\ot f\ot (w,m)).
$$
We also set
\begin{eqnarray}
Y^j(z) & = &
\sideset{}{'}\sum_{s=-\infty}^\infty \frac{\al_s^j}{s z^s},\nn\\
\bar{Y}^j(\bz) & = &
\sideset{}{'}\sum_{s=-\infty}^\infty \frac{\bal_s^j}{s \bz^s},\nn\\
\partial X^j(z) & = &\frac{1}{z} \Gi^{jk}\hp_k -\partial Y^j(z),\label{X}\\
\bpartial X^j(\bz) & = & \frac{1}{\bz} \Gi^{jk}\htp_k -\bpartial \bar{Y}^j(\bz),
\label{bX}\\
\psi^j(z) &=& \sum_{r\in \ZZ+\frac{1}{2}} \frac{\psi^j_r}{z^{r+1/2}},\label{psi} \\
\tpsi^j(\bz) &=& \sum_{r\in \ZZ+\frac{1}{2}} \frac{\tpsi^j_r}{\bz^{r+1/2}},\label{bpsi}
\end{eqnarray}
where a prime on a sum over $s$ means that the term with $s=0$ is omitted,
and $\hp_k$ and $\htp_k$ are defined by
$$
\hp_k=\frac{1}{\sqrt{2}}(\hm_k+\left(-\B_{kj}-G_{kj}\right)\hw^j),\quad
\htp_k=\frac{1}{\sqrt{2}}(\hm_k+\left(-\B_{kj}+G_{kj}\right)\hw^j).
$$
Note that we did not define
$X^j(z,\bz)$ themselves, but only their derivatives. The reason is that the would-be
field $X^j(z,\bz)$ contains terms proportional to $\log  z$ and $\log \bz,$ and therefore
does not belong to $QF_1(V).$

The vacuum vector of $\cV$ is defined by
$$
|vac\rangle=|vac,0,0\rangle.
$$
The operators $T,\bT\in \End(V)$ are defined by
\begin{align}
T&=\hp_j\al^j_{-1}+\sum_{s=1}^\infty
G_{jk}\al^j_{-1-s}\al^k_s+\sum_{r=\frac{1}{2},\frac{3}{2},
\ldots}\left(r+\frac{1}{2}
\right)\psi^j_{-1-r}\psi^k_r,\nn\\
\bT&=\htp_j\bal^j_{-1}+\sum_{s=1}^\infty
G_{jk}\bal^j_{-1-s}\bal^k_s+\sum_{r=\frac{1}{2},
\frac{3}{2},\ldots}\left(r+\frac{1}{2}
\right)\bpsi^j_{-1-r}\bpsi^k_r.\nn
\end{align}

The state-operator correspondence is defined as follows. The state space $V$ is spanned
by vectors of the form
\begin{equation}\label{basisvector}
\al^{j_1}_{-s_1}\ldots\al^{j_n}_{-s_n}\bal^{\bj_1}_{-\bs_1}\ldots \bal^{\bj_\bn}_{-\bs_\bn}
\psi^{i_1}_{-r_1}\ldots\psi^{i_q}_{-r_q}\bpsi^{\bi_1}_{-\br_1}\ldots
\bpsi^{\bi_\bq}_{-\br_\bq}|vac,w,m\rangle,
\end{equation}
where $n,\bn,q,\bq$ are nonnegative integers, $s_1,\ldots,s_n,\bs_1,\ldots,\bs_\bn$
are positive integers, and $r_1,\ldots,r_q,\br_1,\ldots,\br_\bq$ are positive half-integers.
This vector is mapped by $Y$ to the following quantum field:
\begin{multline}\label{Ymap}
\sum_{(w',m')\in \Gamma\oplus \Gamma^*}
\epsilon_{w,m'}\ \cT(w,m)\ pr_{(w', m')}
z^{-2G^{-1}(k, k')}\ \bz^{-2G^{-1}(\bar{k},\bar{k}')}
\exp\left( k_j Y^j(z)_+ + \bar{k}_j\bar{Y}^j(\bz)_+  \right)\\
:\prod_{l=1}^n \frac{\partial^{s_l} X^{j_l}(z) }{(s_l-1)!}
\prod_{\bl=1}^\bn \frac{\partial^{\bs_\bl} X^{\bj_\bl}(\bz) }{(\bs_\bl-1)!}
\prod_{t=1}^q \frac{\partial^{r_t-1/2}\psi^{i_t}(z) }{(r_t-\frac{1}{2})!}
\prod_{\bt=1}^\bq
\frac{\partial^{\br_\bt-1/2}\bpsi^{\bi_\bt}(\bz) }{(\br_\bt-\frac{1}{2})!}:\\
\exp\left( k_j Y^j(z)_- + \bar{k}_j\bar{Y}^j(\bz)_- \right).
\end{multline}

Here $k, \bar{k}, k', \bar{k}'$ are elements of $U^*$
defined by
$$
k_j=\frac{1}{\sqrt{2}}( m_j+\left(-\B_{jk}-G_{jk}\right)w^k),\quad
\bar{k}_j=\frac{1}{\sqrt{2}}( m_j+\left(-\B_{jk}+G_{jk}\right)w^k),
$$
$$
k'_j=\frac{1}{\sqrt{2}}( m'_j+\left(-\B_{jk}-G_{jk}\right)w^{'k}),\quad
\bar{k}'_j=\frac{1}{\sqrt{2}}( m'_j+\left(-\B_{jk}+G_{jk}\right)w^{'k}),
$$
the operator $\cT(w,m)$ is a translation  on the lattice $\Gamma\op\Gamma^*$:
$$
\cT(w,m): (a,b)\mapsto (a+w,b+m),
$$
and the operators $pr_{(w',m')}:V\to V$ are projections
onto the subspace $\cH_b\ot\cH_f\ot(w',m').$
Finally, $\epsilon_{w,m'}$ is a sign equal to $\exp(i\pi l(w,m')).$ We also remind
that for any meromorphic quantum field $a(z)$ the fields $a(z)_+$ and $a(z)_-$ are
defined by~(\ref{posnegfreq}), and there is a similar definition for the anti-meromorphic
fields. Thus $Y^j(z)_\pm$ and $\bY^j(\bz)_\pm$ are given by
\begin{align} \nn
Y^j(z)_-&=\sum_{s>0}\frac{\alpha^j_s}{sz^s}, &
Y^j(z)_+&=\sum_{s<0}\frac{\alpha^j_s}{sz^s}  \\ \nn
\bY^j(\bz)_-&=\sum_{s>0}\frac{\bal^j_s}{s\bz^s}, &
\bY^j(\bz)_+&=\sum_{s<0}\frac{\bal^j_s}{s\bz^s}.
\end{align}

One can easily check that~(\ref{Ymap}) is indeed a well-defined quantum field.
Furthermore, the vector~(\ref{basisvector}) is unchanged when the bosonic oscillators
are permuted, and is multiplied by the parity of the permutation when the fermionic
oscillators are permuted. For the map $Y$ to be well-defined, (\ref{Ymap}) must have
the same property. To see that this is indeed the case, note that the OPE of the
fields $\psi^j$ and $\partial X^j$ is given by
\begin{align}\label{freeOPE}
\partial X^j(z)\partial X^k(w)&\sim \frac{\Gi^{jk}}{(z-w)^2},\nn\\
\psi^j(z)\psi^k(w)&\sim \frac{\Gi^{jk}}{(z-w)},\nn\\
\partial X^j(z)\psi^i(z)&\sim 0,
\end{align}
and similarly for the anti-meromorphic fields. It follows that the singular part of the OPE
for $\psi^j,\bpsi^j,\partial X^j,\bpartial X^j$ and their derivatives is proportional
to the identity operator, and therefore their normal ordered product is supercommutative.

To facilitate the understanding of~(\ref{Ymap}), we list a few special cases of the
state-operator correspondence.

The state $\al^j_{-s}|vac,0,0\rangle$ is mapped by $Y$ to
$$
\frac{1}{(s-1)!}\partial^s X^j(z).
$$
The state $\bal^j_{-s}|vac,0,0\rangle$ is mapped to
$$
\frac{1}{(s-1)!}\bpartial^s X^j(\bz).
$$
The state $\psi^j_{-s}|vac,0,0\rangle$ is mapped to
$$
\frac{1}{(s-\frac{1}{2})!}\partial^{s-1/2}\psi^j(z).
$$
The state $\tpsi^j_{-s}|vac,0,0\rangle$ is mapped to
$$
\frac{1}{(s-\frac{1}{2})!}\bpartial^{s-1/2}\tpsi^j(\bz).
$$
The state $|vac,w,m\rangle$ is mapped to
\begin{multline}
\nn
\sum_{(w',m')\in \Gamma\oplus \Gamma^*}
\!\!\!\!\epsilon_{w,m'}\ \cT(w,m)\
z^{-2G^{-1}(k, k')}\ \bz^{-2G^{-1}(\bar{k},\bar{k}')}
\exp\left( k_j Y^j(z)_+  + \bar{k}_j \bY^j(\bz)_+  \right)\\
\exp\left( k_j Y^j(z)_-  + \bar{k}_j \bY^j(\bz)_-  \right)
pr_{(w', m')} .
\end{multline}

Checking that $(V,\vac,T,\bT,Y)$ satisfies the vertex algebra axioms
is a tedious but straightforward exercise which we leave to the reader.
Implicitly, the axioms are verified in most textbooks on string theory,
for example in~\cite{Polch,LT}.

\subsection{$N=2$ superconformal structure}

We first define an $N=1$ superconformal structure on $\cV$ by setting
\begin{align*}
L(z) &=\frac{1}{2} :G\left(\partial X(z),\partial X(z)\right):-
\frac{1}{2}:G\left(\psi(z),\partial\psi(z)\right):\ ,\nn\\
\bL(\bz)& =\frac{1}{2}:G\left(\bpartial X(\bz),\bpartial X(\bz)\right):
-\frac{1}{2}:G\left(\tpsi(\bz),\bpartial\tpsi(\bz)\right):\ ,\nn \\
Q(z)&=\frac{i}{2\sqrt{2}}:G\left(\psi(z),\partial X(z)\right):\ ,\nn\\
\bQ(\bz)&=\frac{i}{2\sqrt{2}}:G\left(\bpsi(\bz),\bpartial X(\bz)\right):\ . \nn
\end{align*}
It can be easily checked that all these fields are in the image of $Y,$
that $L_{-1}=T,\bL_{-1}=\bT,$ and that they satisfy the OPE's specified in the
Definition~\ref{openone}.
The central charges turn out to be $c=\bc=3d.$

To define an $N=2$ superconformal structure, we need to choose a
complex structure
$I$ on $U$ with respect to which $G$ is a K\"ahler metric. Let $\omega=GI$ be the
corresponding K\"ahler form. Then the left-moving supercurrents and the $U(1)$
current are defined as follows:
\begin{align*}
Q^\pm(z)&=\frac{i}{4\sqrt{2}}:G\left(\psi(z),\partial X(z)\right):\pm
\frac{1}{4\sqrt{2}}:\omega\left(\psi(z),\partial X(z)\right):\ ,\nn\\
J(z)&=-\frac{i}{2}:\omega(\psi(z),\psi(z)):\ .\nn
\end{align*}
The right-moving currents $\bar{Q}^\pm(\bz)$ and $\bar{J}(\bz)$
are defined by the same expressions with $\partial X$ replaced by
$\bpartial X$
and $\psi$ replaced by $\tpsi.$ We omit the check that the OPE's of
these currents are
as specified in the Definition~\ref{ntwo}.
In checking the OPE's the relations~(\ref{freeOPE}) are useful.

\section{Morphisms of toroidal superconformal vertex algebras}\label{morphisms}

\subsection{Isomorphisms of $N=1$ SCVA's}

Let $(T,G,\B)$ and $(T',G',\B')$ be a pair of $2d$-dimensional real tori equipped with
flat metrics and constant B-fields. Given $G$ and $\B,$ we define a flat metric on
$T\times T^*$ by the formula
$$
\cG(G,\B)=2 \begin{pmatrix} G-\B G^{-1}\B & \B G^{-1} \\ -G^{-1}\B & G^{-1} \end{pmatrix}.
$$
The meaning of this formula is that the value of $\cG$ on a pair
of vectors $x_1\op y_1$ and $x_2\op y_2,$ $x_i\in U, y_i\in U^*,i=1,2,$ is
$$
2\begin{pmatrix} x_1 & y_1 \end{pmatrix}
\begin{pmatrix} G-\B G^{-1}\B & \B G^{-1} \\ -G^{-1}\B & G^{-1} \end{pmatrix}
\begin{pmatrix} x_2 \\ y_2 \end{pmatrix}.
$$
$\cG(G,B)$ is obviously a symmetric form on $U\op U^*,$ and its
positive-definiteness follows from the positive-definiteness of $G$
and the identity
\begin{equation}\label{calGvsGB}
\cG=R(G,\B)^t\begin{pmatrix} G & 0 \\ 0 & G \end{pmatrix} R(G,\B),
\end{equation}
where
\begin{equation*}
R(G,\B)=\begin{pmatrix} -1-G^{-1}\B & G^{-1} \\ 1-G^{-1}\B & G^{-1}
\end{pmatrix}. \nn
\end{equation*}
We will use a shorthand $\cG(G,\B)=\cG$ and $\cG(G',\B')=\cG'.$ Recall also that
we have canonical $\ZZ$-valued symmetric bilinear forms on
$\Gamma\op\Gamma^*$ and $\Gamma'\op{\Gamma'}^*$ denoted by $q$ and $q',$ respectively
(see Section~\ref{results}).

In this subsection we prove
\begin{theorem}\label{noneiso}
$N=1$ SCVA's corresponding to $(T,G,\B)$ and $(T',G',\B')$ are
isomorphic if and only if there exists an isomorphism of lattices
$\Gamma\op\Gamma^*$ and
$\Gamma'\op{\Gamma'}^*$ which takes $q$ to $q',$ and $\cG$ to $\cG'.$
\end{theorem}

The ``if'' part of this theorem is proved in many string theory papers, see
for example~\cite{NSW,Zwiebach}. Below we outline a
construction of the isomorphism of $N=1$ SCVA's given an isomorphism of
lattices and then prove the ``only if'' part of the theorem.

Let $g$ be an isomorphism of $\Gamma\op\Gamma^*$ with
$\Gamma'\op{\Gamma'}^*.$ We will write it as follows:
$$
g=\begin{pmatrix} a & b \\ c & d \end{pmatrix},
$$
where $a\in \Hom(\Gamma,\Gamma'),$ $b\in \Hom(\Gamma^*,\Gamma'),$
$c\in \Hom(\Gamma,{\Gamma'}^*),$ $d\in \Hom(\Gamma^*,{\Gamma'}^*).$
The ``realified'' maps from $U,U^*$ to $U',{U'}^*$ will be denoted by the
same letters.
Let us also set $H=G+\B.$ Both $V$ and $V'$ are tensor products of
the group algebra of the respective lattice and bosonic and fermionic
Fock spaces. The vertex algebra isomorphism $f:V\ra V'$ respects this tensor
product structure. $\CC[\Gamma\op\Gamma^*]$ is mapped to
$\CC[\Gamma'\op{\Gamma'}^*]$ in an obvious way:
$$
f:\begin{pmatrix} w \\ m \end{pmatrix}\mapsto \begin{pmatrix} a & b \\ c & d \end{pmatrix}
\begin{pmatrix} w \\ m \end{pmatrix}.
$$
The mapping of Fock spaces is defined by the substitutions
\begin{align}
\begin{pmatrix} a^i_{-s} \\ \ba^i_{-s}\end{pmatrix}
& \mapsto \cM(g,H)^i_j
\begin{pmatrix} a^j_{-s} \\ \ba^j_{-s}\end{pmatrix}, & s=1,2,\ldots, \nn\\
\begin{pmatrix} \theta^i_{-s} \\ \btheta^i_{-s}\end{pmatrix}
& \mapsto \cM(g,H)^i_j
\begin{pmatrix} \theta^j_{-s} \\ \btheta^j_{-s}\end{pmatrix}, & s=\frac{1}{2},\frac{3}{2},\ldots,
\nn
\end{align}
where
$$
\cM(g,H)=\begin{pmatrix} a-bH^t & 0 \\ 0 & a+bH \end{pmatrix}.
$$
In particular, $f$ preserves the bosonic and fermionic vacuum
vectors.

Let us now indicate why this mapping is an isomorphism of $N=1$
SCVA's.
The statement that $g$ takes $q$ to $q'$ is equivalent to
\begin{equation}\label{abcd}
a^t c+c^t a=b^t d+d^t b=0,\quad a^t d+c^t b=id_{\Gamma^*},
\end{equation}
where $a^t$ denotes the conjugate of $a,$ etc.

The statement that
$g$ takes $\cG$ to $\cG'$ is equivalent to
\begin{equation}\label{fractlin}
H'=(c+dH)(a+bH)^{-1},
\end{equation}
where $H'=G'+\B', H=G+\B.$ To show this, let us denote the right-hand
side of the above equation by $H'',$ let $G''$ and $\B''$ be the
symmetric an anti-symmetric parts of $H'',$ and let $\cG''=\cG(G'',\B'').$
In view of~(\ref{calGvsGB}) we have
$$
\cG''=R(G'',\B'')^t\begin{pmatrix} G'' & 0 \\ 0 & G''\end{pmatrix}R(G'',\B'').
$$
Let us multiply this equation by $g^t$ from the left and by $g$ from the
right and use the identity
\begin{equation}\label{RandM}
R(G'',\B'')g=\cM(g,H)R(G,\B),
\end{equation}
which can be easily proved using~(\ref{abcd}).
We get
$$
g^t\cG''g=R(G,\B)^t \cM(g,H)^t \begin{pmatrix} G'' & 0 \\ 0 & G''
\end{pmatrix} \cM(g,H) R(G,\B).
$$
We now use another easily checked identity:
\begin{equation}\label{anotherid}
G''=\left[(a+bH)^t\right]^{-1} G (a+bH)^{-1}=\left[(a-bH^t)^t\right]^{-1}
G (a-bH^t)^{-1},
\end{equation}
and obtain
$$
g^t\cG''g=\cG.
$$
On the other hand, we know that $g^t\cG' g=\cG.$ Thus $\cG''=\cG',$
and hence $G''=G', \B''=\B', H''=H'.$ This proves~(\ref{fractlin}).
As a consequence of $G''=G'$ and (\ref{anotherid}), we obtain a useful
formula relating $G'$ and $G$:
\begin{equation}\label{usefulid}
G'=\left[(a+bH)^t\right]^{-1} G (a+bH)^{-1}=\left[(a-bH^t)^t\right]^{-1}
G (a-bH^t)^{-1}.
\end{equation}

Using these relations, one can easily check that the map $f$ intertwines
$Y$ and $Y',$ i.e.
\begin{equation}\label{Yf}
Y'(f(a),z,\bz)=f\, Y(a,z,\bz) f^{-1}, \quad \forall a\in V.
\end{equation}
In particular, we have
\begin{align}\label{fieldsf}
f^{-1}\begin{pmatrix} {\partial X'}^i(z) \\ {\bpartial X'}^i(z)
\end{pmatrix} f&=
\cM(g,H)^i_j\begin{pmatrix} {\partial X}^j(z) \\ {\bpartial X}^j(z)
\end{pmatrix}, \\ \nn
f^{-1}\begin{pmatrix} {\psi}^{'i}(z) \\ {\bpsi}^{'i}(z) \end{pmatrix} f&=
\cM(g,H)^i_j\begin{pmatrix} {\psi}^j(z) \\ {\bpsi}^j(z) \end{pmatrix}.
\end{align}
These relations and the definition of $L(z),\bL(\bz),Q(z),\bQ(\bz)$
imply that the $N=1$ superconformal structure is
also preserved:
\begin{align}\label{Lf}
L'(z)&=f L(z) f^{-1}, & Q'(z)&=f Q(z) f^{-1}, \\ \nn
\bL'(\bz)&=f \bL(\bz) f^{-1}, &  \bQ'(\bz)&=f \bQ(\bz) f^{-1}.
\end{align}
Hence $f$ is an isomorphism of $N=1$ superconformal vertex algebras.

In the remainder of this subsection we prove the ``only if'' part of the
theorem.
Let $(T,G,\B)$ and $(T',G',\B')$ be two real tori equipped with a flat metric and a
constant B-field. Thus $T=U/\Gamma$ and $T'=U'/\Gamma',$ where $U$ and $U'$ are real
vector spaces and $\Gamma$ and $\Gamma'$ are lattices of maximal rank in the respective spaces.
Clearly, for the $N=1$ SCVA's to be isomorphic, the central charges
of the corresponding super-Virasoro algebras must agree, hence $\dim U=\dim U'.$
We pick an isomorphism of $U$ and $U'$ and a basis in $U.$
Let $\cV=(V,Y,|vac\rangle,L,\bL,Q,\bQ)$ and $\cV'=(V',Y',|vac'\rangle,L',\bL',
Q',\bQ')$ be the corresponding $N=1$ SCVA's.
Let $f:V\ra V'$ be an isomorphism of $N=1$ SCVA's.
This means that the equations~(\ref{Yf}) and (\ref{Lf}) hold
true. In particular, $f$ preserves the form of the OPE.

Consider the ``Hamiltonians'' $L_0,\bL_0\in \End(V).$ A short computation yields:
$$
L_0=\frac{1}{8}\cG(Z,Z)-\frac{1}{4}q(Z,Z)+N_b+N_f, \qquad
\bL_0=\frac{1}{8}\cG(Z,Z)+\frac{1}{4}q(Z,Z)+\bN_b+\bN_f.
$$
Here $Z=(W,M)$ is regarded as an element of $\End(\cH)\ot_\RR (U\op U^*),$
and we defined
\begin{align*}
N_b&=\sum_{s=1}^\infty G\left(\alpha_{-s},\alpha_s\right),&
N_f&=\sum_{r=1/2,3/2,\ldots} rG\left(\psi_{-r},\psi_r\right),\\
\bN_b&=\sum_{s=1}^\infty G\left(\bal_{-s},\bal_s\right),&
\bN_f&=\sum_{r=1/2,3/2,\ldots} rG\left(\bpsi_{-r},\bpsi_r\right).
\end{align*}

The operators $N_b,N_f,\bN_b,\bN_f$ commute with each other.
For what follows it is important to know their spectrum in Fock space. One can show that 
the Fock space decomposes into a tensor sum of the joint eigenspaces of $N_b,N_f,\bN_b,\bN_f,$
and that all the eigenvalues are nonnegative. Furthermore, the spectrum of
$N_b,\bN_b$ is integer, and the spectrum of $N_f,\bN_f$ is half-integer.
Finally, the only vector in $\fH_b\ot\fH_f$ annihilated by all four operators is
$|vac_b\rangle\ot|vac_f\rangle.$ (All of these facts are standard and can be
easily proved using the commutation relations for the oscillators.)

Note also that the spectrum of the operator $\cG(Z,Z)$ is nonnegative
because $\cG$ is a positive-definite form. The only vector in $\CC
\left[\Gamma\op\Gamma^*\right]$ annihilated by $\cG(Z,Z)$ is $(0,0).$

Now let us find all the eigenvectors of $L_0,\bL_0$ with eigenvalues $(1/2,0).$
Suppose $a\in V$ is such an eigenvector.
Since $L_0,\bL_0$ commute with $Z=(W,M),$ we may assume that $a$ is an eigenvector
of $Z$ with an eigenvalue $z=(w,m),$ where $w\in\Gamma, m\in\Gamma^*.$
In view of the above we have three possibilities:
\begin{flalign}
{\rm Case\ 1.}
&&& N_b\, a=\bN_b\, a=N_f\, a=\bN_f\, a=0, &&\nn\\
&&& \frac{1}{2}\cG(z,z)-q(z,z)=2,&\qquad \nn
\frac{1}{2}\cG(z,z)+q(z,z)=0.&\qquad\\
{\rm Case\ 2.}
&&& N_b\, a=\bN_b\, a=N_f\, a=0, &\qquad
\left(\bN_f-\frac{1}{2}\right)a=0,&\qquad\nn\\
&&& \cG(z,z)=q(z,z)=0.&&\nn \\
{\rm Case\ 3.}
\label{three}
&&& N_b\, a=\bN_b\, a=\bN_f\, a=0, &\qquad \left(N_f-\frac{1}{2}\right)a=0,&
\qquad\\
&&& \cG(z,z)=q(z,z)=0.\nn &&
\end{flalign}

The first case is ruled out, because we must have $q(z,z)=-1,$ in
contradiction with the fact that $q$ is an even form.

In the second case, we must have $z=0.$ Then from the formulas for $L_0,\bL_0$
we see that such a vector has eigenvalues $(0,1/2)$ rather than $(1/2,0).$
Hence this case is also ruled out.

In the third case, we must have $z=0.$ Furthermore, it is easy to see that
all vectors satisfying~(\ref{three}) must also satisfy
\begin{align*}
\alpha^i_s a&=\bal^i_s a=0, & i&=1,\ldots,2d,\ s=1,2,\ldots,\\
\bpsi^i_r a&=0, & i&=1,\ldots,2d,\ r=1/2,3/2,\ldots,\\
\psi^i_r a&=0, & i&=1,\ldots,2d,\ r=3/2,5/2,\ldots.
\end{align*}
It follows that $a$ must have the form
$$
a=\left(\sum_{i=1}^{2d} c_i\ \psi^i_{-1/2}\right)|vac,0,0\rangle,
$$
where $c_i,\ i=1,\ldots,2d,$ are arbitrary complex numbers.
A similar argument shows that all eigenvectors of $L_0,\bL_0$ with
eigenvalues $(0,1/2)$ have the form
$$
\left(\sum_{i=1}^{2d} {\bar c}_i\ \bpsi^i_{-1/2}\right)|vac,0,0\rangle,
$$
where ${\bar c}_i,\ i=1,\ldots,2d,$ are arbitrary complex numbers.

Now recall that $L'_0f=fL_0$ and $\bL'_0f=f\bL_0.$ This implies that
$f$ identifies the $(1/2,0)$ eigenspace of $(L_0,\bL_0)$ with the
$(1/2,0)$ eigenspace of $(L'_0,\bL'_0),$ and $(0,1/2)$ eigenspace of
$(L_0,\bL_0)$ with the $(0,1/2)$ eigenspace of $(L'_0,\bL'_0).$
Thus there exist two invertible complex matrices $F^i_j$ and $\bF^i_j$
such that
\begin{align*}
\psi^{'i}_{-1/2}|vac,0,0\rangle &=f\left( F^i_j\psi^j_{-1/2}
|vac,0,0\rangle\right),\\
\bpsi^{'i}_{-1/2}|vac,0,0\rangle &=f\left( \bF^i_j\bpsi^j_{-1/2}
|vac,0,0\rangle\right).
\end{align*}
Applying $Y'$ to both sides of this equation and using~(\ref{Yf}), we obtain:
$$
\psi^{'i}(z)=f\, F^i_j\psi^j(z)\, f^{-1},
\qquad \bpsi^i(\bz)=f\,\bF^i_j\bpsi^j(\bz)\, f^{-1}.
$$

An immediate consequence of this is the transformation law for fermionic
oscillators:
$$
\psi^{'i}_r=f\, F^i_j\psi^j_r\, f^{-1},\qquad \bpsi^{'i}_r=
f\, \bF^i_j\bpsi^j_r\, f^{-1},\qquad r\in\ZZ+\frac{1}{2}.
$$
Compatibility with the commutation relations of the fermionic oscillators
then requires:
\begin{equation*}
F^TG'F=G, \qquad \bF^TG'\bF=G.
\end{equation*}
(Alternatively, one may derive this by comparing the OPE of
$\psi(z),\bpsi(\bz)$ with
themselves and the OPE of $\psi'(z),\bpsi'(\bz)$ with themselves.)

Now let us turn to  bosonic oscillators.
Consider the OPE of $Q(z)$ with $\psi(z)$:
$$
Q(z)\psi^i(w)\sim \frac{i}{2\sqrt{2}}\frac{\partial X^i(w)}{(z-w)}.
$$
Since $f$ preserves the OPE and takes $Q(z)$ to $Q'(z),$ and $\psi(w)$ to
$F^{-1}\psi'(w),$ we infer that
$$
\partial X^{'i}(z)=f\, F^i_j\partial X^j(z)\, f^{-1}.
$$
Similarly, the OPE of $\bQ(\bz)$ with $\bpsi(w)$ implies that
$$
\bpartial\bX^{'i}(\bz)=f\, \bF^i_j\bpartial\bX^j(\bz)\, f^{-1}.
$$
These formulas imply the following transformation
laws for bosonic oscillators:
$$
\alpha^{'i}_n=f\, F^i_j\alpha^j_n\, f^{-1},\qquad \bal^{'i}_n=f\, \bF^i_j
\bal^j_n\, f^{-1},\quad n\in\ZZ.
$$

Another consequence is the transformation law of $Z$:
$$
Z'= f\, gZ\, f^{-1},
$$
where $g\in \Hom_\RR(U\op U^*,U'\op {U'}^*)$ is defined by
\begin{equation}
\nn
g=R(G',\B')^{-1}\begin{pmatrix} F & 0 \\ 0 & \bF \end{pmatrix} R(G,\B),
\end{equation}
and $Z$ and $Z'$ are regarded as elements of
$\End(\CC[\Gamma\op\Gamma^*])\ot_\RR (U\op U^*)$ and
$\End(\CC[\Gamma'\op{\Gamma'}^*])\ot_\RR (U'\op {U'}^*).$ Now note that
$Z$ and $Z'$
are in fact ``realifications'' of some elements in
$\End(\CC[\Gamma\op\Gamma^*])\ot_\ZZ (\Gamma\op\Gamma^*)$ and
$\End(\CC[\Gamma'\op{\Gamma'}^*])\ot_\ZZ (\Gamma'\op{\Gamma'}^*).$
This means that $g$ is a ``realification'' of an element of
$\Hom_\ZZ(\Gamma\op \Gamma^*,\Gamma'\op {\Gamma'}^*),$ which we also denote $g.$

It remains to show that $g$ takes $q$ to $q'$ and $\cG$ to $\cG'.$
To this end notice that the transformation laws for the oscillators imply
$$N'_b=f N_b f^{-1}, \quad N'_f=f N_f f^{-1}, \quad \bN'_b=f \bN_b f^{-1}, \quad
\bN'_f=f \bN_f f^{-1}.
$$
Then it follows from $L'_0=fL_0 f^{-1}$ and $\bL'_0=f\bL_0 f^{-1}$ that
for all $x\in \Gamma\op\Gamma^*$ we have
\begin{align*}
\nn
q'(gx,gx)&=q(x,x),\\
\nn
\cG'(gx,gx)&=\cG(x,x).
\end{align*}
This concludes the proof of the theorem.

\subsection{Isomorphisms of $N=2$ SCVA's}

The goal of this subsection is to prove Theorem~\ref{ntwoiso} which we
restate below. Given a metric $G$ on $U,$ a compatible
complex structure $I$ on $U,$ and $\B\in \Lambda^2 U^*,$ we define a pair of commuting
complex structures on $U\op U^*$ as follows:
\begin{align*}
\cI(I,\B)&=\begin{pmatrix} I & 0 \\ \B I+I^t\B & -I^t \end{pmatrix}, \\
\cJ(G,I,\B)&=\begin{pmatrix} -IG^{-1}\B & IG^{-1} \\
GI-\B IG^{-1}\B & \B IG^{-1}
\end{pmatrix}.
\end{align*}
The complex structure $\cJ$ can be expressed in terms of the K\"ahler form
$\omega=GI$ and $\B$:
$$
\cJ(\omega,\B)= \begin{pmatrix} \omega^{-1}\B & -\omega^{-1} \\ \omega+\B\omega^{-1}\B
& -\B\omega^{-1} \end{pmatrix}.
$$
We will use a simplified notation $\cI(I,\B)=\cI,$ $\cI(I',\B')=\cI',$ etc.
The complex structures $\cI,\cJ$ and the symmetric forms $\cG,q$
are related by an identity
$$\cG=-2q\cI\cJ,$$
where $\cG$ and $q$ are understood as elements of $\Hom_\RR(U,U^*).$

\begin{theorem}\label{thntwoiso}
$Vert(\Gamma,I,G,\B)$ is isomorphic to $Vert(\Gamma',I',G',\B')$ as an $N=2$ SCVA
if and only if
there is an isomorphism of lattices $\Gamma\op\Gamma^*$ and $\Gamma'\op\Gamma^{'*}$
which takes $q$ to $q',$ $\cI$ to $\cI',$ and $\cJ$ to $\cJ'.$
\end{theorem}

To prove this theorem, note that $f:V\ra V'$ is an isomorphism of $N=2$ SCVA's
if and only if it is an isomorphism of the underlying $N=1$ SCVA's, and
maps $J(z)$ to $J'(z)$ and $\bJ(\bz)$ to $\bJ'(\bz).$ Now suppose $f$ is an
isomorphism of $N=1$ SCVA's underlying $Vert(\Gamma,I,G,\B)$ and
$Vert(\Gamma',I',G',\B').$ By Theorem~\ref{noneiso} we know that there exists
$g\in \Hom(\Gamma\op\Gamma^*,\Gamma'\op{\Gamma'}^*)$ which takes $q$ to $q',$
and $\cG$ to $\cG'.$
To prove the theorem, it is sufficient to show that $f$ maps $J(z),\bJ(\bz)$ correctly
if and only if $g$ maps $\cI$ to $\cI'$ and $\cJ$ to $\cJ'.$ In fact, since
$\cG=-2q\cI\cJ$ and $\cG'=-2q'\cI'\cJ',$ it is sufficient to show that $f$
maps $J(z),\bJ(\bz)$ correctly if and only if $g$ maps $\cI$ to $\cI'.$

Using the transformation law~(\ref{fieldsf}) for the fields and the formula
(\ref{usefulid}) relating
$G$ and $G',$ one can easily see that $f$ maps $J(z)$ to $J'(z)$ if and only if
\begin{equation}\label{Ione}
I'=(a-bH^t)I(a-bH^t)^{-1}.
\end{equation}
Similarly, $f$ maps $\bJ(\bz)$ to $\bJ'(\bz)$ if and only if
\begin{equation}\label{Itwo}
I'=(a+bH)I(a+bH)^{-1}.
\end{equation}

On the other hand, $\cI(I,\B)$ can be written as
\begin{equation}\label{calIandI}
\cI(I,\B)=R(G,\B)^{-1}\begin{pmatrix} I & 0\\ 0 & I\end{pmatrix} R(G,\B).
\end{equation}
This and the identity~(\ref{RandM}) imply that $\cI'=g\cI g^{-1}$ if and only
if
$$
\begin{pmatrix} I' & 0 \\ 0 & I' \end{pmatrix}=\cM(g,H)
\begin{pmatrix} I & 0 \\ 0 & I \end{pmatrix} \cM(g,H)^{-1}.
$$
This matrix identity is equivalent to~(\ref{Ione},\ref{Itwo}), which proves the theorem.

Let us also note the following simple corollary of this theorem.
\begin{cor}
Let $(T,I,G,B)$ be a complex torus equipped with a flat K\"ahler metric and a B-field of type
$(1,1).$ Let $T'=U'/\Gamma'$ be another torus of the same dimension and $I'$ be a complex
structure on $T'$. Let $\tilde{\cI}$ and $\tilde{\cI'}$ be the product complex structures
on $T\times T^*$ and $T'\times T^{'*}$. Suppose there exists an isomorphism of 
lattices $g:\Gamma\op\Gamma^*\ra \Gamma'\op\Gamma^{'*}$ mapping $q$ to $q'$ and 
$\tilde{\cI}$ to $\tilde{\cI'}.$ Then on $T'$ there exists a K\"ahler metric $G'$
and a B-field of type $(1,1)$ such that 
$Vert(\Gamma,I,G,\B)$ is isomorphic to $Vert(\Gamma',I',G',\B')$ as an $N=2$ SCVA.
\end{cor}
To show this, we define $H'$ using~(\ref{fractlin}) and set $G'$ and $B'$ to be the
symmetric and skew-symmetric parts of $H',$ respectively. Then it follows from~(\ref{usefulid})
that $G'$ is positive-definite. By Theorem~\ref{noneiso} the $N=1$ SCVA corresponding
to $(T,G,B)$ is isomorphic to $N=1$ SCVA corresponding to $(T',G',B')$. Using that fact
that $g$ intertwines $\tilde{\cI}$ to $\tilde{\cI'}$ it is easy to show that $H'I'+I^{'t}H'=0,$
which means that $G'$ is a K\"ahler metric and $B'$ has type $(1,1)$. In particular,
$\tilde{\cI'}=\cI'.$ Then it follows
from the identity $\cG'=-2q'\cI'\cJ'$ and the fact $g$ intertwines $\cG,q,\cI$ and
$\cG',q',\cI'$ that $g$ also intertwines $\cJ$ and $\cJ'.$ Theorem~\ref{thntwoiso} then implies
that $Vert(\Gamma,I,G,\B)$ is isomorphic to $Vert(\Gamma',I',G',\B')$ as an $N=2$ SCVA.

\subsection{Mirror morphisms of $N=2$ SCVA's}

In this subsection we establish a criterion for the existence of a mirror
morphism between two complex tori equipped with flat K\"ahler
metrics and B-fields.
\begin{theorem}\label{thntwomirror}
$Vert(\Gamma,I,G,\B)$ is mirror to $Vert(\Gamma',I',G',\B')$ if and only if
there is an isomorphism of lattices $\Gamma\op\Gamma^*$ and $\Gamma'\op\Gamma^{'*}$
which takes $q$ to $q',$ $\cI$ to $\cJ',$ and $\cJ$ to $\cI'.$
\end{theorem}

The proof is very similar to that of Theorem~\ref{thntwoiso}. Again it is sufficient
to show that if $f$ is an isomorphism of the underlying $N=1$ SCVA's, and
$g$ the corresponding isomorphism of lattices, then
\begin{equation}
\nn
fJ(z)f^{-1}=-J'(z), \quad f\bJ(\bz)f^{-1}= \bJ'(\bz)
\end{equation}
is equivalent to
\begin{equation}\label{gIJ}
g\cJ g^{-1}=\cI'.
\end{equation}

The first of these is equivalent to
\begin{equation}\label{IIp}
I'=(a+bH)I(a+bH)^{-1}=-(a-bH^t)I(a-bH^t)^{-1}.
\end{equation}
On the other hand, $\cJ(G,I,\B)$ can be written as
$$
\cJ(G,I,\B)=R(G,\B)^{-1}\begin{pmatrix} -I & 0\\ 0 & I\end{pmatrix} R(G,\B),
$$
which together with~(\ref{RandM}) and (\ref{calIandI}) implies that~(\ref{gIJ})
is equivalent to
$$
\cM(g,H) \begin{pmatrix} -I & 0 \\ 0 & I \end{pmatrix} \cM(g,H)^{-1}=
\begin{pmatrix} I' & 0 \\ 0 & I' \end{pmatrix}.
$$
This is obviously equivalent to~(\ref{IIp}). This concludes the proof.

\section{Homological mirror symmetry with B-fields}\label{modHMSC}

\subsection{Mirror symmetry and D-branes}

As explained in Section~\ref{results}, Kontsevich's conjecture must be modified
if the B-field does not vanish. When the image of $\Be$ in $H^2(X,\O_X^*)$ is
torsion, our results on complex tori suggest that
the bounded derived category $D^b(X)$ should be replaced with $D^b(X,\Be),$ the
bounded derived category of coherent modules over an Azumaya algebra. The similarity
class of the Azumaya algebra is determined by the image of $\Be$ in $H^2(X,\O_X^*).$
(Presumably, when $\Be$ does not map to a torsion class, the proper analogue
of $D^b(X)$ is some ``coherent'' subcategory of the derived category of
quasicoherent sheaves on a gerbe over $X,$ see Remark~\ref{gerbe}.)
However, this does not provide any hint as to what the modification of the Fukaya
category might be. In this section we explain some string theory lore which suggests
a particular definition of the Fukaya category in the presence of the B-field.
A similar proposal has been made in~\cite{AP}.

The ordinary $\sigma$-model whose quantization yields an $N=2$ superconformal vertex algebra
is a classical field theory on a two-dimensional manifold $\Sigma=\RR\times\SS^1$
(``the worldsheet''). Let us replace $\SS^1$ with an interval $I=[0,1]$ and consider the same
$\sigma$-model on a worldsheet with boundaries $\RR\times I.$ This procedure
is referred to as passing from closed to open strings.
Now, in order to make
the space of solutions of the Euler-Lagrange equations a symplectic supermanifold, one
has to supply boundary conditions for the fields of the $\sigma$-model on both ends
of the interval. In addition one requires that these boundary conditions preserve $N=2$
superconformal symmetry. To be more precise, while the classical $\sigma$-model on
$\RR\times\SS^1$ has two copies of the $N=2$ super-Virasoro algebra
(with zero central charge) as its classical symmetry, the $\sigma$-model on $\RR\times I$ is 
required to be
symmetric only with respect to a single $N=2$ super-Virasoro algebra. There are
two essentially different classes of such boundary conditions, called A and B boundary
conditions. The B-type boundary conditions preserve the ``diagonal'' super-Virasoro
subalgebra whose generators are given by
$$
L_n+\bL_n,\quad J_n+\bJ_n,\quad Q^+_r+\bQ^+_r,\quad Q^-_r+\bQ^-_r,\quad n\in\ZZ,
r\in\ZZ+\frac{1}{2}.
$$
The A-type boundary conditions preserve a different subalgebra whose generators
are
$$
L_n+\bL_n,\quad -J_n+\bJ_n,\quad Q^-_r+\bQ^+_r,\quad Q^+_r+\bQ^-_r, \quad n\in\ZZ,
r\in\ZZ+\frac{1}{2}.
$$
Superconformally-invariant boundary
conditions for a $\sigma$-model are called supersymmetric (or BPS, for
Bogomolny-Prasad-Sommerfeld) D-branes. Thus we have BPS D-branes of types A and B.
Note that the mirror involution~(\ref{mirrorinv}) exchanges the two types of D-branes.

D-branes are understood best when the B-field is zero. In this case one
can construct examples of the B-type
boundary conditions by starting from a holomorphic submanifold of the Calabi-Yau
manifold $X.$ More generally, one can start from a holomorphic submanifold $M\subset X$
and a holomorphic bundle on $M$ equipped with a compatible connection. On the other
hand, examples of the A-type boundary conditions (with zero B-field)
can be constructed starting from
a Lagrangian submanifold $L\subset X$ (with respect to the K\"ahler form),
a trivial unitary bundle $E$ on $L,$ and a unitary flat connection on $E.$

Note that one can choose different boundary conditions for the Euler-Lagrange equations
on the two ends of the interval $I.$ The only constraint is that both boundary conditions
must be of the same type (A or B). If this condition is violated, then the symmetry of
the corresponding classical field theory is only some subalgebra of the $N=2$
super-Virasoro algebra, namely an $N=1$ super-Virasoro algebra.

After quantization, $\sigma$-model on $\RR\times I$ is supposed to yield
a superconformally invariant quantum field theory on the same manifold. The
axioms of such quantum field theories have not been formulated yet, and we
will not attempt it here. Suffice it to say that physicists expect that any
B-type D-brane can be consistently quantized, while A-type boundary conditions
may lead to ``anomalies,'' i.e. inconsistencies in the quantization procedure. 
One can argue that anomalies are absent if
the A-type D-brane
originates from a {\it special} Lagrangian submanifold. We remind that a special Lagrangian
submanifold in a Calabi-Yau manifold with a K\"ahler metric is defined
by two properties:
it is Lagrangian, and the restriction of a nonzero section of the
canonical bundle to the
submanifold is proportional to its volume form.

Thus to any physicist's Calabi-Yau with zero B-field one can associate two sets: the
set of B-type
D-branes, and the set of (non-anomalous) A-type D-branes. The former
set has many elements in common with the set of coherent sheaves on $X.$
The latter set resembles the set of objects the Fukaya category of $X.$
Moreover, there are heuristic arguments using path integrals showing that either
A or B-type D-branes form an
$A_\infty$\!--category (see~\cite{topstring} and references therein).
Thus, conjecturally, to every physicist's Calabi-Yau with zero B-field
one can canonically associate a pair of $A_\infty$\!--categories, the
categories of A- and B-type D-branes. Assuming there are shift functors on them,
one can define the corresponding triangulated categories as in~\cite{K}.

It is natural to conjecture that for $\Be=0$ the triangulated category associated
with A-type (resp. B-type) D-branes is equivalent to $D\cF(X)$ 
(resp. $D^b(X)$)~\cite{Vafa,Douglas}.
There are several pieces of evidence supporting this conjecture. First, as we have already
remarked, $\cF(X)$ and $Coh(X)$ have many objects in common with the categories
of A and B-type D-branes, respectively. Second, using path integrals one
can argue~\cite{Witten} that the
category of B-type D-branes is independent of the K\"ahler form,
while the category of A-type D-branes is independent of the complex
structure on $X$ if $\omega$ is fixed. For further evidence see~\cite{Douglas}
and references therein.

If this conjecture is true, then Kontsevich's conjecture has a natural
explanation. Suppose we have a mirror pair of physicist's Calabi-Yaus $X$ and $X',$ both
with zero B-field. The corresponding $N=2$ SCVA's are related by a mirror morphism.
Since a mirror morphism of $N=2$ SCVA's acts on the $N=2$ super-Virasoro by the
mirror involution, it exchanges the A and B-type boundary conditions. Hence it
induces an equivalence of $D^b(X)$ with the derived Fukaya category
$D\cF_0(X'),$ and vice versa.

\subsection{Fukaya category with a B-field}

Now let us generalize this to nonzero B-fields. We already know the effect of a
B-field on $D^b(X)$: the sheaf $\O_X$ is replaced with a certain sheaf of
noncommutative algebras.
This agrees with the string theory lore that the B-field makes the D-brane
worldvolume noncommutative~\cite{CDS,DH}.

The effect of the B-field on the Fukaya category seems rather different. Let us
start by recalling the definition of the set of objects of the Fukaya
category~\cite{K}.
Let $(X,\omega)$ be
a symplectic manifold of dimension $2d.$ We fix an almost complex structure $I$ on $X$ compatible
with $\omega$ and thereby obtain a Hermitian metric on $X.$ (If $X$ is
a physicist's Calabi-Yau, it automatically comes equipped with a compatible
complex structure).
Moreover, we assume that $c_1(T^{hol}_X)=0$. In this case the line bundle
$\Lambda^d(\Omega^{hol}_X)$ is trivial and has a nowhere vanishing holomorphic section $\Omega$ 
which is called a calibration. 

Naively, an object of the Fukaya category
should be a triple $(L,E,\nabla),$ where $L$ is a Lagrangian submanifold,
$E$ is a trivial unitary vector
bundle on $L,$ and $\nabla$ is a flat connection on $E.$ From the physical point of view,
such a triple allows one to define an A-type boundary condition for the classical
$\sigma$-model, and therefore it is an A-type D-brane~\cite{Witten,OogOzYin}.

The naive definition of an object does not allow one to define a nontrivial
shift functor and $A_\infty$ structure.
This difficulty can be overcome as follows~\cite{K}.
For any point $x\in L$ the tangent space $T_x L$ is
a Lagrangian subspace of $T_x X.$ The Grassmannian of Lagrangian
subspaces has fundamental group equal to $\ZZ.$
Each Lagrangian submanifold comes with a Gauss map
from $L$ to $\L G,$ where $\L G\to X$ is a fibration whose fiber over $x$
is the Grassmannian of Lagrangian subspaces of $T_x X.$
Consider a fibration
$\wt{\L G}\to X$ covering $\L G\to X$ such that its fiber is 
the universal cover of the fiber of $\L G\to X.$
(As mentioned in \cite{K}, there is a canonical choice of such a fibration if
$c_1(T^{hol}_X)=0.$)
Instead of $L,$ we will consider pairs $(L, i),$ where $i$ is a lift of the Gauss
map to $\wt{\L G}.$ Not every Lagrangian $L$ admits such a lift, so not any 
Lagrangian submanifold can be extended to an object of the Fukaya
category. Note that any Lagrangian $L$ comes equipped with two natural
$d$-forms: the volume form and the restriction of the calibration $\Omega$.
The latter is defined up to a multiplicative constant.
Their quotient is a nowhere vanishing function $f$ which maps $L$ to $\CC^*$.
One can show that the Gauss map admits a lift to $\wt{\L G}$ if and only if 
the image $f(L)$ is contractible.
For example, any special Lagrangian $L$ has a lift, because
by definition of speciality the function $f$ is constant for any such $L$.

To summarize, we can define an object of the Fukaya category in the
absence of the
B-field as a quadruple $(L,i,E,\nabla),$ where $L$ and $i$ are as above,
and $(E,\nabla)$ is a trivial complex vector bundle on $L$ with a unitary flat connection.
The natural fiberwise action
of $\ZZ$ on
$\wt{\L G}\to X$ induces an action of $\ZZ$ on such quadruples. One hopes
that this action extends to a shift functor from the Fukaya category to itself.

Now let us try to guess how the definition of the Fukaya category
 should be modified when
$\Be\neq 0.$ Let $\B$ be a closed 2-form on $X$ representing $\Be\in H^2(X,\RR/\ZZ).$ 
(Since we assumed that
$\Be$ is in the kernel of the Bockstein homomorphism $H^2(X,\RR/\ZZ)\ra H^3(X,\ZZ),$
such a 2-form exists.)
Let $F_\nabla$ be the curvature of
a connection $\nabla$ on a bundle $E$ on $L.$ If $\B=0,$ the condition
on $\nabla$ is
\begin{equation}\label{naive}
F_\nabla=0.
\end{equation}
On the other hand, it is a general principle of string theory that
the equations
of motion must be invariant with respect to a substitution
\begin{equation}\label{symmetry}
\B\ra \B+d\lambda, \quad \nabla\ra \nabla+2\pi i\ id_E\,\lambda\vert_L,
\end{equation}
where $\lambda$ is any real 1-form on $X.$ This must be true because the
action of the $\sigma$-model on $\RR\times I$ is invariant with respect to
such transformations~\cite{Polch}.
This requirement is sufficient to fix the generalization of~(\ref{naive}) to
arbitrary $\B$:
\begin{equation}\label{FB}
F_\nabla=2\pi i\ id_E \B\vert_L.
\end{equation}
We propose that an object of the Fukaya category for $\Be\neq 0$ is
a quadruple $(L,i,E,\nabla),$ where $L$ and $i$ are the same as above, $E$ is a complex
vector bundle on $L,$ and $\nabla$ is a connection on $E$ satisfying~(\ref{FB}).

We can make some checks of this proposal. First, our definition of an object
depends on how one lifts $\Be\in H^2(X,\RR/\ZZ)$ to a 2-form $\B.$ However, given
two different 2-forms $\B_1$ and $\B_2$ representing $\Be,$ there is a one-to-one
map between the corresponding sets of objects. Indeed, let $f=\B_2-\B_1.$ It is easy
to see that $f$ has integral periods, and therefore there exists a line bundle
$\cN$ on $X$ and a connection $\nabla_0$ on $\cN$ such that the curvature
of $\nabla_0$ is equal to $2\pi i f.$ The bijection between the set of objects
corresponding to $\B_1$ and the set of objects corresponding to $\B_2$ is given by
\begin{equation}\label{newB}
L\mapsto L,\quad i\mapsto i,\quad E\mapsto E\ot \cN\vert_L,\quad \nabla_1\mapsto
\nabla_1\otimes id_{\cN} + id_{E}\otimes \nabla_0.
\end{equation}
Second, from the equation~(\ref{FB}) we see that $c_1(E)=\rank(E) b\vert_L,$ where
$b$ is the de Rham cohomology class of $\B.$
Since $c_1(E)$ is integral, we infer that
$$
\rank(E)\Be\vert_L=0.
$$
In particular, for $\rank(E)=1,$ we get that the restriction of $\Be$ to $L$
must vanish. This is consistent with the results of Hori et al.~\cite{HoriIqVafa},
who analyzed the A-type boundary conditions in the rank-one case. Hori et al.
find that the restriction of $B$ to $L$ must be zero if one wants to make
an A-type D-brane out of $L.$ We found that it is sufficient to require
$\Be\vert_L=0.$

We need to address one more subtlety. The original HMSC required $E$ to be a unitary
vector bundle and $\nabla$ to be a unitary connection~\cite{K}. This requirement
naturally arises in the string theory context as well. Nevertheless, this condition is
much too strong. Even in the case of the elliptic curve one has to allow for
non-unitary connections on the A-side if
one wants to account for all bundles
on the B-side~\cite{PoliZas}. In that case,
the right thing to do is to require
the holonomy representation of $\nabla$ to have
eigenvalues with unit modulus. It is natural to
conjecture that this is also the right requirement for
$\dim_\CC X>1$ or $\Be\neq 0.$

In the absence of the B-field, any pair $(L,i)$ can be extended
(in many
different ways) to an object of the Fukaya category. The situation
is more complex
for $\Be\neq 0.$ Recall that
to any flat connection on a manifold $L$ one can canonically associate
a finite-dimensional representation of $\pi_1(L)$
(or, equivalently, a finite-dimensional representation
of the group algebra of $\pi_1(L)$),
and vice versa.
In fact,
this map is a one-to-one correspondence.  Similarly,
given a bundle $E$ on $L$ and a connection $\nabla$ on $E$ such that
$F_\nabla$ satisfies~(\ref{FB}),
one can construct a finite-dimensional representation
of a twisted group algebra of $\pi_1(L)$ in the following way.
To $(E, \nabla)$ we can associate a projective representation $R$
of $\pi_1(L).$
To any such $R$ one can attach an element $\psi_R$ of
$H^2(\pi_1(L),U(1)).$
Acting on it with the natural embedding
\begin{equation}\label{Hopf}
H^2(\pi_1(L),U(1))\stackrel{j}{\ra} H^2(L,U(1))
\end{equation}
we obtain an element $j(\psi_R)\in H^2(L,U(1)).$
One can show that
$j(\psi_R)=\Be\vert_L$(we identify $\RR/\ZZ$ with $U(1)$).

To any 2-cocycle $\psi$ one can associate a twisted group algebra
$\CC_{\psi}[\pi_1(L)],$ which is a vector space generated by the elements
of $\pi_1(L)$ with the following multiplication law:
$$
g\cdot h=\psi(g,h)gh, \qquad g,h\in\pi_1(L).
$$

The correspondence between
pairs $(E,\nabla)$ satisfying (\ref{FB}) and finite-dimensional
representations of the twisted group algebra
$\CC_{\psi}[\pi_1(L)]$
is one-to-one. A proof of this fact is given in Appendix~\ref{projflat}.
The eigenvalues of the holonomy representation of $\nabla$ have unit
modulus if an only if the eigenvalues of $g\in\pi_1(L)$ have unit modulus.
In particular this means that a Lagrangian submanifold $L$ can be
extended to an object of the Fukaya category only if $\Be\vert_L$
is in the
image of the homomorphism (\ref{Hopf}).

As a by-product, we obtained an equivalent definition of
an object of the Fukaya category: it is a triple $(L,i,R),$
where $L,i$ are the same
as above, and $R$ is a finite-dimensional representation of
the twisted group algebra $\CC_{\psi}[\pi_1(L)]$
such that $j(\psi_R)=\Be\vert_L$ and
all the eigenvalues of $R(g)$ have unit modulus for all $g\in \pi_1(L).$

Morphisms in the modified Fukaya category $\cF(X,\Be)$ are defined in analogy
with~\cite{Fukaya,K}. Let $\cU_1=(L_1,i_1,E_1,\nabla_1)$ and $\cU_2=
(L_2,i_2,E_2,\nabla_2)$ be two
objects such that $L_1$ and $L_2$ intersect transversally.
Morphisms from $\cU_1$ to $\cU_2$ in $\cF(X)$ form a
complex of vector spaces defined by the rule
\begin{equation}
\Hom^{\cdot}(\cU_1, \cU_2)=
\bigoplus_{x\in L_1\cap L_2}\Hom^i(E_1\vert_x, E_2\vert_x)
\end{equation}

It is graded in the following way. For any point $x\in L_1 \cap L_2$
we have two points $i_1(x)$ and $i_2(x)$ on the universal cover
of the Lagrangian Grassmannian of $T_x X.$
To these two points we can associate an integer
$\mu(i_1(x),i_2(x))$ which is called the Maslov index of $i_1(x),i_2(x)$
(see for example \cite{Ar}).
By definition, the space
$\Hom(E_{1}\vert_x , E_{2}\vert_x)$ has a grading $\mu(i_1(x),i_2(x)).$

The differential on $\Hom(\cU_1, \cU_2)$ is defined by the  rule
$$
d(u)=\sum_{z\in L_1\cap L_2} m_1(u; z),
$$
where $u\in \Hom(E_1\vert_x, E_2\vert_x),$ and
$m_1(u;z)\in \Hom(E_1\vert_z, E_2\vert_z)$ is given by
$$
m_1(u;z)=\sum_{\phi: D\to X} \pm\exp(2\pi i\int_D \phi^*(-\B+i\omega))
\cdot P\exp(\oint_{\partial D}\phi^*\nabla).
$$
Here $\phi$ is an (anti)-holomorphic map from the disk
$D=\{|w|\le 1,w\in\CC\}$ to $X$ such that
$\phi(-1)=x, \phi(1)=z$
and $\phi([x, z])\subset L_{2}$ and $\phi([z,x])\subset L_1.$
The path-ordered integral is defined by the following rule
$$
P\exp(\oint_{\partial D}\phi^*\nabla):=P\exp(\int_{x}^{z}\phi^*\nabla_2) \cdot u\cdot
P\exp(\int_{z}^{x}\phi^*\nabla_1)
$$
This homomorphism from $E_1\vert_z$ to $E_2\vert_z$ can be described
as follows.
We take a vector $e\in E_1\vert_z,$
use the connection $\nabla_1$ transport it to $E_1\vert_x,$
apply the map $u,$ and obtain an element of $E_2\vert_x.$
Then we transport this element to $E_2\vert_z$ using the connection
$\nabla_2.$

The $\pm$ sign indicates the natural
orientation on the space of (anti)-holomorphic maps.
One expects that there are finitely many such maps if
$\mu_z - \mu_x =1.$

To define the composition of morphisms, let us take
$u\in \Hom(E_1\vert_x, E_2\vert_x)$
and $v\in \Hom(E_2\vert_y, E_3\vert_y),$ where $x\in L_1\cap L_2$ and
$y\in L_2\cap L_3.$
Then the composition of $u$ and $v$ is defined as
$$
v\circ u= \sum_{z\in L_1\cap L_3} m_2(v,u; z),
$$
where $m_2(v,u;z)\in \Hom(E_1\vert_z, E_3\vert_z)$ is given by
$$
m_2(v,u;z)=\sum_{\phi: D\to X} \pm\exp(2\pi i\int_D \phi^*(-\B+i\omega))
\cdot P\exp(\oint_{\partial D}\phi^*\nabla)
$$
Here we sum over (anti)-holomorphic maps $\phi$ from a two-dimensional disk
$D$ to $X,$ such that
three fixed points  $p_1, p_2, p_3\in \partial D$ are mapped to $x,y,z$ respectively,
and $\phi([p_i, p_{i+1}])\in L_{i+1}.$
The path-ordered integral here  is calculated by the rule
$$
P\exp(\oint_{\partial D}\phi^*\nabla):=P\exp(\int_{p_2}^{p_3}\phi^*\nabla_3)\cdot v
\cdot P\exp(\int_{p_1}^{p_2}\phi^*\nabla_2) \cdot u\cdot
P\exp(\int_{p_3}^{p_1}\phi^*\nabla_1)
$$

In the same manner we can define higher order compositions using
zero-dimensional components of spaces of maps $\phi$ from
the disk $D$ to $X$ with $\phi(\partial D)$ sitting in the union
of Lagrangian submanifolds.

It is easy to check that the above definition of morphisms and
their compositions
does not change if we replace $\B$ with another 2-form with the same image in
$H^2(X,\RR/\ZZ).$ The check makes use of~(\ref{FB}) and~(\ref{newB}). This
confirms our claim that the Fukaya category depends only on $\Be.$

The rules for computing morphisms and their compositions can be explained
heuristically using the path integral for the $\sigma$-model on a worldsheet
with boundaries~\cite{Witten}.

The category $\cF_0(X)$ has the same objects as $\cF(X),$
but the morphisms are the degree zero cohomology groups of the complexes
defined above. Note that different objects of $\cF(X)$ often
become isomorphic in $\cF_0(X)$.
For example, in the case when $X$ is a real symplectic 2-torus, any one-dimensional
submanifold is Lagrangian. Many of them admit a lift of the Gauss map.
Thus the category $\cF(X)$ contains many more objects
than the derived category of the elliptic curve (an elliptic curve with a flat metric
is self-mirror).
But in $\cF_0(X)$ any object becomes isomorphic to some other object
associated with a special Lagrangian submanifold (see \cite{PoliZas}).
More generally, it appears likely that working in the category $\cF_0(X)$
one may restrict the set of objects
of the Fukaya category and consider only special Lagrangian submanifolds
with respect to a holomorphic calibration. For different $L$ the calibrations may differ
by a multiplicative constant. This restriction is also
natural from the string theory point of view, since, as explained above, 
non-anomalous A-type D-branes are associated
with special Lagrangian submanifolds in a Calabi-Yau~\cite{OogOzYin}.

\appendix
\section{Supersymmetric $\sigma$-model of a flat torus}
\label{sigmamodel}

In this section we define the classical field theory known in the physics literature
as the $N=1$ supersymmetric $\sigma$-model. The data needed to specify a $\sigma$-model
consist of a $C^\infty$ manifold $M$ (``the target space''), a Riemannian metric $G$ on
$M,$ and a 2-form $\B$ on $M.$
We then discuss
the problem of the quantization of the $\sigma$-model in the case when the target
space is a flat torus.
The superconformal vertex algebra constructed in Section~\ref{torusSCVA} can
be regarded as a solution of the quantization problem. A detailed discussion of
supersymmetric $\sigma$-models can be found in~\cite{IASbook}.

Let $\cW$ be a two-dimensional $C^\infty$ manifold $\RR\times\SS^1$ (``the worldsheet'').
We parametrize $\cW$ by $(\tau,\sigma)\in \RR\times \RR/(2\pi\ZZ).$ The
coordinate $\tau$ is regarded as ``time.''
We endow $\cW$ with a Minkowskian metric $ds^2=d\tau^2-d\sigma^2$
and orientation $d\tau\wedge d\sigma.$ Thus $*d\sigma=d\tau, *d\tau=d\sigma.$
The symmetric tensor corresponding to the metric will be denoted $g.$
General coordinates on $\cW$ will be denoted $(y^0,y^1).$
The invariant volume element $d\tau\wedge d\sigma=d^2y\sqrt{-\det g}$ will be denoted $d\Sigma.$
We denote by $S^+$ and $S^-=S^{+*}$ the complexified semi-spinor representations of $SO(1,1)$
and by $V$ its complexified fundamental representation.
Complexified semi-spinor representations are one-dimensional complex vector spaces
endowed with $SO(1,1)$-invariant nondegenerate morphisms
\begin{equation}
\gamma: S^-\ra V\ot S^+, \qquad  \bgamma:S^+\ra V\ot S^-.
\end{equation}
These morphisms are determined up to a scalar factor, and we assume that
they satisfy the Clifford algebra relation
$$
\gamma \bgamma+\bgamma\gamma=2g^{-1}\cdot \id_{S^+\op S^-}.
$$
Here $g^{-1}$ is regarded as map $\CC\ra V^*\ot V^*.$
In a suitable basis, one has
$$\gamma=\begin{pmatrix} 1 \\ -1\end{pmatrix},\qquad
\bgamma=\begin{pmatrix} 1 \\ 1\end{pmatrix}.$$

Since $H^1(\cW,\ZZ_2)=\ZZ_2,$ there are two inequivalent spinor structures on $\cW.$
The trivial one is called the periodic, or Ramond, spin structure in the physics
literature. The nontrivial one is known as the anti-periodic, or Neveu-Schwarz, spin
structure. Both spin structures play a role in string theory, but for our purposes
it will be sufficient to consider the Neveu-Schwarz spin structure.
The corresponding semi-spinor bundles on $\cW$ will be denoted by the same
letters $S^+,S^-.$
The parity-reversed (i.e. odd) semi-spinor bundles will be denoted by $\Pi S^+,
\Pi S^-.$ More generally, $\Pi$ will denote the parity-reversal functor.
The vector space morphisms $\gamma$ and $\bgamma$ give rise to a pair of
bundle morphisms $S^-\ra T\cW\ot S^+$ and $S^+\ra T\cW\ot S^-$
which we denote by the same letters.

Let $M$ be a $C^\infty$ manifold endowed with a Riemannian metric $G$ and a
real 2-form $\B.$ At this stage we do not require $\B$ to be closed.
The indices of the tangent bundle $TM$ will be denoted by
$j,k,l,\ldots$ in the upper position. The indices of the cotangent bundle
$T^*M$ will be denoted by the same letters in the lower position. Summation over
repeating indices is always implied.

Let $X$ be a $C^\infty$ map from $\cW$ to $M.$
Let $\psi$ and $\bpsi$ be $C^\infty$ sections
of $X^*TM\op \Pi S^+$ and $X^*TM\op \Pi S^-,$ respectively.
$N=1$ supersymmetric $\sigma$-model with worldsheet
$\cW$ and target $(X,G,\B)$ is a classical field theory on $\cW$
defined by the action
\begin{multline}\label{action}
\frac{1}{4\pi}\int_\cW  G_{jk}(X) \left(dX^j\wedge *dX^k\right)+
\frac{1}{4\pi}\int_\cW  \B_{jk}(X) \left(dX^j\wedge dX^k\right)+\\
\frac{1}{4\pi}\int_\cW \left(G_{jk}(X)\psi^j i\bgamma\cdot \nabla\psi^k+
G_{jk}(X)\bpsi^j i\gamma\cdot \nabla\bpsi^k+\frac{1}{2}
R_{jklm}(X)\psi^j\psi^k\bpsi^l\bpsi^m\right)d\Sigma.
\end{multline}
Here the covariant derivatives $\nabla\psi$ and $\nabla\bpsi$ are sections of $X^*TM\ot
\Pi S^\pm\ot T^*\cW$ defined as follows:
\begin{align}\nn
\nabla\psi^j&=D\psi^j+\left(\left\{\begin{array}{l} j \\ kl\end{array}\right\}
+ \frac{3}{2}\left(G^{-1}\right)^{jm}(d\B)_{klm}\right) dX^k \psi^l,\\ \nn
\nabla\bpsi^j&=D\bpsi^j+\left(\left\{\begin{array}{l} j \\ kl\end{array}\right\}
- \frac{3}{2}\left(G^{-1}\right)^{jm}(d\B)_{klm}\right) dX^k \bpsi^l,
\end{align}
where $\{j,kl\}$ are the Christoffel symbols constructed from $G,$ and $D:S^\pm\ra S^\pm\ot
T^*\cW$ is the Levi-Civita covariant derivative constructed from $g.$
$R_{jklm}(X)$ is the curvature corresponding to the following connection 1-form on $M$
$$
\left(\left\{\begin{array}{l} j \\ kl\end{array}\right\}
+ \frac{3}{2}\left(G^{-1}\right)^{jm}(d\B)_{klm}\right) dx^l.
$$
In the last term in the action we used twice the natural $SO(1,1)$-invariant pairing
$S^+\ot S^-\ra \CC.$

This complicated-looking action has an elegant reformulation in terms of superfields,
i.e. maps from a super-Riemann surface to $M$~\cite{IASbook}.

The extrema of the action~(\ref{action}) are given by the solutions of the
Euler-Lagrange equations.
In the case when all the fields are even, it is well known that the space of
solutions of the Euler-Lagrange equations is a manifold with a natural symplectic
structure. This statement remains true in the supersymmetric context
(see e.g.~\cite{HenTeit}). In the present case the symplectic structure is given by
\begin{multline}\label{sympl}
\frac{1}{2\pi}\int_{\tau=\tau_0}\left(\delta X^j\wedge \delta \left(
G_{jk}(X)\frac{\partial X^k}{\partial\tau}+\B_{jk}(X)
\frac{\partial X^k}{\partial\sigma}\right)\right.\\
\left.+iG_{jk}(X) \delta\psi^j\wedge\delta\psi^k+
iG_{jk}(X)\delta\bpsi^j\wedge\delta\bpsi^k\right)d\sigma.
\end{multline}
Here we used the fact the Euler-Lagrange equations are second-order in time derivatives
of $X$ and first-order in time derivatives of $\psi,\bpsi,$ and therefore
a solution is completely determined by the values of $X,\ \partial X/\partial \tau,
\ \psi,$
and $\bpsi$ on any circle $\tau=\tau_0.$ One can check that the symplectic structure thus
defined does not depend on $\tau_0.$
The space of solutions endowed with this symplectic structure is called the phase space
of the $\sigma$-model.

We are interested in the case when $M$ is a torus $T^{2d}=\RR^{2d}/\Gamma,$
$\Gamma\cong \ZZ^{2d},$ with a constant
metric $G$ and a constant 2-form $\B.$ We will fix an isomorphism between $\Gamma$
and $\ZZ^{2d}.$ Without loss of generality we may assume that the action of $\Gamma$ on
$\RR^{2d}$ is
$$
x^j\mapsto x^j+2\pi n^j,\quad n^j\in \ZZ,\quad j=1,2,\ldots,2d.
$$
In this special case the $\sigma$-model action becomes
\begin{multline}
\frac{1}{4\pi}\int_\cW\left(G_{jk}\left(\frac{\partial X^j}{\partial\tau}
\frac{\partial X^k}{\partial\tau}-\frac{\partial X^j}{\partial\sigma}
\frac{\partial X^k}{\partial\sigma}\right)+2B_{jk}\frac{\partial X^j}{\partial\tau}
\frac{\partial X^k}{\partial\sigma}\right.\\
\left. +iG_{jk}\psi^j\left(\frac{\partial}{\partial\tau}+
\frac{\partial}{\partial\sigma}\right)\psi^k
+iG_{jk}\bpsi^j\left(\frac{\partial}{\partial\tau}-
\frac{\partial}{\partial\sigma}\right)\bpsi^k\right)d\tau d\sigma.
\end{multline}
The Euler-Lagrange equations have a simple form:
\begin{equation}\label{EL}
\left(\frac{\partial^2}{\partial \sigma^2}-\frac{\partial^2}{\partial \tau^2}\right)
X^j=0, \qquad
\left(\frac{\partial}{\partial\sigma}+\frac{\partial}{\partial\tau}\right)\psi^j=0,
\qquad \left(\frac{\partial}{\partial\sigma}-
\frac{\partial}{\partial\tau}\right)\bpsi^j=0.
\end{equation}
In what follows we will use the notation
$$
\partial_-=\frac{1}{2}\left(\frac{\partial}{\partial\sigma}-
\frac{\partial}{\partial\tau}\right),\qquad
\partial_+=\frac{1}{2}\left(\frac{\partial}{\partial\sigma}+
\frac{\partial}{\partial\tau}\right).
$$
The Poisson brackets of the fields evaluated at equal times follow from~(\ref{sympl}):
\begin{align}\label{poisson}
\left\{ X^j(\tau,\sigma),X^k(\tau,\sigma')\right\}_{P.B.}&=0, \\ \nn
\left\{ X^j(\tau,\sigma),\frac{\partial X^k}{\partial\tau}(\tau,\sigma')\right\}_{P.B.}&=
2\pi\left(G^{-1}\right)^{jk}\delta\left(\sigma-\sigma'\right), \\ \nn
\left\{\psi(\tau,\sigma),\bpsi(\tau,\sigma')\right\}_{P.B.}&=0,\\ \nn
\left\{\psi(\tau,\sigma),\psi(\tau,\sigma')\right\}_{P.B.}&=-2\pi i
\left(G^{-1}\right)^{jk}\delta\left(\sigma-\sigma'\right), \\ \nn
\left\{\bpsi(\tau,\sigma),\bpsi(\tau,\sigma')\right\}_{P.B.}&=-2\pi i
\left(G^{-1}\right)^{jk}\delta\left(\sigma-\sigma'\right).
\end{align}
The Poisson brackets between even and odd fields vanish.

Note that the neither the Euler-Lagrange equations~(\ref{EL}) nor the symplectic
structure corresponding to~(\ref{poisson}) depend on $\B.$ This happens whenever $\B$
is closed, because in this case the $\B$-dependent terms in the action are locally total
derivatives. We will see below that quantization of the $\sigma$-model introduces
arbitrariness which is parametrized by a class in $H^2(M,\RR/\ZZ).$ The usual
interpretation is that while the classical $\sigma$-model does not detect a closed B-field,
the quantized $\sigma$-model detects the image of $\B$ in $H^2(M,\RR/\ZZ).$

The Euler-Lagrange equations~(\ref{EL}) can be rewritten in the Hamiltonian form:
\begin{gather}
\begin{align*}
\frac{\partial X^j}{\partial \tau}(\tau,\sigma)&=
\left\{X^j(\tau,\sigma),H(\tau)\right\}_{P.B.},\nn\\
\frac{\partial}{\partial\tau}\left(\frac{\partial X^j}{\partial \tau}\right)(\tau,\sigma)&=
\left\{\left(\frac{\partial X^j}{\partial \tau}\right)(\tau,\sigma),H(\tau)
\right\}_{P.B.},
\\
\frac{\partial\psi^j}{\partial\tau}(\tau,\sigma)&=
\left\{\psi^j(\tau,\sigma),H(\tau)\right\}_{P.B.},\nn\\
\frac{\partial\bpsi^j}{\partial\tau}(\tau,\sigma)&=
\left\{\bpsi^j(\tau,\sigma),H(\tau)\right\}_{P.B.}.\nn
\end{align*}
\end{gather}
The Hamiltonian $H$ is a function on the phase space given by
\begin{equation*}
H(\tau_0)=\frac{1}{4\pi}\int_{\tau=\tau_0}
G_{jk}\left(\frac{\partial X^j}{\partial\tau}\frac{\partial X^k}{\partial\tau}
+\frac{\partial X^j}{\partial\sigma}\frac{\partial X^k}{\partial\sigma}
-i\psi^j\frac{\partial\psi^k}{\partial\sigma}+
i\bpsi^j\frac{\partial\bpsi^k}{\partial\sigma}\right)d\sigma.
\end{equation*}
As a consequence of the equations of motion, we have $\frac{dH(\tau_0)}{d\tau_0}=0.$

Hamiltonian vector fields on the phase space are those vector fields which preserve the
symplectic form. They obviously form a Lie (super-)algebra with respect to the Lie bracket.
We will now exhibit a subalgebra in this super-algebra which is isomorphic to the direct
sum of two copies of the $N=1$ super-Virasoro algebra.

Recall that given a function $W$ on the phase space, we can define a Hamiltonian vector field
$v_W$ as follows:
$$
v_W(\cdot)=\left\{\ \cdot\ ,W\right\}_{P.B.}
$$
One has an identity
$$
\left[v_W,v_U\right]_{Lie}=v_{\{W,U\}_{P.B.}}.
$$
We will define a set of functions on the phase space which forms a super-Virasoro algebra
with respect to the Poisson bracket; then the corresponding set of Hamiltonian vector fields
forms a super-Virasoro algebra with respect to the Lie bracket.

The set of functions we want to define is a vector space generated over $\CC$ by the
following elements:
\begin{align}\label{neqone}
L_n&\ds =\frac{1}{2\pi}\int_{\tau=\tau_0} e^{-in\sigma}
G_{jk}\left(\partial_- X^j\partial_- X^k-\frac{i}{2}\psi\partial_-\psi\right)d\sigma,
&{}&\qquad n\in\ZZ,&{}\nn\\
\bL_n&\ds =\frac{1}{2\pi}\int_{\tau=\tau_0} e^{in\sigma}G_{jk}
\left(\partial_+ X^j\partial_+ X^k+\frac{i}{2}\bpsi\partial_+\bpsi\right)d\sigma,
&{}&\qquad n\in\ZZ,&{}\\
Q_r&\ds =\frac{-i}{4\pi}\int_{\tau=\tau_0} e^{-ir\sigma} G_{jk}\psi^j
\partial_- X^k\ d\sigma,
&{}&\ds\qquad r\in\ZZ+\frac{1}{2},&{}\nn\\
\bQ_r&\ds =\frac{i}{4\pi}\int_{\tau=\tau_0} e^{ir\sigma}
G_{jk}\bpsi^j \partial_- X^k\ d\sigma,
&{}&\ds\qquad r\in\ZZ+\frac{1}{2}.&{}\nn
\end{align}
Two remarks are in order concerning these expressions. First, all these
functions on the phase space implicitly depend on $\tau_0$ as a parameter.
Second, since we picked the anti-periodic spin structure on $\cW,$ the
lift of $\psi$ to the universal cover of $\cW$ is an anti-periodic function of $\sigma.$
This is the reason the index $r$ runs over half-integers.

The Poisson brackets of the generators can be easily computed using~(\ref{poisson}), and the
nonvanishing ones turn out to be
\begin{align}\label{poissonone}
\ds\left\{L_m,L_n\right\}_{P.B.}&\ds =-i(m-n)L_{m+n},
&\ds\qquad\left\{\bL_m,\bL_n\right\}_{P.B.}&\ds =-i(m-n)\bL_{m+n},\nn\\
\ds\left\{L_m,Q_r\right\}_{P.B.}&\ds =-i\left(\frac{m}{2}-r\right)Q_{m+r},
&\ds\qquad\left\{\bL_m,\bQ_r\right\}_{P.B.}&\ds =-i\left(\frac{m}{2}-r\right)\bQ_{m+r},\\
\ds\left\{Q_r,Q_s\right\}_{P.B.}&\ds=-\frac{i}{2}L_{r+s},
&\ds\qquad \left\{\bQ_r,\bQ_s\right\}_{P.B.}&\ds =-\frac{i}{2}\bL_{r+s},\nn
\end{align}
Thus the space spanned by the generators is a Lie super-algebra isomorphic to the direct
sum of two copies of the $N=1$ super-Virasoro algebra (with zero central charge).

Note that $L_0+\bL_0=H.$ Recalling that the $\tau$-dependence of any function $F$ on the
phase space is determined by $$\frac{dF}{d\tau}=\{F,H\}_{P.B.},$$
and using~(\ref{poissonone}), one can show that all the generators
have a very simple dependence on $\tau_0$:
\begin{align*}
L_n(\tau_0)=e^{-in\tau_0}L_n(0), & \qquad \bL_n(\tau_0)=e^{-in\tau_0}\bL_n(0),\\ \nn
Q_r(\tau_0)=e^{-ir\tau_0}Q_r(0), & \qquad \bQ_r(\tau_0)=e^{-ir\tau_0}\bQ_r(0).
\end{align*}
Thus the space spanned by the generators does not depend on $\tau_0.$

The presence of two copies of the $N=1$ super-Virasoro algebra acting on the
phase space is a feature of the supersymmetric $\sigma$-model with an arbitrary
target $(M,G,B).$ This fact is crucial for string theory
applications of the $\sigma$-model, see~\cite{Polch} for details.

Now let us choose a constant complex structure $I$ on $M$ such that $G$ is a
Hermitian metric. This makes $M$ a K\"ahler manifold. Let $\omega=GI$ be the
corresponding K\"ahler form.
It turns out that we can embed each of the two $N=1$ super-Virasoro algebras in a bigger
$N=2$ super-Virasoro algebra.
The additional generators are given by
\begin{align}\label{neqtwo}
Q^\pm_r&=\frac{-i}{8\pi}\int_{\tau=\tau_0} e^{-i(r+1/2)\sigma}
\left(G_{jk}\mp i\omega_{jk}\right)\psi^j \partial_- X^k\ d\sigma,&{}&
 \qquad r\in \ZZ+\frac{1}{2},\nn\\
\bQ^\pm_r&=\frac{i}{8\pi}\int_{\tau=\tau_0} e^{i(r+1/2)\sigma}
\left(G_{jk}\mp i\omega_{jk}\right)\bpsi^j \partial_+ X^k\ d\sigma,
&{}&\qquad r\in \ZZ+\frac{1}{2},\\ \nn
J_n&=\frac{-i}{4\pi}\int_{\tau=\tau_0}
e^{-in\sigma}\omega_{jk}\psi^j\psi^k\ d\sigma,
&{}&\qquad
n\in \ZZ, \\
\nn
\bJ_n&=\frac{-i}{4\pi}\int_{\tau=\tau_0} e^{in\sigma}\omega_{jk}\bpsi^j\bpsi^k\ d\sigma,
&{}&\qquad n\in\ZZ.
\end{align}
Note that $Q_r=Q^+_r+Q^-_r$ and $\bQ_r=\bQ^+_r+\bQ^-_r$ for all $r.$
The Poisson brackets between $L_n,Q^\pm_r,$ and $J_n$ are given by
\begin{alignat*}{2}
{}&\left\{L_m,Q^\pm_r\right\}_{P.B.}&=
& -i\left(\frac{m}{2}-r\right)Q^\pm_{r+m},
\nn\\
{}&\left\{L_m,J_n\right\}_{P.B.}&=&\, i n J_{n+m},\nn\\
{}&\left\{Q^+_r,Q^+_s\right\}_{P.B.}&=& \left\{Q^-_r,Q^-_r\right\}_{P.B.}=0,\\
{}&\left\{Q^+_r,Q^-_s\right\}_{P.B.}&=&
-\frac{i}{4}L_{r+s}-\frac{i}{8}(r-s)J_{r+s},\nn\\ \nn
{}&\left\{J_m,Q^\pm_r\right\}_{P.B.}&=&
\mp iQ^\pm_{r+m}.
\end{alignat*}
The Poisson brackets between the barred generators have the same form. The Poisson
brackets between barred and unbarred generators are trivial, as usual.

Again, the emergence of the $N=2$ super-Virasoro is not limited to the particular
situation we are considering: one can prove that the phase space of the
supersymmetric $\sigma$-model is acted upon by the
$N=2$ super-Virasoro if $(M,G)$ is an arbitrary K\"ahler manifold,
and $\B$ is closed~\cite{AGF}. The statement can be further generalized to
B-fields which are not closed~\cite{GatesHullRocek}.

Let us now look more closely at the space of solutions of the Euler-Lagrange
equations. Note that any map $X:\cW\ra M$ induces a homomorphism of the homology
groups $H_1(\cW)\ra H_1(M).$ The group $H_1(\cW)\cong \pi_1(\cW)\cong \ZZ$ has
a preferred generator, namely the loop winding the $\SS^1$ in the direction of
increasing $\sigma.$
Since $H_1(T^{2d})=\Gamma,$ we see that to any map $X:\cW\ra M$ we can
assign an element $w(X)$ of $\Gamma.$ The components of $w$ are the so-called winding
numbers of the map $X.$ Thus the phase space of the $\sigma$-model is a
disconnected sum
$$
\cM=\bigsqcup_{w\in\Gamma} \cM_w.
$$
We will see in a moment that $\cM_w$ is connected for all $w.$

The Euler-Lagrange equations~(\ref{EL}) are linear and can be solved by Fourier
transform. The general solution in $\cM_w$ is given by
\begin{align}
X^j&=x^j+\sigma w^j+\tau\left(G^{-1}\right)^{jk}p_k+
\frac{i}{\sqrt 2}\label{clasX}
\sideset{}{'}\sum_{s=-\infty}^\infty \frac{1}{s}\left(\alpha^j_s e^{is(\sigma-\tau)}+
\bal^j_s e^{-is(\sigma+\tau)}\right), \\
\psi^j&=\sum_{r\in \ZZ+1/2}\psi^j_r e^{ir(\sigma-\tau)},\label{claspsi}\\
\bpsi^j&=\sum_{r\in \ZZ+1/2}\bpsi^j_r e^{-ir(\sigma+\tau)}.\label{clasbpsi}
\end{align}
Here $\alpha^j_s,\bal^j_s$ are complex numbers satisfying
$(\alpha^j_s)^*=\alpha^j_{-s},
(\bal^j_s)^*=\bal^j_{-s}$; $\psi^j_r,\ \bpsi^j_r$ are elements of $\Pi \CC$;
$x^j,\ j=1,\ldots,2d,$ take values in $\RR/(2\pi\ZZ)$; and $p_j,\ i=1,
\ldots,2d,$ take values in $\RR.$
The variables $\alpha^j_s,\bal^j_s,\psi^j_r,\bpsi^j_r$ will be
referred to as ``the oscillators.'' The variables $(x^j,p_j),\ j=1,\ldots,2d,$ together
parametrize a copy of $T^*M\cong T^{2d}\times\RR^{2d}.$

Thus for any $w\in\Gamma$ the supermanifold $\cM_w$ is a product of the vector
superspace spanned by
$\alpha_n,\bal_n,n\in\ZZ,\psi_r,\bpsi_r,r\in\ZZ+1/2,$ and the cotangent bundle of
$M.$

The Poisson brackets of the coordinates on $\cM_w$ can be computed from~(\ref{poisson})
and (\ref{clasX}-\ref{clasbpsi}). The non-vanishing ones are given by
\begin{gather}
\begin{align*}
\left\{\alpha^j_n,\alpha^k_m\right\}_{P.B.}&
=-in\left(G^{-1}\right)^{jk}\delta_{m+n},
&\left\{\bal^j_n,\bal^k_m\right\}_{P.B.}&
=-in\left(G^{-1}\right)^{jk}\delta_{m+n},\nn\\
\left\{\psi^j_r,\psi^k_s\right\}_{P.B.}&
=-i\left(G^{-1}\right)^{jk}\delta_{r+s},
&\left\{\bpsi^j_r,\bpsi^k_s\right\}_{P.B.}&
=-i\left(G^{-1}\right)^{jk}\delta_{r+s}.
\end{align*} \\
\left\{x^j,p_k\right\}_{P.B.} = \delta^j_k,\nn
\end{gather}
Thus the symplectic supermanifold $\cM_w$ decomposes into a product of a symplectic
vector superspace spanned by the oscillators and $T^*M$ with the standard symplectic
structure.

It is customary to continue analytically the time variable $\tau$ to the imaginary axis.
If we set $\tau=it,$ then the combination $v=\sigma+\tau=\sigma+it$ becomes a complex
variable. Since we identify $\sigma\sim \sigma+2\pi,$ it
is convenient to set $v=i\log z$ where $z\in \CC^*.$ After analytic continuation
$\partial_-$ and $\partial_+$ become $\partial_v=-iz\partial_z$ and $\bpartial_v=i\bz\bpartial_z,$
respectively. The functions $X^j(v(z))$
are multi-valued functions of $z$ if $w\neq 0.$ But their derivatives
with respect to $z$ and $\bz$ are single-valued, and moreover are holomorphic
and anti-holomorphic, respectively:
\begin{align*}
\frac{\partial X^j}{\partial z}&=-\frac{i}{2z}\left(
\left(G^{-1}\right)^{jk}p_k-w^j\right)-\frac{i}{\sqrt 2}
\sideset{}{'}\sum_{s=-\infty}^\infty \frac{\al_s^j}{z^{s+1}}, \\ \nn
\frac{\partial X^j}{\partial\bz}&=-\frac{i}{2\bz}\left(
\left(G^{-1}\right)^{jk}p_k+w^j\right)-\frac{i}{\sqrt 2}
\sideset{}{'}\sum_{s=-\infty}^\infty \frac{\bal_s^j}{\bz^{s+1}}.
\end{align*}
Note that after rescaling $X^j\ra (i{\sqrt 2})X^j$ these expressions become formally the
same as~(\ref{X},\ref{bX}), except that
in~(\ref{X},\ref{bX}) the coordinates on the phase space $w^k,p_k,\alpha^k_s,\bal^k_s$
are replaced with the operators $W^k,M_k-\B_{kl}W^l,\alpha^k_s,\bal^k_s,$ respectively.
This replacement is the quantization map discussed in more detail below.

Similarly, after analytic continuation to imaginary $\tau,$ the sections $\psi^j$
and $\bpsi^j$ become holomorphic and anti-holomorphic, respectively. One additional
subtlety arises due to the fact that $\psi$ and $\bpsi$ are sections of semi-spinor
bundles. Thus the coordinate change $v\mapsto z=e^{-iv}$ must be accompanied
by $\psi^j\mapsto z^{-1/2} \psi^j,$ and $\bpsi^j\mapsto \bz^{-1/2}\bpsi^j.$
This accounts for the shift $r\mapsto r+\frac{1}{2}$ between
(\ref{claspsi},\ref{clasbpsi}) and (\ref{psi},\ref{bpsi}).

Let us now turn to the quantization of the $\sigma$-model. This discussion
provides a motivation for the constructions of Section~\ref{torusSCVA}.

Since the classical phase space is a disconnected sum of identical pieces labeled by
$w\in\Gamma,$ the quantum-mechanical Hilbert space will be a tensor sum of
identical Hilbert spaces labeled by $w\in\Gamma.$ Thus we only need to understand
how to quantize the supermanifold $\cM_w.$ In turn, $\cM_w$ decomposes as a product
of $T^*M$ with the standard symplectic structure, and a vector superspace
spanned by the oscillators.

The vector superspace spanned by the oscillators can
be quantized using the well-known Fock-Bargmann prescription. The resulting Hilbert superspace is
the so-called Fock space, i.e. the completion with respect to a suitable norm
of the space of polynomials of even variables $a^i_{-s},\ba^i_{-s},
s=1,2,\ldots,$ and odd variables $\theta_{-r},\btheta_{-r}, r=1/2,3/2,\ldots.$ We will
denote this space of polynomials $\cH_{Fock}.$

The quantization of $T^*M$ is also standard and yields the Hilbert
space which is the completion of the space $C^\infty(M)$ of smooth functions on
$M=\RR^{2d}/\Gamma$ with respect to
an $L^2$ norm. Using Fourier transform, this Hilbert space can be identified with the
completion of the group algebra of $\Gamma^*$ with respect to an $\ell^2$ norm.

Thus the quantization procedure sketched above leads to the Hilbert space which
is a suitable completion of an infinite-dimensional superspace
$$
\op_{w\in\Gamma} \CC[\Gamma^*]\ot \cH_{Fock}
$$
This can be written in a more symmetric form:
$$
\CC[\Gamma\op\Gamma^*]\ot \cH_{Fock}.
$$
For our purposes, only the superspace structure, and not the Hilbert space structure,
is important. Thus we need not perform the completion procedure, and can take
the above superspace as the state space of the $N=2$ superconformal vertex algebra
corresponding to the supersymmetric $\sigma$-model. We will call this vector superspace
the {\it state space} of the quantized $\sigma$-model.

Finding a suitable state space is but a part of the quantization
problem. Quantizing a classical field theory usually requires finding a sufficiently
large subset of functions on the phase space closed under the Poisson brackets, and a
map from this subset to the set of linear operators on the state space, such
that the Poisson brackets are mapped to $-i$ times the graded commutator. The choice of the
subset of functions on the phase space is dictated by physical considerations.
For example, for string theory applications
it is imperative to have an $N=1$ super-Virasoro algebra acting on the state space.
Thus the distinguished subset must include the generators of the $N=1$ super-Virasoro
algebra~(\ref{neqone}) and their linear combinations. We will also require that the
subset include the generators of the $N=2$ super-Virasoro~(\ref{neqtwo}).
Usually one also requires that the distinguished subset include the fields in terms of
which the classical action is written. In our case these are
$X^j(\sigma,\tau),\psi^j(\sigma,\tau),\bpsi^j(\sigma,\tau).$
One also wants the operator corresponding to
the Hamiltonian $H=L_0+\bL_0$ to have nonnegative spectrum.

To quantize the fields $X^j,\psi^j,$ and $\bpsi^j$ it is sufficient to quantize
the oscillators and $(x^j,p_j)$ (the coordinates on $T^*M$). The Fock-Bargmann
quantization map sends oscillators with negative subscripts to multiplication
operators on the space of polynomials:
\begin{align*}
\al^j_s&\mapsto a^j_s, & \bal^j_s&\mapsto\ba^j_s, & s&=-1,-2,\ldots, \\ \nn
\psi^j_r&\mapsto \theta^j_r, & \bpsi^j_r&\mapsto\btheta^j_r, &
r&=-\frac{1}{2},-\frac{3}{2},
\ldots .
\end{align*}
The oscillators with positive subscripts are mapped
to differentiation operators on the space of polynomials:
\begin{align*}
\al^j_s&\mapsto s\left(G^{-1}\right)^{jk}\frac{\partial}
{\partial a^k_{-s}}, &
\bal^j_s&\mapsto s\left(G^{-1}\right)^{jk}\frac{\partial}
{\partial \ba^k_{-s}}, &
s&=1,2,\ldots, \\ \nn
\psi^j_r&\mapsto \left(G^{-1}\right)^{jk}\frac{\partial}{\partial \theta^k_{-r}}, &
\bpsi^j_r&\mapsto \left(G^{-1}\right)^{jk}\frac{\partial}{\partial \btheta^k_{-r}}, &
r&=\frac{1}{2},\frac{3}{2}, \ldots .
\end{align*}
It is easy to see that the (graded) commutators between these operators
are equal to $i$ times the Poisson brackets of their classical counterparts, as
required.

The quantization of $(x^j,p_j)$ proceeds as follows. The function $x^j$ is a
multi-valued function on the phase space and cannot be quantized. But any
smooth function $f(x^1,\ldots,x^{2d})$ which is periodic, i.e. invariant with
respect to shifts $x^j\ra x^j+2\pi n^j,\ n^j\in\ZZ,$ is a univalued function on the
phase space. The standard quantization of $T^*M$ maps such a function to
a multiplication operator on $C^\infty(M)$:
$$
f(x^1,\ldots,x^{2d})\mapsto f(x^1,\ldots,x^{2d}).
$$

Actually, the vector space we are dealing with is not just $C^\infty(M),$ but
a $\Gamma$-graded vector space
$$
{\mathfrak F}=\op_{w\in\Gamma} C^\infty(M),
$$
and therefore we should quantize a pair $(f,w)$ rather than $f.$
This leads to an important subtlety.
If $w=0,$ we can assign to $(f,w)$ a multiplication
operator which acts on each of the $\Gamma$-homogeneous components of ${\mathfrak F}$
in an identical manner.
On the other hand, if $w\neq 0,$ it does not seem right to assign to it multiplication by $f,$
since such a quantization procedure would map different classical functions to the
same quantum-mechanical operator. A natural guess for the operator corresponding to
$(f,w)$ is multiplication by $f$ followed by an operator $T_w,$ where
$T_w$ shifts the $\Gamma$-grading by $w.$ This guess will be justified below.

Under the standard quantization of $T^*M,$ the function $p_j$ is mapped to a
differentiation operator on ${\mathfrak F}$:
\begin{equation}\label{momentumop}
p_j\mapsto \hat{p}_j=-i\,\frac{\partial}{\partial x^j}.
\end{equation}
If $\hat{f}_w$ is the quantum operator corresponding to the function $(f,w)\in {\mathfrak F},$
we have the commutation relation
$$
[\hat{f}_w,\hat{p}_j]=i\,\widehat{\left(\frac{\partial f}{\partial x^j}\right)_w}.
$$
This should be compared with the classical relation
$$
\{f(x),p_j\}_{P.B.}=\frac{\partial f(x)}{\partial x^j}.
$$
The Fourier transform which identifies the completion of ${\mathfrak F}$ with the
completion of $\CC[\Gamma\op\Gamma^*]$
sends $\hat{p}_j$ to the following operator $M_j$ on $\CC[\Gamma\op\Gamma^*]$:
\begin{equation}\label{pFourier}
M_j:(w,m)\mapsto m_j(w,m),\qquad \forall (w,m)\in \Gamma\op\Gamma^*.
\end{equation}

Putting all this together, we obtain the quantization map for $\partial X^j,$
$\bpartial X^j,$ $\psi^j,$ and $\bpsi^j.$ It is easy to check that this yields
the expressions~(\ref{X}-\ref{bpsi}) of Section~\ref{torusSCVA} with $\B=0$
(after we rescale $X^j$ by a factor $i\sqrt 2$).

Now we can also motivate the state-operator correspondence postulated in
Section~\ref{torusSCVA}. The main idea that the quantization map should send local
classical observables to local quantum fields belonging to the image of $Y.$
For example, $\partial X^j, \bpartial X^j, \psi^j, \bpsi^j$ and their
derivatives are local classical observables, so the corresponding quantum fields must
lie in the image of $Y.$ These considerations explain the mapping of
the states $\alpha^j_{-s}\vac,$ $\bal^j_{-s}\vac,$ $\psi^j_{-r}\vac,$
and $\bpsi^j_{-r}\vac.$ Together with the axioms of vertex algebra, this uniquely
fixes the mapping of other states in the subspace $w=m=0.$ Other natural
local classical observables are suitable exponentials of $X^j(z,\bz).$ (The classical
field $X^j(z,\bz)$ itself is multi-valued and therefore should not be quantized.)
Requiring that they
map to local quantum fields fixes the form of $Y$ for all $(w,m)\in\Gamma\op\Gamma^*.$
An interested reader is referred to~\cite{Polch} for details.

Another important
ingredient is the quantization of the $N=2$ super-Virasoro algebra.
Naively, one would like to define the quantum generators by the same
formulas~(\ref{neqone},\ref{neqtwo}), but with the classical fields
replaced by the quantum fields. This idea runs into an immediate
problem since the products of quantum fields at the same point are not
well-defined. The normal ordering prescription resolves this problem
and leads to well-defined operators. One can easily check that this definition
of the generators of the $N=2$ super-Virasoro is equivalent to the one given in
Section~\ref{torusSCVA}. The operators thus
defined form an infinite-dimensional Lie super-algebra which
is a central extension of the classical $N=2$ super-Virasoro~(\ref{neqone},\ref{neqtwo}).
One can also check that the spectrum of $H=L_0+\bL_0$ is nonnegative.

It remains to explain how to include the effect of the B-field.
As remarked above, a closed B-field does not affect the classical
$\sigma$-model. However, the above quantization procedure admits a modification
which depends on a class in $H^2(M,\RR/\ZZ).$ We wish to interpret this
class as the cohomology class of the B-field.

The modification affects the quantization of $T^*M$ and consists
in replacing the space of smooth functions on $\RR^{2d}/\Gamma$
with the space of smooth functions on $\RR^{2d}$ satisfying
the following quasi-periodicity condition:
$$
f(x^1+2\pi n^1,\ldots,x^{2d}+2\pi n^{2d})=
e^{-2\pi i \B_{jk}n^j w^k}f(x^1,\ldots,x^{2d}),
$$
where $\B_{jk}$ is a real skew-symmetric matrix which we can interpret as
an element of $H^2(M,\RR)$ in a natural manner.
We will denote the space of such functions $C^\infty_w(M,\B).$ It is clear that
$C^\infty_w(M,\B)$ depends only on the image of $\B$ in $H^2(M,\RR/\ZZ).$
Thus the modification consists of replacing $\mathfrak F$ with the space
$$
{\mathfrak F}(B)=\op_{w\in\Gamma} C^\infty_w(M,\B).
$$
Fourier transform identifies a completion of $C^\infty_w(M,\B)$ with a completion of
$\CC[\Gamma^*],$ as before, so the
Hilbert space of the quantum theory is unaffected by $\B.$
But the map of the classical functions on the phase space to operators is affected.

First, the product of two quasi-periodic functions $f\in C^\infty_w(M,\B)$ and $f'\in
C^\infty_{w'}(M,\B)$ belongs to the space $C^\infty_{w+w'}(M,\B).$
Hence the multiplication operators
do not preserve the $\Gamma$-grading on ${\mathfrak F}(B).$
Rather, multiplication by $f\in C^\infty_w(M,\B)$
shifts the grading by $w.$ If we want the limit $\B\ra 0$ to be smooth, we
have to postulate that even for $\B=0$ multiplication by $f\in C^\infty_w(M,\B)$ shifts the
grading by $w.$ This provides a justification for the guess made above.
Second, while the function $p_j$ is still mapped according to~(\ref{momentumop}), the
Fourier transform of $\hat{p}_j$ is different from~(\ref{pFourier}). Namely, it is easy to
see that the Fourier transform of the differentiation operator on $C^\infty_w(M,\B)$ is given by
$M_j-\B_{jk}w^k.$ Putting these facts together, one obtains the quantization
map for all classical fields in agreement with~(\ref{X}-\ref{bpsi}).

\section{The relation between vertex algebras and chiral algebras}\label{vertexandchiral}

In this appendix we describe some properties of vertex algebras in
the sense of Definition~\ref{VAdef}. Let $(V,\vac,T,\bT,Y)$ be a
vertex algebra. We prove that the subspace of $V$ spanned by
vectors which are mapped by $Y$ to meromorphic fields has a
natural structure of a chiral algebra. Furthermore,
anti-meromorphic fields form another chiral algebra, and these two
chiral algebras supercommute with each other. We also describe an
analogue of the Borcherds (or associativity) formula for vertex
algebras. Finally, we show that any chiral algebra is a vertex
algebra.

We start with the following useful lemma.
\begin{lemma}\label{OPEuniqueness}
Let $N,M$ be integers and let $h_j, j=1,\ldots,K$ be distinct real
numbers belonging to $[0,1).$ Suppose the following relation holds
\begin{equation}\label{first}
\sum_{j=1}^K i_{z,w}\frac{1}{(z-w)^{N+h_j}}\ i_{\bz,\bw}
\frac{1}{(\bz-\bw)^{M+h_j}}\ C_j(z,\bz,w,\bw)=0,
\end{equation}
where $C_j(z,\bz,w,\bw)\in QF_2(V).$ Then $C_j(z,\bz,w,\bw)\equiv
0$ for all $j.$
\end{lemma}
It is sufficient to prove the statement for $M=N=0.$ Let $v\in V$
be an arbitrary vector. We are going to prove that the value of
$C_j$ on $v$ vanishes for all $j.$ To this end let us evaluate
both sides of~(\ref{first}) on $v$ and set $w=zx$ and
$\bw=\bz\bx.$ Since $C_j\in QF_2(V),$ the expression
$C_j(z,\bz,zx,\bz\bx)(v)$ can be written as
\begin{equation}\label{nser}
\sum_{\alpha, \beta} f_{\alpha\beta}(x,\bx)
z^{-\alpha}\bz^{-\beta},
\end{equation}
where each $f_{\alpha\beta}$ is a finite sum of fractional powers
of $x,\bx$ with coefficients in $V.$ Hence the value of the
left-hand side of Eq.~(\ref{first}) on $v$ is a sum
$$
\sum_{\alpha,\beta}z^{-\alpha}\bz^{-\beta}
\sum_{(\gamma,\delta)\in J_{\alpha\beta}}x^{-\gamma}\bx^{-\delta}
T_{\alpha\beta\gamma\delta},
$$
where $J_{\alpha\beta}\subset\RR^2$ is a finite set for each
$(\alpha,\beta).$ Each $T_{\alpha\beta\gamma\delta}$  has the form
\begin{equation}\label{pol}
\sum_{j=1}^K i_{x}\frac{1}{(1-x)^{h_j}}\ i_{\bx}
\frac{1}{(1-\bx)^{h_j}}\ f_{j}(x,\bx),
\end{equation}
where all $f_{j}$ are polynomials in $x,\bx$ with coefficients in
$V,$ and $h_j\in [0,1)$ are distinct real numbers. The symbol
$i_x$ (resp. $i_\bx$) means ``expand in a Taylor series around
$x=0$'' (resp. $\bx=0$). To prove the lemma it sufficient to show
that if the expression Eq.~(\ref{pol}) is zero, then $f_j\equiv 0$
for all $j.$ To prove this, we rewrite $f_j$ as a polynomial in
$1-x$ and $1-\bx.$ Then Eq.~(\ref{pol}) takes the form
$$
\sum_{l=1}^{L} i_{x}\frac{1}{(1-x)^{t_l}}\ i_{\bx}
\frac{1}{(1-\bx)^{s_l}}\ a_{l},
$$
where $(t_l,s_l)$ are distinct pairs of real numbers, and each
$a_l\in V$ is a coefficient of some $f_j$. Let us denote this
expression by $T.$ We will show by induction in $L$ that if $T$ is
equal to $0$ then $a_l=0$ for all $l.$ This will imply that
$f_j\equiv 0$ for all $j.$ The base of induction is evident.
Suppose $a_1\ne 0.$ Multiply $T$ by $i_{x}(1-x)^{t_1} i_x
(1-\bx)^{s_1}$ and apply to the resulting expression an operator
$$
A(1-x)\partial_x +B(1-\bx)\partial_{\bx},
$$
where $A,B$ are arbitrary real numbers. We obtain a sum with $L-1$
terms:
$$
\sum_{l=2}^{L} i_{x}\frac{1}{(1-x)^{t_l}}\ i_{\bx}
\frac{1}{(1-\bx)^{s_l}}\ (A(t_l-t_1)+B(s_l-s_1)) a_{l}
$$
which is equal to $0$ whenever $T=0.$ Since $A$ and $B$ are
arbitrary, by the induction hypotesis we get $a_l=0$ for
$l=2,\ldots,L.$ Consequently, $a_1$ is  equal to $0$ as well. This
proves the lemma.

\begin{theorem}{\rm (Uniqueness theorem)}
Let ${\mathcal V}$ be a subspace in $QF_1(V)$ which satisfies the
following conditions:
\begin{enumerate}
\item any field $A(z, \bar{z})\in {\mathcal V}$ is mutually local
with all fields $Y(a), a\in V$;
\item all fields are creative, i.e.
$A(z, \bar{z})|0\rangle\in V[[z, \bar{z}]]$.
\end{enumerate}
Then the map
$$
\begin{array}{l}
s: {\mathcal V}\to V[[z, \bar{z}]],\\
A(z, \bar{z})\mapsto A(z, \bar{z})|0\rangle
\end{array}
$$
is injective.
\end{theorem}
Suppose $A(z, \bar{z})|0\rangle=0$. Take a vector $a\in V$ and
consider $Y(a)$. From locality we know that
$$
Y(a)(z, \bar{z})A(w, \bar{w})=\sum_{j=1}^M
i_{z,w}\frac{1}{(z-w)^{h_j+N}}\ i_{\bar{z},\bar{w}}
\frac{1}{(\bar{z}-\bar{w})^{h_j+N}}\ C_j(z,\bar{z},w,\bar{w}).
$$
Hence we have
$$
\sum_{j=1}^M i_{z,w}\frac{1}{(z-w)^{h_j+N}}\ i_{\bar{z},\bar{w}}
\frac{1}{(\bar{z}-\bar{w})^{h_j+N}}\
C_j(z,\bar{z},w,\bar{w})|0\rangle=0.
$$
Using the arguments of Lemma B.1, we get 
$C_j(z,\bar{z},w,\bar{w})|0\rangle=0$ for all $j$. Now from
locality we obtain
$$
A(w, \bar{w})Y(a)(z, \bar{z})|0\rangle=0.
$$
This implies that $A(w, \bar{w})a=0$ for any $a\in V$. Hence $A(w,
\bar{w})=0,$ and the theorem is proved.

\begin{cor}\label{corfirst}
For any $a\in V$ the following identities hold:
$$
Y(Ta)=\partial Y(a),\qquad Y(\bar{T}a)=\bar{\partial}Y(a).
$$
\end{cor}
Both fields $Y(Ta)$ and $\partial Y(a)$ are mutually local with
all fields $Y(b)$. Moreover we have
$$
Y(Ta)|0\rangle=\partial Y(a)|0\rangle=T e^{Tz+ \bar{T}\bar{z}}a.
$$
Hence by the uniqueness theorem
$$
Y(Ta)=\partial Y(a).
$$
The other identity is proved similarly.

We call a vector $a\in V$ meromorphic (resp. anti-meromorphic) if $Y(a)$
is meromorphic (resp. anti-meromorphic).
To show that meromorphic and anti-meromorphic vectors form two
supercommuting chiral algebras, it is sufficient to prove the
following proposition.
\begin{proposition}\label{propspecial}
Let $V$ be a vertex algebra.
Then
\begin{enumerate}
\item the subspace of meromorphic vectors is closed with respect to $Y$
and $T$, i.e. $T(a)$ and $a_{(n)}b$ are meromorphic when
$a\in V$ and $b\in V$ are meromorphic,
\item the OPE of two meromorphic fields $a(z)$ and $b(w)$
can be
written in the form
\begin{align*}
a(z)b(w)=i_{z,w}\frac{1}{ (z-w)^{N}}\ C(z,w),&\quad\\
(-1)^{p(a)p(b)}b(w)a(z)=i_{w,z}\frac{1}{ (z-w)^{N}}\ C(z,w),&
\qquad C(z,w)\in QF_2(V),
\end{align*}
where $N$ is an integer,
\item
If $a\in V$ is meromorphic and $b\in V$ is
anti-meromorphic, then their OPE has the form
$$
a(z)b(\bw)=C(z,\bw),\quad
(-1)^{p(a)p(b)}b(\bw)a(z)=C(z,\bw),\qquad C(z,\bw)\in QF_2(V).
$$
\end{enumerate}
\end{proposition}
Let us prove statement (1) of the proposition.
From Corollary~\ref{corfirst} we infer that $a$ is meromorphic if
and only if $\bar{T}a=0$. Since $T$ and $\bar{T}$ commute, this
immediately implies that $Ta$ is meromorphic when $a$ is meromorphic.
Further, consider $Y(a)b$, where both $a$ and $b$ are
meromorphic. We have
$$
\bar{T}Y(a)b=Y(a)(\bar{T}b)=0.
$$
Hence  $\bar{T}(a_{(n)}b)=0,$ and all $a_{(n)}b$ are meromorphic as
well.

Statements (2) and (3) of the proposition
are special cases of a more general
statement which we are going to prove.
\begin{proposition} Let $a,b\in V.$ If $a$ is meromorphic,
then the OPE of $a(z)$ and $b(w,\bw)$ can be written in the form
\begin{align}\label{want}
a(z)b(w,\bw)=i_{z,w}\frac{1}{ (z-w)^{N}}\ D(z,w,\bw),&\quad\\
(-1)^{p(a)p(b)}b(w,\bw)a(z)=i_{w,z}\frac{1}{ (z-w)^{N}}\
D(z,w,\bw), &\quad
        \nonumber
\end{align}
where $D(z,w,\bw)\in QF_2(V),$ and $N$ is an integer.
\end{proposition}
This means that if a certain variable does not appear on the
left-hand-side of the OPE, it does not appear on the
right-hand-side either.

The general form of the OPE of $a(z)$ and $b(w,\bw)$ is
$$
a(z)b(w,\bw)= \sum_{i=1}^M i_{z,w}\frac{1}{(z-w)^{N+h_i}}\
i_{\bz,\bw} \frac{1}{(\bz-\bw)^{N+h_i}}\ C_i(z,\bz,w,\bw),
$$
where $N\in \ZZ,$ $h_i, i=1,\ldots,M,$ are distinct real numbers
which belong to $[0,1),$ and $C_i\in QF_2(V).$

Let us act on both sides with an operator
$(\bz-\bw)\frac{\partial}{\partial \bz}.$ We get
$$
0=\sum_{i=1}^M i_{z,w}\frac{1}{(z-w)^{N+h_i}}\ i_{\bz,\bw}
\frac{1}{(\bz-\bw)^{N+h_i}}
\left(-(N+h_i)+(\bz-\bw)\frac{\partial}{\partial\bz}\right)C_i.
$$
By Lemma~\ref{OPEuniqueness} we may conclude that for all $i$ we
have
\begin{equation}\label{difeq}
\left(-(N+h_i)+(\bz-\bw)\frac{\partial}{\partial\bz}\right)C_i=0.
\end{equation}
Now let us show that $C_i\equiv 0$ if $h_i\ne 0.$ Assume the
converse. Then there is a vector $v\in V$ such that
$$
C_i(z,\bz,w,\bw)(v)=\sum_{\alpha,\beta,\gamma,\delta}
c_{(\alpha\beta\gamma\delta)}z^{-\alpha}\bz^{-\beta}
w^{-\gamma}\bw^{-\delta}\ne 0.
$$
Eq.~(\ref{difeq}) implies
\begin{equation}\label{difeq1}
(N+h_i+\beta)c_{(\alpha\beta\gamma\delta)}=
(\beta-1)c_{(\alpha,\beta-1,\gamma,\delta+1)}
\end{equation}
Since $C_i\in QF_2(V),$ we can choose $\alpha,\beta,\gamma,\delta$
so that $c_{(\alpha,\beta,\gamma,\delta)}\ne 0$ and
$c_{(\alpha,\beta-1,\gamma,\delta+1)}=0.$ {}From
Eq.~(\ref{difeq1}) we find that $\beta=-(N+h_i).$ Furthermore,
(\ref{difeq1}) implies that
$$
c_{(\alpha,\beta+k,\gamma,\delta-k)}= \binom{\beta+k-1}{k}
c_{(\alpha,\beta,\gamma,\delta)}=\binom{-(N+h_i)+k-1}{k}
c_{(\alpha,\beta,\gamma,\delta)}
$$
for all $k\in \NN.$ If $h_i\not\in \ZZ$ then the vector
$c_{(\alpha,\beta+k,\gamma,\delta-k)}\in V$ is nonzero for all
$k\in \NN.$ But this contradicts the condition $C_i\in QF_2(V).$

Since $h_i\in [0,1)$ for all $i,$ and $h_i\neq h_j$ for $i\neq j,$
we conclude that $C_i=0$ for all $i$ except maybe one, and for the
latter value of $i$ we have $h_i=0.$ In addition, for
$c_{(\alpha,\beta+k,\gamma,\delta-k)}$ to be zero for $k>>0,$ as
required by the condition $C_i\in QF_2(V),$ the integer $N$ must
be nonnegative. Thus the OPE of $a(z)$ and $b(w,\bw)$ has the form
$$
a(z)b(w,\bw)=
 i_{z,w}\frac{1}{(z-w)^{N}}\ i_{\bz,\bw}
\frac{1}{(\bz-\bw)^{N}}\ C(z,\bz,w,\bw),
$$
where $C(z,\bz,w,\bw)\in QF_2(V)$ and $N\geq 0.$

Applying Eq.~(\ref{difeq}) to $C(z,\bz,w,\bw)$ and differentiating
it with respect to $\bz,$ we infer that
$$
C(z,\bz,w,\bw)=\frac{1}{N!}(\bz-\bw)^N \partial_{\bz}^N
C(z,\bz,w,\bw) \qquad \text{and} \qquad
\partial_{\bz}^{N+1} C(z,\bz,w,\bw)=0.
$$
For this reason the element $\frac{1}{N!}\partial_{\bz}^{N}
C(z,\bz,w,\bw)\in QF_2(V)$ does not depend on $\bz.$ Let us denote
it by $D(z,w,\bw).$ Then the OPE of $a(z)$ and $b(w,\bw)$ takes
the form
\begin{align}
a(z)b(w,\bw)& =i_{z,w} \frac{1}{(z-w)^{N}}\ D(z,w,\bw),\nn \\
(-1)^{p(a)p(b)}b(w,\bw)a(z)& =i_{w,z}\frac{1}{(z-w)^{N}}\
D(z,w,\bw).\nn
\end{align}
This completes the proof of Proposition~\ref{propspecial}.
As a corollary, we have:
\begin{cor}
Meromorphic and anti-meromorphic vectors form two supercommuting
chiral algebras.
\end{cor}

In the theory of chiral algebras an important role is played by
the so-called Borcherds formula which expresses the OPE of any two
fields $a(z)$ and $b(z)$ in the image of $Y$ through their normal
ordered product and the Borcherds products $a_{(n)}b.$ We will prove
an analogue of the Borcherds formula for vertex
algebras.

Note that any field $D(z,w,\bw)\in QF_2(V)$ meromorphic in the
first variable can be expanded in a Taylor series in $(z-w)$ to an
arbitrarily high order. This means that for any integer $K>0$
there exists a field $D_K(z,w,\bw)\in QF_2(V)$ such that
$$
D(z,w,\bw)=\sum_{j=0}^{K-1} \left.\frac{(z-w)^j}{j!}
\frac{\partial^j D(z,w,\bw)}{\partial z^j}\right|_{z=w}+ (z-w)^K
D_K(z,w,\bw).
$$
To prove this, it is sufficient to show that for any
$D(z,w,\bw)\in QF_2(V)$ we have
$$
D(z,w,\bw)-D(w,w,\bw)=(z-w) D_1(z,w,\bw)
$$
for some $D_1(z,w,\bw)\in QF_2(V).$ This fact is trivial. Note
also that if $D\in QF_2(V)$ contains fractional powers of $z$ (and
therefore also depends on $\bz$), the Taylor formula need not
hold.

Using the Taylor formula, the OPE~(\ref{want}) can be rewritten in
the following form
$$
a(z)b(w,\bw)=\sum_{j=1}^{N} i_{z,w}\frac{1}{ (z-w)^{j}}\
C_j(w,\bw)+ D_N(z,w,\bw),
$$
where $C_j(w,\bw)\in QF_1(V)$ for all $j,$ $D_N(z,w,\bw)\in
QF_2(V).$ It is easy to see that $C_j$ and $D_N$ are uniquely
defined by this formula.

Moreover it can be easily checked that $C_n(w, \bar{w})$ coincides
with
$$
a(w)_{(n)}b(w, \bar{w}):=
Res_{z}((z-w)^{n-1} (a(z)b(w, \bar{w})- b(w, \bar{w})a(z)))
$$

The analogue of the Borcherds formula provides explicit
expressions for $C_j$ and $D_N$ in terms of $a$ and $b$:
\begin{align}\label{borch}
C_j(w,\bw)=Y\left(a_{(j)}b\right)(w,\bw),& \quad j=1,\ldots,N,
&\qquad D_N(z,w,\bw)=:a(z)b(w,\bw):.
\end{align}
Here the normal ordered product $:a(z)b(w,\bw):\in QF_2(V)$ is
defined as follows. Let
$$
a(z)_+=\sum_{n\leq 0} a_{(n)} z^{-n},\quad a(z)_-=\sum_{n>0}
a_{(n)} z^{-n}.
$$
Then the normal ordered product of $a(z)$ and $b(w,\bw)$ is
defined by
$$
:a(z)b(w,\bw):=a(z)_+ b(w,\bw)+(-1)^{p(a)p(b)}b(w,\bw) a(z)_-.
$$
Thus the OPE of a meromorphic field and an arbitrary field takes
the form
\begin{equation}\label{borchone}
a(z)b(w,\bw)=\sum_{j=1}^N i_{z,w}\frac{1}{(z-w)^{j}}\
Y\left(a_{(j)}b\right)(w,\bw)+ :a(z)b(w,\bw):.
\end{equation}
Similarly, the OPE of an anti-meromorphic field and an arbitrary
field is given by
\begin{equation}\label{borchtwo}
a(\bz)b(w,\bw)=\sum_{j=1}^N i_{\bz,\bw}\frac{1}{(\bz-\bw)^{j}}\
Y\left(a_{(j)}b\right)(w,\bw)+ :a(\bz)b(w,\bw):.
\end{equation}

To prove the analogue of the Borcherds formula it is sufficient to show that
$a(w)_{(n)}b(w, \bar{w})$ is mutually local with any $Y(c)$.
Indeed, it can be easily checked that
$$
a(w)_{(n)}b(w, \bar{w})|0\rangle= Y(a_{(n)}b)(w, \bar{w})|0\rangle,
$$
and hence by the uniqueness theorem we obtain
$$
a(w)_{(n)}b(w, \bar{w})= Y(a_{(n)}b)(w, \bar{w}).
$$

\begin{lemma}
If $a\in V$ is meromorphic, then
$a(z)_{(n)}b(z, \bar{z}), n\ge 1$ is mutually local with any $Y(c)$.
\end{lemma}

We have to prove that
$$
a(w)_{(n)}b(w, \bar{w})=
Res_{z}((z-w)^{n-1} (a(z)b(w, \bar{w})- b(w, \bar{w})a(z)))
$$
is mutually local with any $Y(c)=c(z, \bar{z})$.

Let us consider
$$
A=(z_1 -z_2)^{n-1}( a(z_1)b(z_2, \bar{z}_2)c(z_3, \bar{z}_3)-
b(z_2, \bar{z}_2) a(z_1)c(z_3, \bar{z}_3))
$$
and
$$
B=(z_1 -z_2)^{n-1}( c(z_3, \bar{z}_3)a(z_1)b(z_2, \bar{z}_2)-
c(z_3, \bar{z}_3)b(z_2, \bar{z}_2) a(z_1)).
$$

We know that for some sufficiently large $r\in\NN$ the following identities hold:
$$
\begin{array}{l}
(z_1 -z_2)^r a(z_1)b(z_2, \bar{z}_2)=
(z_1-z_2)^r b(z_2, \bar{z}_2) a(z_1),\\
(z_1-z_3)^r a(z_1)c(z_3, \bar{z}_3)=
(z_1 -z_3)^r c(z_3, \bar{z}_3)a(z_1).
\end{array}
$$
Now let us consider $(z_2-z_3)^M$. We have
$$
(z_2 -z_3)^M=\sum_{s=0}^{M} \binom{M}{s}(z_2 -z_1)^{M-r}(z_1 - z_3)^s.
$$
Let us multiply $A$ with $(z_2-z_3)^M$, where $M\ge 2r$. We get
$$
\sum_{s=0}^{M} \binom{M}{s}(z_2 -z_1)^{M-r}(z_1 - z_3)^s A.
$$
For $0\le s\le r$ the s-th summand in this expression is $0$,
because $(z_1-z_2)^{M-s} (z_1 -z_2)^{n-1} =(z_1 -z_2)^{r'}$ where
$r'\ge r$.
Hence the expression is equal to
$$
\sum_{s=r+1}^{M} \binom{M}{s}(z_2 -z_1)^{M-r}(z_1 - z_3)^s A=\\
$$
$$
\sum_{s=r+1}^{M} \binom{M}{s}(z_2 -z_1)^{M-r}(z_1 - z_3)^s
(z_1 -z_2)^{n-1}( a(z_1)b(z_2, \bar{z}_2)c(z_3, \bar{z}_3)-
b(z_2, \bar{z}_2) a(z_1)c(z_3, \bar{z}_3))=\\
$$
$$
\sum_{s=r+1}^{M} \binom{M}{s}(z_2 -z_1)^{M-r}(z_1 - z_3)^s
(z_1 -z_2)^{n-1}( a(z_1)b(z_2, \bar{z}_2)c(z_3, \bar{z}_3)-
b(z_2, \bar{z}_2) c(z_3, \bar{z}_3)a(z_1))=\\
$$
$$
\sum_{s=r+1}^{M} \binom{M}{s}(z_2 -z_1)^{M-r}(z_1 - z_3)^s
(z_1-z_2)^{n-1}[a(z_1), b(z_2, \bar{z}_2)c(z_3, \bar{z}_3)].
$$
In the same way we find that
$$
(z_2 -z_3)^M B=
\sum_{s=r+1}^{M} \binom{M}{s}(z_2 -z_1)^{M-r}(z_1 - z_3)^s
[a(z_1), c(z_3, \bar{z}_3)b(z_2, \bar{z}_2)].
$$
{}From our definition of a vertex algebra we know that
\begin{align*}\label{OPEaxiomone}
b(z_2,\bar{z}_2)c(z_3,\bar{z}_3)=
\sum_{j} i_{z_2, z_3}\frac{1}{(z_2-z_3)^{h_j+N}}\
i_{\bar{z}_2,\bar{z}_3}
\frac{1}{(\bar{z}_2-\bar{z}_3)^{h_j+N}}\ E_j(z_2,\bar{z}_2,z_3,\bar{z_3}),\\
c(z_3,\bar{z}_3)b(z_2,\bar{z}_2)=\sum_{j} i_{z_3,z_2}\frac{1}{(z_2-z_3)^{h_j+N}}\
i_{\bar{z}_3,\bar{z}_2}\frac{1}{(\bar{z}_2-\bar{z}_3)^{h_j+N}}\
E_j(z_2,\bar{z}_2,z_3,\bar{z}_3)
\end{align*}
for some $E_j$ from $QF_2(V).$
Substituting these expressions into the formulas above we find that
\begin{multline*}
(z_2-z_3)^M (a(z_2)_{(n)}b(z_2, \bar{z}_2)) c(z_3, \bar{z}_3)=
Res_{z_1}(\sum_{s=r+1}^{M} \binom{M}{s}(z_2 -z_1)^{M-r}(z_1 - z_3)^s
(z_1-z_2)^{n-1}\\
\sum_{j} i_{z_2, z_3}\frac{1}{(z_2-z_3)^{h_j+N}}\
i_{\bar{z}_2,\bar{z}_3}
\frac{1}{(\bar{z}_2-\bar{z}_3)^{h_j+N}}\
[a(z_1),E_j(z_2,\bar{z}_2,z_3,\bar{z_3})]),
\end{multline*}
and
\begin{multline*}
(z_2-z_3)^M c(z_3, \bar{z}_3)a(z_2)_{(n)}b(z_2, \bar{z}_2)=
Res_{z_1}(\sum_{s=r+1}^{M} \binom{M}{s}(z_2 -z_1)^{M-r}(z_1 - z_3)^s
(z_1-z_2)^{n-1}\\
\sum_{j} i_{z_3,z_2}\frac{1}{(z_2-z_3)^{h_j+N}}\
i_{\bar{z}_3,\bar{z}_2}\frac{1}{(\bar{z}_2-\bar{z}_3)^{h_j+N}}\
[a(z_1),E_j(z_2,\bar{z}_2,z_3,\bar{z_3})].
\end{multline*}
To prove mutual locality of $a(z)_{(n)}b(z, \bar{z})$
with any $Y(c)$ one only needs to show that one can divide both
sides of the above equations by $(z_2-z_3)^M$. In fact, it is sufficient
to show this for $M=1,$ and then use induction on $M.$ 

To show
that one can divide both sides by $z_2-z_3,$ we note that the kernel
of multiplication by $z-w$ consists of expressions of the form 
$$
\sum_{n\in\ZZ} \left(\frac{z}{w}\right)^n D(z,\bz,w,\bw),
$$
where $D(z,\bz,w,\bw)$ is a formal fractional power series with coefficients
in $End(V)$ (but not necessarily an element of $QF_2(V)$).
If $D(z,\bz,w,\bw)$ is not identically zero, then there exists $v\in V$ such that when this expression is applied to $v,$ one gets a fractional power
series with coefficients in $V$ containing arbitrarily large negative powers
of $w$ and $z.$ 
On the other hand, applying any element of $QF_1(V)$ or $QF_2(V)$
to any $v\in V$ one always obtains a fractional power series with powers
bounded from below. This implies that one can divide both sides of
the above equations by $z_2-z_3.$  The Borcherds formulas are proven.

Three remarks are in order here. First, it seems that there is no
analogous way to rewrite the OPE of two fields when neither of
them is meromorphic or anti-meromorphic. Consequently, the normal
ordered product of two general fields is not a very useful
concept.

Second, given two meromorphic fields, one can define {\it two}
normal ordered products:
\begin{align*}
:a(z)b(w): & =a(z)_+b(w)+(-1)^{p(a)p(b)}b(w)a(z)_-, \\
:b(w)a(z): & =b(w)_+a(z)+(-1)^{p(a)p(b)}a(z)b(w)_- .
\end{align*}
Correspondingly, there are two different OPEs that one can write
down. The first one is
\begin{align*}
a(z)b(w)&=\sum_{j=1}^N i_{z,w}\frac{1}{(z-w)^{j}}\
Y\left(a_{(j)}b\right)(w)+
:a(z)b(w):,\\
(-1)^{p(a)p(b)}b(w)a(z)&=\sum_{j=1}^N i_{w,z}\frac{1}{(z-w)^{j}}\
Y\left(a_{(j)}b\right)(w)+:a(z)b(w):,
\end{align*}
and the second one is
\begin{align*}
b(w)a(z)&=\sum_{j=1}^N i_{w,z}\frac{1}{(w-z)^{j}}\
Y\left(b_{(j)}a\right)(z)+
:b(w)a(z):,\\
(-1)^{p(a)p(b)}a(z)b(w)&=\sum_{j=1}^N i_{z,w}\frac{1}{(w-z)^{j}}\
Y\left(b_{(j)}a\right)(z)+:b(w)a(z):.
\end{align*}
In general, the two normal ordered products are not related in any
simple way.

Third, given a meromorphic and an anti-meromorphic field, one can
also define two normal ordered products. However, in this case
they always coincide up to a sign:
$$
:a(z)b(\bw):=(-1)^{p(a)p(b)}:b(\bw)a(z):.
$$
Indeed, the OPE formulas~(\ref{borchone},\ref{borchtwo}) read in
this case
\begin{align*}
a(z)b(\bw)&=(-1)^{p(a)p(b)}b(\bw)a(z)=:a(z)b(\bw):,\\
b(\bw)a(z)&=(-1)^{p(a)p(b)}a(z)b(\bw)=:b(\bw)a(z):.
\end{align*}
This fact also follows directly from the definition of the normal
ordered product and the fact that meromorphic and anti-meromorphic
fields in the image of $Y$ supercommute.

Finally, let us show that any chiral algebra is a special case of
a vertex algebra with $\bar{T}=0$ and the image of $Y$ consisting
of meromorphic fields only. The only thing which needs to be
checked is the OPE axiom. For a chiral algebra, the OPE of any two
fields in the image of $Y$ has the form
\begin{align}\nn
a(z)b(w)&=\sum_{n=1}^N i_{z,w}\frac{1}{(z-w)^{n}}\
Y\left(a_{(n)}b\right)(w)+ :a(z)b(w):,\\ \nn
(-1)^{p(a)p(b)}b(w)a(z)&=\sum_{n=1}^N i_{w,z}\frac{1}{(z-w)^{n}}\
Y\left(a_{(n)}b\right)(w)+:a(z)b(w):
\end{align}
Obviously, $a_{(n)}b(w)$ belongs to $QF_2(V).$ It is also easy to
check that $:a(z)b(w):$ also belongs to $QF_2(V).$ Hence, the
above OPE can be rewritten as
$$
a(z)b(w)=i_{z,w}\frac{1}{(z-w)^{N}}\ C(z,w)
$$
where $C(z,w)\in QF_2(V).$ Therefore the OPE axiom is satisfied.

\section{Projectively flat connections and the fundamental group}\label{projflat}

In this appendix we establish a relation between projectively flat connections on
complex vector bundles on a connected manifold and finite representations of a
twisted group algebra of the fundamental group.
This relation is a generalization of the well-known statement that flat connections on
complex vector bundles are in one-to-one correspondence with representations of the
fundamental group.

Let $M$ be a paracompact connected $C^{\infty}$\!-manifold.
Let us fix a closed real 2-form $B$ on $M.$
Consider a complex vector bundle $E$ on $M$ with a connection $\nabla$
such that its curvature
$F_{\nabla}\in \Omega^2\otimes \End(E)$ is equal to
\begin{equation}\label{prflat}
F_\nabla=2\pi i B\otimes id_{E}
\end{equation}
Such a connection is called projectively flat, and it is flat
if and only if $B=0.$
When $B$ is non-zero, we can consider the condition
(\ref{prflat}) as a "twisted" variant of the flatness condition.

We will prove that the set of such connections is in one-to-one correspondence
with finite representations of a twisted group algebra of $\pi_1(M)$
defined below.

Let us fix a point $x\in M.$
Since $(E, \nabla)$ is projectively flat,
for any contractible closed  path $c$ starting at $x$ the holonomy
operator $H_c: E_{x}\lto E_x$ is equal to $t_c \cdot id,$
where $t_c$ is a nonzero complex number.
By the Reduction Theorem (see \cite{KN})
$(E,\nabla)$
can be reduced locally to a $\CC^*$\!--bundle, and therefore
by Stockes' theorem
$$
t_c=\exp(2\pi i\int_D \phi^*{B}),
$$
where $\phi$ is a map from the two dimensional disk $D$ to $M$
satisfying $\phi(\partial D)=c.$ Since $B$ is a real 2-form,
$(E,\nabla)$ in fact locally reduces to a $U(1)$\!-bundle.

The above formula for $t_c$ is independent of the choice of $\phi$
only if
\begin{equation}\label{triv}
\exp(2\pi i\int_{S^2} \phi^* {B})=1
\end{equation}
for any map $\phi$ from the 2-dimensional sphere $S^2$ to $M.$
Thus a vector bundle $(E, \nabla)$ with curvature
$F_{\nabla}=2\pi i B\otimes id_{E}$
can exist only
if the de Rham cohomology class of ${B}$ belongs to the kernel of
the composition homomorphism
$$
H^2(M,\RR)\ra H^2(M, U(1))\ra \Hom(\pi_2(M), U(1)).
$$

Let us consider the Hopf sequence
$$
\pi_2(M)\lto H_2(M, \ZZ)\lto H_2(K(G,1), \ZZ)\lto 0,
$$
where $G:=\pi_1(M).$
This sequence induces an injective map
\begin{equation}\label{cocycle}
0\lto H^2(K(G,1), U(1))\lto H^2(M, U(1)).
\end{equation}
Denote by $\cB$ the image of $B$ in $H^2(M, U(1)).$
We showed that if $\cB$ does not belong to the image of the  map (\ref{cocycle})
then the set of vector bundles $(E, \nabla)$ with curvature
$F_{\nabla}=2\pi i B\otimes id_{E}$ is empty.

Assume now that $\cB$ is in the image of the map (\ref{cocycle}).
Let us fix a point $x\in M$ and for each element $g\in G$
choose a closed path $c_g$ beginning at $x$ and representing
$g$ such that the closed path
$c_{g^{-1}}$ coincides with the inverse of $c_{g}$ for any $g.$
Let $c_{(g,h)}$ be a loop which is the union of
the loops $c_h, c_g,$ and $c_{(gh)^{-1}}$
This loop is contractible.
Define a function $\psi: G\times G\to U(1)$
by the rule
\begin{equation}\label{psic}
\psi(g,h)=\exp(2\pi i\int_D  \phi^*{B}),
\end{equation}
where $\phi$ is a map from the two dimensional disc $D$ to $M$
satisfying $\phi(\partial D)=c_{(g,h)}.$
It is easy to see that this function is a 2-cocycle on the group $G.$
Moreover, if we choose the representatives $c_g$ differently, we obtain
a
cocycle which is cohomologous to $\psi.$

The holonomy operators along the loops $c_g, c_h,$ and $c_{gh}$
satisfy the following relation
$$
H_{c_g}\cdot H_{c_h}=\psi(g,h)H_{c_{gh}}.
$$
This identity has the following representation-theoretic meaning.
With any 2-cocycle $\psi$ one can associate a twisted group algebra
$\CC_{\psi}[G],$ which is a vector space generated by the elements
$g\in G$ with the following multiplication law:
$$
g\cdot h=\psi(g,h)gh
$$
(Note that if two 2-cocycles are cohomological to each other, then
the corresponding twisted group algebras are isomorphic.)
The holonomy operators $H_{c_g}$ define a representation of
the twisted group algebra $\CC_{\psi}[G]$ on the vector space
$E_x.$

An equivalent definition of the algebra $\CC_{\psi}[G]$ goes as follows.
Let $Lp_x$ be the loop space of $M$ with the well-known
composition of loops (which is associative only up to a homotopy).
Let us consider the corresponding non-associative ``group''
algebra $\CC[Lp_x].$ Then the algebra $\CC_{\psi}[G]$ is a
factor-algebra
of $\CC[Lp_x]$ modulo all relations of the form
$$
c - \exp(2\pi i \int_D \phi^*B)\cdot 1 =0
$$
where $c$ is a contractible loop, and $\phi$ is a map from the disc $D$ to
$M$
such that $\phi(\partial D)=c.$ By (\ref{triv}) this definition does not
depend on the choice of $\phi.$ For any loop $c\in Lp_x$ we denote by
$r(c)$ the element of the twisted group algebra which is the image of
$c$ with respect to this factorization.

In this way to any vector bundle $(E,\nabla)$ satisfying the condition
(\ref{prflat}) we can associate a finite-dimensional
representation of the twisted group algebra.
We assert that this is a one-to-one correspondence.
To show this, we describe how to construct $(E,\nabla)$
starting from a representation $R$ of the twisted group algebra.

Let $C_{M}$ be the sheaf of algebras of complex-valued
$C^{\infty}$\!--functions on $M.$
Let $\cA$ be a sheaf of algebras on $M$ defined as
$\CC_{\psi}[G]\ot_{\CC} C_M.$
If $R$ is a representation of the twisted group algebra,
then the sheaf $\cR=R\ot_{\CC} C_M$
has a natural left module structure over the sheaf of algebras
$\cA.$
Below we construct a sheaf $\cP$ of right $\cA$\!--modules
with a connection $\nabla_{\cP}$ and set $E=\cP\ot_{\cA} \cR.$
This sheaf is the sheaf of sections of a complex vector bundle on $M,$
and $\nabla_P$ induces a natural connection $\nabla$ on it.

Let $\wt{M}\stackrel{\tau}{\lto}M$ be a universal covering.
Denote by $\wt{B}$ the pull-back of the form $B$ to $\wt{M}.$
It is easy to check that $B$ belongs to
the image of the map (\ref{cocycle}) if and only if
$\wt{B}$ is an exact form.
Let us choose a 1-form $\eta$ on $\wt{M}$ such that
$d\eta=\wt{B}.$

Consider a sheaf of algebras
$\wt{\cA}=\CC_{\psi}[G]\ot_{\CC} C_{\wt{M}}$ on $\wt{M}.$ The
tautological
action of $G$ on $\wt{M}$ can be lifted to a left action on $\wt{\cA}$
as follows.
Let $c_g$ be a loop in $M$ based at a fixed point $x\in M$ and
representing
the element $g\in G,$ and let $r(c_g)$ be the corresponding element of
the twisted group algebra of $G$ (see above).
Let $x_0$ be a lift of $x$ to $\wt{M}.$
Let $\wt{c}_{g}$ be a path on $\wt{M}$ which covers $c_g,$ begins
at $g^{-1}(x_0)$ and ends at $x_0.$
For any point $y\in \wt{M}$ let us choose some path $d_y$ from $y$ to
$x_0.$ Let $\wt{c}_{g,y}$ be a path from $g^{-1}(y)$ to $y$
which is a composition
of $g^{-1}(d_y),$ $\wt{c}_g,$ and $d_y^{-1}.$
The left action of the group
$G$ on the sheaf $\wt{\cA}$ is defined by the rule:
$$
g( a\ot f)(y)=\exp(-2\pi i \int_{\wt{c}_{g,y}}\eta)
( r(c_{g})a\ot f(g^{-1}y)),
$$
where $a\in \CC_{\psi}[G]$ and $f$ is a $C^\infty$\!--function
on $\wt{M}.$

This definition does not depend on the choice
of $d_y,$ because the form $\wt{B}$ is $G$-invariant.
Nor does it depend on the choice of $c_g,$ because
for any other loop $c'_g$ representing $g$ we have
$$
\exp(-2\pi i \int_{\wt{c}'_{g,y}}\eta)r(c'_{g})=
\exp(-2\pi i \int_{\wt{c}'_{g,y}}\eta + 2\pi i \int_D
\phi^* \wt{B}) r(c_{g})=\exp(-2\pi i \int_{\wt{c}_{g,y}}\eta) r(c_{g}),
$$
where $\phi$ is a map from $D$ to $\wt{M}$ such that
$\phi(\partial D)$ is the composition of $\wt{c}'_{g}$
and the inverse of $\wt{c}_{g}.$

Furthermore, we can define a connection on $\wt{\cA}$ by the formula
$$
\wt{\nabla}(a\ot f)= a\ot (df+2 \pi i f\eta ).
$$
This connection is $G$-invariant.
Indeed, let us regard $\int_{\wt{c}_{g,y}}\eta$ as a function
on $\wt{M}$ and denote it by $h(y).$ Then we have
\begin{align*}
g\wt{\nabla}(a\ot f)(y)&= g(a\ot (df + 2\pi i f\eta))(y)\\
&=
\exp(-2\pi i h(y))
r(c_g)a\ot (df(g^{-1}y)+ 2\pi i f(g^{-1}y)\eta(g^{-1}y)).
\end{align*}
On the other hand, since $dh(y)= \eta(y)- \eta(g^{-1}y)$ we obtain
\begin{align*}
\wt{\nabla}g(a\ot f)(y)&=
\wt{\nabla}(r(c_g)a\ot \exp(-2\pi i h(y))f(g^{-1}y))\\
&=\exp(-2 \pi i h(y))r(c_g)a\ot ( df(g^{-1}y) - 2\pi i f(g^{-1}y)dh(y) +
2\pi i f(g^{-1}y)\eta(y))\\
&=
\exp(-2\pi i h(y))
r(c_g)a\ot (df(g^{-1}y)+ 2\pi i f(g^{-1}y)\eta(g^{-1}y)).
\end{align*}

The definitions of the connection $\wt{\nabla}$ and
the action of the group $G$ on $\wt{\cA}$ depend on the choice of $\eta.$
However, if we take another form $\eta'=\eta +df$
then the data $(\wt{\cA}, \wt{\nabla})$ and $(\wt{\cA}, \wt{\nabla}')$
are isomorphic
under the multiplication by the function $\exp(-2\pi i f).$
Moreover, this isomorphism is compatible with the action of the group $G.$

We define a sheaf $\cP$ on $M$ as
the sheaf of invariants $\tau_*(\wt{\cA})^{G}$ with a
connection $\nabla_{\cP}$ induced by $\wt{\nabla}.$

The sheaf $\cP$ has a right module structure over $\cA.$
It is locally free of rank 1 as an $\cA$\!-module.
It follows from the preceding discussion that the datum $(\cP, \nabla_{\cP})$
is unique and depends only on the form $B.$

To any representation $R$ of the twisted group algebra of $G$
we attach a complex vector bundle
$E=\cP\ot_{\cA}\cR$ with the connection $\nabla$ induced
by $\nabla_{\cP}.$
It is easy to see that the representation of the twisted group algebra
on the space $E_x$ corresponding to $\nabla$ is isomorphic to $R.$ Thus pairs $(E,\nabla)$
satisfying~(\ref{prflat}) are in one-to-one correspondence with
finite-dimensional representations of $\CC_\psi[G],$ where the cocycle
$\psi$ is defined by~(\ref{psic}).

\section*{Acknowledgements}
We are grateful to Maxim Kontsevich for valuable comments and to Markus
Rosellen for pointing out a gap in the reasoning of Appendix B in the
first version of the paper.
We also wish to thank the Institute for Advanced Study, Princeton, NJ,
for a very stimulating atmosphere.
The first author was supported by DOE grant DE-FG02-90ER40542.
The second author was supported in part by RFFI grant 99-01-01144 and
a grant for support of leading scientific groups N 00-15-96085.
The research described in this publication was made possible in part
by Award No RM1-2089 of the U.S. Civilian Research and Development
Foundation for the Independent States of the Former Soviet Union (CRDF).

\end{document}